\documentclass[11pt]{article}
\usepackage{kotex}
\usepackage{enumerate}
\usepackage{amsmath,amsthm,amssymb}
\allowdisplaybreaks[1]
\usepackage{authblk}
\usepackage{natbib}

\usepackage{minted}
\usepackage{bbm}
\usepackage{xcolor}
\usepackage{graphicx,psfrag,epsf}
\usepackage{subcaption}
\usepackage{epstopdf}
\usepackage{epsfig}
\usepackage[toc]{appendix}
\usepackage{bbm}
\usepackage{enumitem}
\usepackage{xcolor}
\usepackage{url} 
\usepackage[ruled,vlined]{algorithm2e}
\usepackage{makecell}
\usepackage{cancel}
\usepackage{soul}
\usepackage{dsfont}
\usepackage{booktabs}
\usepackage{multirow}
\usepackage{makecell}
\usepackage{rotating}
\usepackage{longtable}
\usepackage{array}
\usepackage{siunitx}
\graphicspath{{arxiv/}}

\usepackage{pdfpages}

\usepackage{datatool}
\usepackage{booktabs} 

\RequirePackage[hyperindex,breaklinks,colorlinks,citecolor=blue,urlcolor=black]{hyperref} 

\theoremstyle{plain}

\newcommand{\blind}{0}

\begin{document}

\def\spacingset#1{\renewcommand{\baselinestretch}%
{#1}\small\normalsize} \spacingset{1}

\if0\blind
{
\title{Adaptive Functional Clustering with Structured Dependence \\ via Variational Inference}
\author[1]{Seojin Lee}
\author[1]{Neulpum Jeong}
\author[1,2]{Seonghyun Jeong\thanks{ Corresponding author: \texttt{sjeong@yonsei.ac.kr}}}
\affil[1]{Department of Statistics and Data Science, Yonsei University}
\affil[2]{Department of Applied Statistics, Yonsei University}
\maketitle
} \fi
	
\if1\blind
{
\bigskip
\bigskip
\bigskip
\begin{center}
{\LARGE\bf Title}
\end{center}
\medskip
} \fi
\spacingset{1.1}

\begin{abstract}
Functional clustering is an important tool for identifying latent heterogeneity in functional data and has been widely applied across various scientific fields. However, many existing methods are not fully adaptive, as they may require the number of clusters to be prespecified and may lack automatic control over the smoothness of the underlying functions. They also commonly assume independent and identically distributed errors, thereby overlooking additional within-curve dependence. We propose a fully adaptive Bayesian procedure for functional clustering that addresses these limitations through Dirichlet process priors, adaptive smoothness control, and flexible covariance modeling. For computational scalability, the proposed method employs variational inference as an efficient alternative to Markov chain Monte Carlo. Together, these features provide a unified Bayesian framework for functional clustering and cluster-specific mean function estimation in the presence of structured within-curve dependence.
\end{abstract}

\noindent\textbf{Keywords}: Functional clustering; Dirichlet process mixtures; Variational inference; Bayesian nonparametrics; Spline smoothing; Shrinkage prior.

\section{Introduction}
With rapid advancements in data collection technologies and computational power, analysis of functional data, defined as continuous random functions over a compact domain \citep{Ramsay_intro, weather_data}, has become increasingly prevalent across various scientific and industrial fields. Owing to its inherently infinite-dimensional nature, functional data analysis (FDA) demands complex modeling approaches to effectively represent intricate phenomena, such as longitudinal measurements \citep{yao2005functional, hall2006properties}, brain imaging \citep{zhu2012multivariate, zhu2011fadtts, wang2014regularized}, and numerous other digital recordings \citep{crainiceanu2009generalized, shou2015structured}. Among the various branches of FDA, clustering of functional data has attracted significant attention as an effective exploratory tool for capturing and summarizing variability, providing a concise representation of underlying patterns among subjects \citep{Chiou_2007, wang2016functional}.

To address the challenges posed by high volume and dimensionality in functional data, various scalable and flexible methodologies for functional clustering have been developed. Many approaches project functional data into a finite-dimensional space using functional principal component decomposition or basis expansion techniques, such as B-splines and wavelets \citep{Zhang_intro}. Clusters are then commonly identified based on criteria such as centroid-based, density-based, and model-based clustering techniques. Centroid-based methods are frequently used owing to their straightforward approach of assigning functional data to the nearest cluster centroid, often employing $k$-means or $k$-medoids techniques \citep{abraham2003, Garcia, Garcia_intro, Antoniadis}. Density-based clustering, in contrast, identifies clusters by detecting regions with a high density of functional data, helping to capture the inherent structure of the data \citep{Tobin, Ren, Mattia}. Model-based methods, meanwhile, determine clusters based on underlying probabilistic modeling structures \citep{Fraley_Raftery_2002}.

One of the most prevalent probabilistic structures in the model-based approach is the mixture model framework, which allows for flexible treatment of uncertainty in cluster assignments \citep{Fraley_Raftery_2002, chamroukhi2019model}.
Despite their flexibility, existing mixture-based approaches to functional clustering share several common limitations. First, finite-mixture formulations require the number of clusters to be specified in advance. In frequentist implementations, mixture parameters are typically estimated via the expectation-maximization algorithm under a finite mixture model \citep{Bouveyron, Same, Jacques, Giacofci, Chamroukhi, Centofanti}. Information criteria, such as the Bayesian information criterion (BIC), can be used to select among candidate models, but this requires repeated model fitting and a separate model-selection step. Second, appropriate smoothness control is crucial for accurately estimating cluster-specific functions, yet it is often overlooked in existing functional clustering methods. In basis expansion approaches, smoothness may be governed by the choice of basis representation, the number and placement of knots, or the strength of a roughness penalty. However, these choices are not always incorporated into the clustering procedure in a data-adaptive manner, making smoothness control an important but frequently neglected aspect of functional clustering \citep{Serban2005, abraham2003, Centofanti}. Third, within-curve dependence is often simplified by assuming conditionally independent and identically distributed measurement errors \citep{Petrone2009, Suarez, yang2022non}. Although this assumption simplifies estimation and computation, it may be inappropriate for real-world functional data, in which observations collected from the same unit are generally correlated because of their ordered or sequential structure \citep{yao2005functional, laird1982random}.

Bayesian modeling provides a natural framework for addressing some of these limitations, and a substantial body of work has developed Bayesian methods for mixture-based functional clustering \citep{Gelfand, Ray, Chib, Rodriguez, Suarez, Sohn,Rigon,Xian}. 
However, none of these methods simultaneously address all of the practical limitations outlined above. Moreover, existing Bayesian approaches typically rely on Markov chain Monte Carlo (MCMC) methods for posterior computation, which can be computationally burdensome. This issue is particularly relevant for functional data, as the corresponding models often involve a large number of within-curve observations and high-dimensional latent representations. Variational inference (VI) offers a computationally efficient alternative by approximating the posterior distribution through optimization rather than sampling. Indeed, VI has been applied across various areas of FDA, including scalar-on-function regression \citep{Goldsmith}, functional registration \citep{Zhang, Earls}, functional additive models \citep{Mclean}, functional mixed models \citep{Huo}, and random allocation models \citep{Nguyen}.
Nevertheless, its use for fully adaptive functional clustering remains limited.

Among Bayesian nonparametric approaches, the Dirichlet process (DP) prior \citep{DP_Ferguson} is widely used for clustering because it allows the number of occupied clusters to be inferred from the data rather than fixed in advance. Combining a DP prior with variational inference can therefore address both the automatic determination of the number of clusters and the computational cost of posterior inference. However, relatively few studies have considered this combination in functional clustering.
For example, \citet{Rigon} introduced an enriched Dirichlet process mixture model and developed a corresponding variational inference algorithm. \citet{Xian} proposed a variational inference-based functional clustering method that combines basis expansion with a DP prior and uses a random effect to account for within-curve correlation.
These methods eliminate the need to prespecify the number of clusters and reduce the computational burden of posterior inference. However, neither method includes a data-driven mechanism for controlling the smoothness of the cluster-specific functions. Moreover, \citet{Rigon} assumes conditionally independent measurement errors, whereas \citet{Xian} represents within-curve dependence using only a random intercept. Although the latter structure is more flexible and captures subject-specific variation, the resulting equal correlation among all within-curve observations may still be too restrictive for densely observed functional data.

In this paper, we propose a functional clustering framework that is fully adaptive to both the number of clusters and function smoothness. Specifically, the proposed method combines a DP prior for mixture-based clustering with a shrinkage prior that adaptively controls the smoothness of basis-expanded functions. The method additionally accommodates within-curve dependence through two covariance models: a random-effects structure for persistent subject-specific correlation and an Ornstein--Uhlenbeck process for time-decaying serial correlation. We derive a coordinate-ascent VI algorithm for each covariance specification.
The resulting method is computationally efficient and outperforms existing approaches in both clustering and function estimation while addressing their key practical limitations. The method is implemented in Python, and the source code is available at \url{https://github.com/sjlee246/adaptive-functional-clustering}.

The remainder of the paper is organized as follows. Section~\ref{sec:Proposed model} introduces the proposed modeling framework. Section~\ref{sec:Variational update} presents the corresponding VI algorithms and variational parameter updates. Section~\ref{sec:Simulation study} reports simulation studies evaluating the proposed methods across various settings. Section~\ref{sec:Application} presents applications to real functional datasets. Finally, Section~\ref{sec:discussion} concludes the paper with a discussion of the findings and future directions.

\section{Modeling framework and prior specification}
\label{sec:Proposed model}

We consider a functional mean model expressed as \begin{align*}
y_i(t) = f_i(t) + \epsilon_{i}(t),\quad i=1,\dots,n,\quad t\in\mathcal T,
\end{align*}
where $\mathcal T\subset \mathbb R$ is a closed interval denoting the index set, $y_i:\mathcal T\rightarrow \mathbb R$ denotes the functional response for unit $i$, $f_i:\mathcal T\rightarrow \mathbb R$ represents the mean function for unit $i$, and $\epsilon_i:\mathcal T\rightarrow \mathbb R$ is a mean-zero stochastic process. 
We assume that $\epsilon_i$ has a zero-mean Gaussian process with $E[\epsilon_i(t)\epsilon_i(t')]=G_i(t,t')$ for $t,t'\in\mathcal T$, where $G_i:\mathcal T\times\mathcal T\rightarrow [0,\infty)$ is a positive definite covariance kernel for unit $i$.
In practice, for each $i=1,\dots, n$, we observe discretized realizations of $y_i$ at points $t_{ij}\in\mathcal T$ for $j=1,\dots,m_i$, where $t_{i,j-1}< t_{ij}$. 
The number of realizations $m_i$ for each unit $i$ may be small or large, leading to either sparse or dense functional data \citep{zhang2016sparse}. The points $t_{ij}$ are typically equally spaced; however, we allow for unevenly spaced observations to provide additional flexibility. This latter setup is particularly useful in the context of sparse functional data. In what follows, we write
$\mathbf y_i=(y_i(t_{i1}),\dots,y_i(t_{im_i}))^T\in\mathbb R^{m_i}$
and
$\boldsymbol\epsilon_i=(\epsilon_i(t_{i1}),\dots,\epsilon_i(t_{im_i}))^T
\in\mathbb R^{m_i}$.
The error vector $\boldsymbol\epsilon_i$ then has a multivariate normal distribution with mean zero and a covariance matrix whose $(j,j')$th element is $G_i(t_{ij},t_{ij'})$. Without loss of generality, we assume that $\mathcal T=[0,1]$, which can be achieved by a simple rescaling.

In this study, we focus on clustering the unit-specific mean functions $f_1,\dots,f_n$ and the covariance functions $G_1,\dots,G_n$ into a few groups based certain shared inherent characteristics.
To develop an effective automatic clustering procedure, three key modeling aspects must be carefully considered:
(1) parameterizing and estimating the infinite-dimensional functions $f_i$ without prior knowledge of their smoothness levels, (2) modeling covariance kernels $G_i$ that account for within--Unit dependency, and (3) devising a clustering procedure without prespecifying the number of clusters. 
This section outlines the framework addressing these modeling aspects. Specifically, the first aspect is achieved through basis expansion with an appropriate regularization prior, which reduces the influence of a predetermined number of basis terms; as detailed in Section~\ref{subsec:mean function}. For the second aspect, we consider two modeling assumptions for the Gaussian process---specifically, namely, the random-effects process and the Ornstein--Uhlenbeck process, as discussed in Section~\ref{subsec:errors}.
Lastly, the third aspect is effectively handled by using DP mixtures owing to their nature as infinite mixture models; an overview of DP mixture models is provided in Section~\ref{subsec:DP mixture}.

\subsection{Basis expansion and shrinkage estimation}
\label{subsec:mean function}

Appropriate modeling of the mean functions is essential not only for accurate function estimation but also for achieving reliable clustering results. Ideally, the model should be flexible enough to capture both local and global characteristics of the functions while avoiding overfitting by adapting to their unknown properties. Among various approaches, basis expansion combined with data-driven knot
determination provides a flexible way to achieve this objective
\citep{smith1996nonparametric,kohn2001nonparametric}.

Let equally spaced knot points $\kappa_s$, $s=1,\dots,D-4$, satisfy $0<\kappa_1<\dots<\kappa_{D-4}<1$. The bounds 0 and 1 are chosen based on the range of the index set $\mathcal T=[0,1]$ but can be replaced by the minimum and maximum of $\{t_{ij},\, i=1,\dots,n,\, j=1,\dots,m_i\}$. The equal spacing can also be replaced by the quantiles of the $t_{ij}$ values based on the empirical measure. For each cluster $k$, we use radial cubic spline functions to express $f_k$ as
\begin{align}
f_k(t;\boldsymbol{\beta}_k) = \beta_{k0} + \sum_{s=1}^3 \beta_{ks} t^s + \sum_{s=4}^{D-1}\beta_{ks} \vert t-\kappa_{s-3} \vert^3, \label{regression_spline}
\end{align}
where $\boldsymbol{\beta}_k = (\beta_{k0}, \cdots, \beta_{k,D-1})^T$ is the spline coefficient vector. The basis functions used in \eqref{regression_spline} generate the cubic spline space, similar to other basis systems such as the B-splines and the truncated power basis \citep{smith1996nonparametric}. We adopt this form because it ensures that the resulting basis terms are on similar scales, making it more convenient to place a reasonable informative prior on $\boldsymbol{\beta}_k$ for shrinkage estimation.

Letting $\mathbf X_i\in\mathbb R^{m_i\times D}$ denote the matrix with the $j$th row given by $(1,t_{ij},t_{ij}^2,t_{ij}^3 , |t_{ij}-\kappa_1|^3,\dots,|t_{ij}-\kappa_{D-4}|^3)^T$, we can express the $i$th observation vector as $\mathbf{y}_i = \mathbf{X}_i\boldsymbol{\beta}_i + \boldsymbol{\epsilon}_i$. To improve numerical stability and facilitate prior specification, each non-intercept basis column and the whole response values are standardized to have mean zero and unit variance before model fitting. This places all regression coefficients on a comparable scale, allowing a common prior specification for $\boldsymbol\beta_k$ and ensuring that the informativeness of the hyperparameters remains essentially invariant to the scale of the data.

Estimation quality can depend strongly on the specification of knots, making it important to choose knot locations that adequately capture both local and global features of the underlying functions. To reduce this sensitivity, we place a dense set of candidate knots and use adaptive shrinkage to suppress the contributions of unnecessary knot-specific basis functions.
To this end, we adopt a hyper-lasso prior, a hierarchical extension of the Bayesian lasso \citep{Lasso_laplace_park}. This prior mitigates the shrinkage bias of the lasso while retaining convenient closed-form updates under the VI algorithm for the proposed functional clustering model. A VI algorithm for this hyper-lasso prior for splines was recently studied by \citet{spline_lasso}.

To fully specify the prior for $\boldsymbol\beta_k$, we first reparameterize the covariance kernel as $G_k=\phi_k^{-1}\tilde G_k$ using a scalar parameter $\phi_k$ and a scaled covariance kernel $\tilde G_k$. This reparametrization provides scale invariance and semi-conjugacy for the prior of $\boldsymbol\beta_k$ when conditioned on $\phi_k$, with a gamma prior on $\phi_k$. 
We aim to prevent shrinkage of the first four components of $\boldsymbol\beta_k$ by assigning a diffuse prior, as it is more reasonable to always include the cubic part in the model.
The resulting joint prior of $(\boldsymbol\beta_k,\phi_k)$ for cluster $k$ is given by
\begin{align}
\begin{split}
    \boldsymbol{\beta}_k \vert \boldsymbol{\tau}_k,\phi_k &\sim \text{N}_D(\mathbf{0},\phi_k^{-1}\mathbf{C}_{\boldsymbol{\tau}_k}),\\
    \tau_{kj} \vert \lambda_k &\sim \text{Exp}(\lambda_k),\quad j=1,\dots,D-4,\\
    \lambda_k &\sim \text{Gamma}(g_0, h_0),\\
      \phi_k &\sim \text{Gamma}(a_0, b_0),
\end{split}
\label{eqn:NEGprior}
\end{align}
where $\boldsymbol{\tau}_k = (\tau_{k1}, \cdots, \tau_{k,D-4})^T$, 
$\mathbf{C}_{\boldsymbol{\tau}_k} = \text{blkdiag}(\rho \mathbf{I}_4,\text{diag}(\boldsymbol{\tau}_k))$ 
with a sufficiently large $\rho>0$,
$\mathbf{I}_d$ denotes the $d\times d$ identity matrix,
$\text{Exp}(\lambda)$ denotes an exponential distribution with rate parameter $\lambda$, 
$\text{Gamma}(a,b)$ denotes a gamma distribution with shape parameter $a$ and rate parameter $b$, 
and $\text{blkdiag}$ and $\text{diag}$ refer to the block diagonal and diagonal matrices formed from the specified components.

There are several possible alternatives to the prior in \eqref{eqn:NEGprior}. For example, a standard Bayesian P-spline prior \citep{lang2004bayesian} controls function roughness by penalizing differences between adjacent spline coefficients. Although it admits convenient closed-form VI updates, its standard formulation with a single smoothing parameter controls only global smoothness and does not allow local adaptation without further extension. Other alternatives include global-local shrinkage priors, such as the horseshoe prior \citep{horseshoe}. Their heavy tails can alleviate the shrinkage bias of the lasso. However, straightforward VI can perform poorly because of strong posterior dependence among latent scale variables \citep{neville2014mean}. In contrast, the prior in \eqref{eqn:NEGprior} retains closed-form coordinate updates and performs well for spline modeling, as also supported by \citet{spline_lasso}.

\subsection{Covariance kernels for the Gaussian process}
\label{subsec:errors}

While reparameterizing the covariance kernel as $G_k=\phi_k^{-1}\tilde G_k$ with a gamma prior on $\phi_k$, we still need to determine the parameterization of $\tilde G_k$ and a suitable prior distribution on its indexing parameters. As discussed earlier, establishing a reasonable dependence structure for the error process is crucial for achieving effective clustering results. We consider two parameterizations that allow closed-form expressions for variational updates: the random-effects process and the Ornstein--Uhlenbeck process with its discrete approximation.

\subsubsection{Random-effects process}
Our first approach uses the bilinear kernel with a nugget effect:
\begin{align}
\tilde G_k(t,t') =\mathbf w(t)^T \mathbf Q_k^{-1} \mathbf w(t')+ \delta_{t,t'} ,\quad t,t'\in\mathcal T,
\label{eqn:raneffcov}
\end{align}
where $\mathbf w(\cdot)$ is a finite-dimensional known basis, $\mathbf Q_k$ is a positive definite matrix, and $\delta_{t,t'} $ is the Kronecker delta, which equals one if $t=t'$ and zero otherwise. A common choice for $\mathbf w(\cdot)$ includes a polynomial, that is, $\mathbf w(t)=(1,t,t^2,\dots,t^{\ell-1})^T$ for $\ell\ge1$.
The parameterization in \eqref{eqn:raneffcov} is motivated by random-effects models \citep{laird1982random}. 
For unit $i$ in cluster $k$, the covariance kernel in \eqref{eqn:raneffcov} is expressed by decomposing $\epsilon_i$ into two independent processes as $\epsilon_i(t)=u_i(t)+e_i(t)$, where $u_i(t) = \mathbf w(t)^T \boldsymbol \xi_i$ is a random process defined
with a unit-specific random effect 
\begin{align}
\boldsymbol \xi_i\vert \phi_k,\mathbf Q_k\sim\text{N}(0,\phi_k^{-1}\mathbf Q_k^{-1})
\label{eqn:re_random_effect}
\end{align}
and $e_i(t)$ is a Gaussian white noise with $E[e_i(t)^2]=\phi_k^{-1}$ and $E[e_i(t)e_i(t')]=0$, $t\ne t'\in\mathcal T$. This decomposition enables closed-form variational updates (see Section~\ref{subsubsec:updating rule}) and makes the covariance kernel in \eqref{eqn:raneffcov} interpretable with a suitably chosen $\mathbf w(\cdot)$. For example, we obtain a random-intercept model with $\mathbf w(t)=1$ and a random-slope model with $\mathbf w(t)=(1,t)^T$.
To complete the prior specification, we assign a semi-conjugate Wishart prior to $\mathbf{Q}_k$, 
\begin{align}
    \mathbf{Q}_k &\sim \text{Wishart}(\mathbf{S}_{0}, r_{0}),
    \label{eqn:wishart}
\end{align}
where $r_0$ represents the degrees of freedom and $\mathbf{S}_0$ is the scale matrix.
In what follows, we refer to this modeling framework as the RE model.

\subsubsection{Ornstein--Uhlenbeck process}
Another convenient parameterization for $\tilde G_k$ is obtained by using the Ornstein--Uhlenbeck process:
\begin{align}
    \tilde G_k(t,t')
    =
    \exp(-\tilde\zeta_k|t-t'|),
    \quad t,t'\in\mathcal T,
    \label{eqn:OUcov}
\end{align}
where $\tilde\zeta_k>0$ is a cluster-specific decay parameter. This covariance structure is widely used because it provides a simple interpretation of within--Unit dependence: the correlation decreases toward zero as the distance between $t$ and $t'$ increases
\citep{maller2009ornstein}. The covariance structure in \eqref{eqn:OUcov} admits an equivalent first-order Markov representation at the observed time points. Specifically,
$\epsilon_i(t_{i1}) \sim N(0,\phi_k^{-1}),$
and, for $j=2,\ldots,m_i$,
\begin{align}
\epsilon_i(t_{ij})
&=
\zeta_{ij,k}\epsilon_i(t_{i,j-1})
+
e_i(t_{ij}),
\qquad
\zeta_{ij,k}
=
\exp[-\tilde\zeta_k(t_{ij}-t_{i,j-1})],
\label{eqn:ou_markov}
\end{align}
where the innovations $\{e_i(t_{ij})\}_{j=2}^{m_i}$ are mutually independent and satisfy
\begin{align*}
e_i(t_{ij})
\sim
N\!\left(
0,
\phi_k^{-1}(1-\zeta_{ij,k}^2)
\right).
\end{align*}
Under this representation, $\phi_k^{-1}$ corresponds to the stationary variance of the Ornstein--Uhlenbeck process. Unlike the RE specification, the variational update for $\tilde\zeta_k$ does not admit a closed-form expression because of its nonlinear appearance in the covariance structure. We address this issue
using the approximation strategy described in
Section~\ref{subsubsec:ou_approx} and Appendix~B.

Although we assume $\mathcal T=[0,1]$ after rescaling, the observed time points may originally be recorded on an arbitrary scale. Therefore, the interpretation of $\tilde\zeta_k$ depends on the scale of the time variable and changes under rescaling. Under the present parameterization, $\tilde\zeta_k$ represents the rate of correlation decay over the rescaled observation domain. We assign the gamma prior
\begin{align}
    \tilde\zeta_k
\sim
\mathrm{Gamma}(p_{0},q_{0}),
\label{eqn:zetagamma}
\end{align}
where $p_{0}$ and $q_{0}$ denote the shape and rate hyperparameters, respectively. Henceforth, we refer to this specification as the OU model.

\subsection{Dirichlet process mixtures for functional clustering}
\label{subsec:DP mixture}

The DP is an infinite-dimensional extension of the Dirichlet distribution and is commonly used as a prior distribution over the space of probability measures \citep{DP_Ferguson}. Suppose that a random measure $\mathcal{H}$ follows a DP, parameterized by a base probability distribution $\mathcal{H}_0$ and a positive concentration parameter $\alpha$. This implies that, for any $T$-partitioned set $\{A_1, \cdots, A_T\}$ of the sample space with an integer $T>0$, we obtain
\begin{align*}
(\mathcal{H}(A_1),\mathcal{H}(A_2), \cdots, \mathcal{H}(A_T)) \sim \text{Dir}(\alpha \mathcal{H}_0(A_1), \alpha \mathcal{H}_0(A_2), \cdots, \alpha \mathcal{H}_0(A_T)),
\end{align*}
where $\text{Dir}(\cdot)$ denotes the Dirichlet distribution.
This is commonly denoted as $\mathcal H\sim \text{DP}(\alpha,\mathcal H_0)$.

The stick-breaking process (SBP) provides an explicit characterization of the DP \citep{Stick_breaking_Sethuraman}. Specifically, with $v_k\sim \text{Beta}(1,\alpha)$ and $\boldsymbol{\eta}_k\sim \mathcal H_0$ for $k=1,2,\dots$, the SBP defines $\mathcal H\sim \text{DP}(\alpha,\mathcal H_0)$ as
\begin{align}
    \mathcal{H}(\cdot) &= \sum_{k=1}^{\infty} \pi_k(\mathbf{v})\delta_{\boldsymbol{\eta}_k}(\cdot), \quad \pi_k(\mathbf{v}) =  v_k \prod_{j=1}^{k-1}(1-v_j) ,\quad k=1,2,\dots,\label{Stick-breaking}
\end{align}
where $\delta_a$ denotes the Dirac (point mass) measure at $a$ and $\mathbf{v} = \{v_k, k=1,2,\dots\}$ is a sequence of stick-breaking proportions that define the mixing weights. Thus, the mixing weights $\pi_k(\mathbf{v})$ are obtained from successively broken fractions of a unit-length `stick' that is represented by each $v_k$. This construction illustrates the discrete nature of the DP, as $\mathcal H$ is almost surely a discrete random measure with atoms $\boldsymbol\eta_k$ drawn from $\mathcal H_0$. The concentration parameter $\alpha$ influences the clustering behavior: as 
$\alpha$ approaches zero, the model becomes more likely to retain the current number of clusters, reducing the possibility of forming new clusters. A widely used default value is $\alpha = 1$, which provides a convenient baseline specification for the prior clustering behavior \citep{gelman2013bayesian}.
Moreover, although the expression in \eqref{Stick-breaking} provides an infinite mixture framework, it is common to impose a conservative upper bound $K$ on the number of mixture components and set $v_K=1$ for computational reasons \citep{Stick_breaking_Ishwaran}. 
This prespecification of $K$ has a negligible affect the clustering results, provided it is sufficiently large, as it merely serves as an upper bound. 
A typical choice is $K=30$ \citep{gelman2013bayesian}.

DP mixture models leverage the discrete nature of DP priors to define infinite mixtures \citep{DP_mixture}. In this context, $\boldsymbol{\eta}_k$ acts as a cluster-specific parameter that determines the distribution for the $k$th cluster. 
In our functional clustering framework, we have $\boldsymbol\eta_k=\{\boldsymbol\beta_k,\phi_k,\mathbf Q_k\}$
for \eqref{eqn:raneffcov} and
$\boldsymbol\eta_k=\{\boldsymbol\beta_k,\phi_k,\tilde\zeta_k\}$ for \eqref{eqn:OUcov}. Since $\mathcal H_0$ acts as a prior for $\boldsymbol\eta_k$, the base measure is specified by \eqref{eqn:NEGprior} and \eqref{eqn:wishart} for the modeling assumption in \eqref{eqn:raneffcov}, and by \eqref{eqn:NEGprior} and \eqref{eqn:zetagamma} for assumption \eqref{eqn:OUcov}.
Let $\mathbf z=(z_1,\dots,z_n)^T$ represent clustering assignment variables, which indicate the membership of each unit, that is, $z_i\in\{1,\dots,K\}$. The hierarchical specification with the SBP is given by
\begin{align*}
\begin{split}
        \mathbf{y}_i \vert z_i, \boldsymbol{\eta}_{z_i}&\sim \text{N}_{m_i}(\mathbf f_{z_i},\phi_{z_i}^{-1}\tilde{\mathbf G}_{z_i}), \quad i=1,\dots,n, \\
    z_i \vert \mathbf{v} &\sim \text{Discrete}(\pi_1(\mathbf{v}), \dots,\pi_K(\mathbf{v})), \quad i=1,\dots,n,\\ 
    v_k &\sim \text{Beta}(1, \alpha), \quad k=1,\dots,K-1,
\end{split}
\end{align*}
where $\mathbf f_{z_i}=(f_{z_i}(t_{i1};\boldsymbol\beta_{z_i}),\dots, f_{z_i}(t_{im_i};\boldsymbol\beta_{z_i}))^T\in\mathbb R^{m_i}$ is the mean vector parameterized by \eqref{regression_spline}, $\tilde{\mathbf G}_{z_i}=(\!(\tilde G_{z_i}(t_{ij},t_{ij'}))\!)_{j,j'}\in\mathbb R^{m_i\times m_i}$ is the scaled covariance matrix parameterized either by \eqref{eqn:raneffcov} or \eqref{eqn:OUcov}, and $\text{Discrete}(\pi_1,\dots,\pi_K)$ denotes a discrete distribution with probability $\pi_k$ for the $k$th category.
Unlike finite mixture models, the number of clusters in a DP mixture is not fixed in advance but is determined in a data-driven manner, even with a prespecified upper bound $K$. 

\section{Variational inference for the proposed method}
\label{sec:Variational update}

\subsection{Mean-field variational inference}
VI approximates the target distribution with a distribution that minimizes Kullback-Leibler (KL) divergence \citep{VI_intro}. 
Let $\boldsymbol{\theta}$ be a collection of parameters. The KL divergence between the posterior distribution $\mathcal P(\, \cdot \, \vert \, \mathbf{y})$ and the variational distribution $\mathcal Q(\cdot)$ is defined as
\begin{align*}
    KL[ \mathcal Q(\cdot) ,\mathcal P(\, \cdot\, \vert \, \mathbf{y})] = \int \log \frac{q (\boldsymbol{\theta})}{p(\boldsymbol{\theta} \vert \mathbf{y})} d \mathcal Q (\boldsymbol{\theta})= \mathbb{E}_{\mathcal Q}[\log q(\boldsymbol{\theta})] - \mathbb{E}_{\mathcal Q}[\log p(\boldsymbol{\theta}, \mathbf{y})] + \log p(\mathbf{y}), 
\end{align*}
where $p(\cdot\vert \mathbf y)$ and $q(\cdot)$ are the densities of $\mathcal P(\, \cdot \, \vert \, \mathbf{y})$ and $\mathcal Q(\cdot)$, respectively, $p(\boldsymbol{\theta},\mathbf{y})$ is the joint density of the parameters and the observations, $p(\mathbf{y})$ is the marginal likelihood, and $\mathbb{E}_{\mathcal Q}$ is the expectation operator under the measure $\mathcal Q$.
Since the KL divergence is always nonnegative, minimizing the KL divergence is equivalent to maximizing the evidence lower bound (ELBO), defined as $\mathcal{L}(\mathcal Q)=\mathbb{E}_{\mathcal Q}[\log p(\boldsymbol{\theta}, \mathbf{y})] - \mathbb{E}_{\mathcal Q}[\log q(\boldsymbol{\theta})]$.
To simplify the structure of the variational distribution, a common approach is to assume that it factorizes into several mutually independent marginal distributions, known as mean-field variational inference (MFVI) \citep{mean_field}. Specifically, MFVI assumes that $\mathcal Q$ is factorized into $L$ independent components as $\mathcal Q(\boldsymbol\theta)=\prod_{l=1}^L \mathcal Q_l(\boldsymbol\theta_l)$.

Under the mean-field assumption, the fully factorized variational distribution enables iterative optimization of each parameter to convergence while keeping other components fixed. This approach is known as coordinate ascent variational inference (CAVI) \citep{CAVI}. 
Specifically, it is evident that maximizing $\mathcal{L}(\mathcal Q)$ with respect to the $l$th component is achieved by maximizing
\begin{align}
     \mathbb{E}_{\mathcal Q}[\log p(\boldsymbol\theta_l \vert \boldsymbol{\theta}_{-l}, \mathbf{y})] - \mathbb{E}_{\mathcal Q}[\log q_l(\boldsymbol\theta_l)] = \int \log \frac{\exp(\mathbb{E}_{-\boldsymbol\theta_l}[\log p(\boldsymbol\theta_l \vert \boldsymbol{\theta}_{-l}, \mathbf{y})])}{q_l(\boldsymbol\theta_l)} \, d \mathcal Q_l(\boldsymbol\theta_l) ,
     \label{ELBO}
\end{align}
where $\boldsymbol{\theta}_{-l}$ denotes all parameters $\boldsymbol\theta$ except for the $l$th component $\boldsymbol\theta_l$, $q_l$ is the density of $\mathcal Q_l$, and $\mathbb{E}_{-\boldsymbol\theta_l}$ denotes the expectation under $\prod_{k\ne l}\mathcal Q_{k}$ (with a slight abuse of notation).
Using the Lagrange multiplier with the constraint $\int d\mathcal Q_l(\boldsymbol\theta_l) = 1$, the solution maximizing \eqref{ELBO} leads to the coordinate ascent update of $q_l$ as 
\begin{align*}
    q_l(\boldsymbol\theta_l) 
    \propto \exp(\mathbb{E}_{-\boldsymbol\theta_l}[\log p(\boldsymbol{\theta}\vert\mathbf{y})]) ,\quad l=1,\dots,L.
\end{align*}
This updating rule shows that we can work with the original joint posterior $p(\boldsymbol{\theta}\vert\mathbf{y})$ in deriving the coordinate ascent algorithm. The CAVI algorithm is similar to Gibbs sampling in that $q_l(\boldsymbol\theta_l)$ is based on the full conditional distributions. However, rather than sampling from the posterior, CAVI achieves tractable solutions by optimizing the variational parameters. For a systematic overview of MFVI and CAVI, readers may refer to \citet{blei2017variational}.

\subsection{Updating the variational posterior}
We provide the variational distribution for the proposed method and its updating rule based on CAVI with the mean-field approximation. 
Let $\boldsymbol\theta_{\mathrm{RE}}$ denote the collection of parameters under the RE covariance specification in
\eqref{eqn:raneffcov}. The variational distribution $\mathcal Q_{\mathrm{RE}}$ for $\boldsymbol\theta_{\mathrm{RE}}$ is expressed as a product of the  distributions for each parameter component, with variational parameters optimized by the coordinate ascent algorithm. 
Specifically, we obtain
\begin{align}
\mathcal Q_{\mathrm{RE}}(\boldsymbol\theta_{\mathrm{RE}}) = \prod_{k=1}^{K-1}\mathcal Q_k^1(v_k)\prod_{k=1}^K \Big[\mathcal Q_k^2 (\boldsymbol\beta_k,\phi_k)\prod_{j=1}^{D-4}\mathcal 
Q_{kj}^3(\tau_{kj})\mathcal Q_k^4 (\lambda_k)\mathcal Q_k^5 (\mathbf Q_k)\Big] \prod_{i=1}^n \Big[\mathcal Q_i^6(\boldsymbol\xi_i)\mathcal Q_i^7(z_i)\Big].    
\label{eqn:REqdistn}
\end{align}
For the model specification with the OU process in \eqref{eqn:OUcov}, let $\boldsymbol \theta_{\mathrm{OU}}$ be the collection of the associated parameters. The variational posterior $\mathcal Q_{\mathrm{OU}}$ of $\boldsymbol \theta_{\mathrm{OU}}$ is given by
\begin{align}
\mathcal Q_{\mathrm{OU}}(\boldsymbol\theta_{\mathrm{OU}}) = \prod_{k=1}^{K-1} \tilde{\mathcal{Q}}_k^1(v_k)  \prod_{k=1}^K \Big[\tilde{\mathcal{Q}}_k^2 (\boldsymbol\beta_k,\phi_k)\prod_{j=1}^{D-4}\tilde{\mathcal{Q}}_{kj}^3(\tau_{kj})\tilde{\mathcal{Q}}_k^4 (\lambda_k)\tilde{\mathcal{Q}}_k^5(\tilde{\zeta}_k)\Big] \prod_{i=1}^n \tilde{\mathcal{Q}}_i^6(z_i).
\label{eqn:OUqdistn}
\end{align}
The specific form of each variational distribution and the corresponding updating rules for both the RE and OU covariance specifications are provided in Section~\ref{subsubsec:ou_approx}. The coordinate ascent algorithm is run until convergence with a specified stopping criterion. We terminate the algorithm when the relative changes in all parameters are sufficiently small.

\subsubsection{Multiple-start optimization}
\label{subsubsec:multistart}
The proposed RE model appears to be particularly sensitive to local optima during optimization. Estimation of covariance components in RE models often leads to multimodal or weakly identified objective surfaces \citep{Sun2007, Williams2015}. Furthermore, VI for DP mixtures is known to be sensitive to local optima and the choice of starting values \citep{DPGMM_Blei}. To improve robustness against these optimization issues, we perform multiple runs from different starting values for the variational parameters, and retain the solution that achieves the highest ELBO.

For the OU model, optimization appears to be less susceptible to these issues. In our numerical experiments, a single set of starting values was generally sufficient, and we therefore use a single starting point throughout this paper. Nevertheless, multiple starts may still be used as a more conservative strategy to improve robustness or solution quality. Since the OU model is computationally more demanding than the RE model, however, we do not recommend using a large number of starting points in practice.

\subsubsection{Gamma approximation for the decay parameter $\tilde\zeta_k$}
\label{subsubsec:ou_approx}
Under the OU model, the variational factor for the decay parameter $\tilde\zeta_k$ does not admit a closed-form update because of the non-conjugate form of the OU likelihood. Therefore, an approximate update is required only for $\tilde\zeta_k$.

We approximate the variational distribution of $\tilde\zeta_k$ by a gamma distribution. This choice is particularly advantageous computationally, since many of the nonlinear expectations required for updating the remaining variational parameters can be reduced to Laplace transforms of $\tilde\zeta_k$. Among common distributions supported on the positive real line, the gamma distribution is especially convenient because its Laplace transform has a remarkably simple closed form, which allows these expectations to be evaluated through infinite-series representations without repeated numerical integration. The resulting series is either evaluated in closed form using standard special functions or approximated directly. The gamma parameters are determined by matching the mode and curvature of the gamma log-density to those of the target log-density, with the latter obtained numerically. Detailed derivations and implementation are provided in Appendix~B.1.5.

\subsubsection{Variational distributions and updating rules}
\label{subsubsec:updating rule}
We now specify the variational family for each factor of
$\mathcal Q_{\mathrm{RE}}$ and $\mathcal Q_{\mathrm{OU}}$ in
\eqref{eqn:REqdistn} and \eqref{eqn:OUqdistn}, respectively. We write
$\operatorname{NG}(\boldsymbol\nu,\boldsymbol\Omega,a,b)$ for the
normal-gamma distribution under which $\boldsymbol\beta\mid\phi\sim
\text{N}(\boldsymbol\nu,\phi^{-1}\boldsymbol\Omega)$ and
$\phi\sim\operatorname{Gamma}(a,b)$, and $\operatorname{GIG}(c,d,f)$ for the
generalized inverse Gaussian distribution with density proportional to
$x^{c-1}\exp\{-(dx+f/x)/2\}$, $x>0$. For the RE model, each factor of $\mathcal Q_{\mathrm{RE}}$ is given by
\begin{align*}
\mathcal Q_k^1(v_k)
&=
\operatorname{Beta}(\gamma_{k,1},\gamma_{k,2}),
\\
\mathcal Q_k^2(\boldsymbol\beta_k,\phi_k)
&=
\operatorname{NG}
(\boldsymbol\nu_k,\boldsymbol\Omega_k,a_k,b_k),
\\
\mathcal Q_{k,j}^3(\tau_{k,j})
&=
\operatorname{GIG}
(1/2,c_{\tau_k},f_{\tau_{k,j}}),
\\
\mathcal Q_k^4(\lambda_k)
&=
\operatorname{Gamma}(g_0+D-4,h_k),
\\
\mathcal Q_k^5(\mathbf Q_k)
&=
\operatorname{Wishart}(\mathbf S_k,r_k),
\\
\mathcal Q_i^6(\boldsymbol\xi_i)
&=
\text{N}(\boldsymbol\mu_i,\boldsymbol\Sigma_i),
\\
\mathcal Q_i^7(z_i)
&=
\operatorname{Discrete}(\varphi_{i,1},\ldots,\varphi_{i,K}),
\end{align*}
where the variational parameters are updated according to the CAVI procedure in Algorithm~\ref{alg:cavi_re}.
For the OU model, each factor of $\mathcal Q_{\mathrm{OU}}$ takes the same
distributional form as its RE counterpart for $v_k$,
$(\boldsymbol\beta_k,\phi_k)$, $\tau_{k,j}$, and $\lambda_k$:
\begin{align*}
\tilde{\mathcal Q}_k^1(v_k)
&=
\operatorname{Beta}(\tilde\gamma_{k,1},\tilde\gamma_{k,2}),
\\
\tilde{\mathcal Q}_k^2(\boldsymbol\beta_k,\phi_k)
&=
\operatorname{NG}
(\tilde{\boldsymbol\nu}_k,\tilde{\mathbf\Omega}_k,\tilde a_k,\tilde b_k),
\\
\tilde{\mathcal Q}_{k,j}^3(\tau_{k,j})
&=
\operatorname{GIG}
(1/2,\tilde c_{\tau_k},\tilde f_{\tau_{k,j}}),
\\
\tilde{\mathcal Q}_k^4(\lambda_k)
&=
\operatorname{Gamma}(g_0+D-4,\tilde h_k),
\\
\tilde{\mathcal Q}_k^5(\tilde\zeta_k)
&\approx
\operatorname{Gamma}(\tilde r_k,\tilde s_k),
\\
\tilde{\mathcal Q}_i^6(z_i)
&=
\operatorname{Discrete}(\tilde\varphi{i,1},\ldots,\tilde\varphi{i,K}),
\end{align*}
where the variational parameters are updated according to the CAVI procedure in Algorithm~\ref{alg:cavi_ou}.
The full derivation of each update is given in Appendix~A.1 for the RE model and Appendix~B.1 for the OU model; here we only record the resulting distributional forms and parameter expressions. The updates for
$(\gamma_{k,1},\gamma_{k,2})$,
$(c_{\tau_k},f_{\tau_{k,j}})$, and $h_k$ have the same
algebraic form under the two covariance specifications, with tildes used for the corresponding OU variational parameters. Here, $K_\nu(\cdot)$ denotes the modified Bessel function of the second
kind of order $\nu$, and $\Psi(\cdot)$ denotes the digamma function. In
Algorithm~\ref{alg:cavi_re}, define
$\mathbf W_i\in\mathbb R^{m_i\times\ell}$ as the matrix whose $j$th row
is $\mathbf w(t_{ij})^T$. In Algorithm~\ref{alg:cavi_ou}, let
$\mathbf x_{ij}$ denote the column vector such that the $j$th row of
$\mathbf X_i$ is $\mathbf x_{ij}^T$, with
$r_{ij,k}=y_{ij}-\mathbf x_{ij}^T\tilde{\boldsymbol\nu}_k$.
For $\zeta_{ij,k}$ defined in \eqref{eqn:ou_markov}, we additionally set
$\zeta_{i1,k}=\zeta_{i,m_i+1,k}=0$ for notational convenience. We then define
\begin{align*}
w_{ij,k}
&=
\frac{1}{1-\zeta_{ij,k}^2},
\qquad j=1,\ldots,m_i+1,
\\
d_{ij,k}
&=
w_{ij,k}+w_{i,j+1,k}-1,
\qquad j=1,\ldots,m_i,
\\
o_{ij,k}
&=
-\zeta_{ij,k}w_{ij,k},
\qquad j=1,\ldots,m_i.
\end{align*}
Throughout both algorithms, all required expectations are evaluated with
respect to the current variational distribution.
For the non-conjugate decay-parameter update in
Algorithm~\ref{alg:cavi_ou},
$q^*(\tilde\zeta_k)$ denotes the exact coordinate-optimal density satisfying
$q^*(\tilde\zeta_k)\propto\exp\{\ell_k(\tilde\zeta_k)\}$, and
$\hat\zeta_k
=
\arg\max_{\tilde\zeta_k>0}\ell_k(\tilde\zeta_k)$.
The numerical evaluation of the expectations above, together with the
function $\ell_k(\cdot)$ and its second derivative, is provided in
Appendix~B.1.5.

\begin{algorithm}[t!]
\footnotesize
\SetAlgoSkip{}
\SetInd{0.5em}{0.5em}
\caption{CAVI algorithm for the RE model}
\label{alg:cavi_re}
\KwIn{$\{\mathbf y_i,\mathbf X_i,\mathbf W_i\}_{i=1}^n$; hyperparameters
$\alpha,a_0,b_0,g_0,h_0,\rho,\mathbf S_{0},r_{0}$; truncation level $K$}
\Repeat{\textnormal{the relative change in all variational parameters falls below a prespecified tolerance}}{
\For{$k=1,\ldots,K-1$}{
    $\gamma_{k,1} \leftarrow 1+\sum_{i=1}^n\varphi_{i,k}$\;
    $\gamma_{k,2} \leftarrow \alpha+\sum_{i=1}^n\sum_{h=k+1}^{K}\varphi_{i,h}$\;
}
\For{$k=1,\ldots,K$}{
    $\boldsymbol\Omega_k \leftarrow
    \Big(\operatorname{blkdiag}\big\{\rho^{-1}\mathbf I_4,\operatorname{diag}(\mathbb E_{\mathcal Q_{\mathrm{RE}}}[\tau_{k,1}^{-1}],\ldots,\mathbb E_{\mathcal Q_{\mathrm{RE}}}[\tau_{k,D-4}^{-1}])\big\} +\sum_{i=1}^n\varphi_{i,k}\mathbf X_i^T\mathbf X_i\Big)^{-1}$
    $\hphantom{\boldsymbol\Omega_k \leftarrow{}}\text{where } \mathbb E_{\mathcal Q_{\mathrm{RE}}}[\tau_{k,j}^{-1}] = \dfrac{\sqrt{c_{\tau_k}}\,K_{3/2}(\sqrt{c_{\tau_k}f_{\tau_{k,j}}})}{\sqrt{f_{\tau_{k,j}}}\,K_{1/2}(\sqrt{c_{\tau_k}f_{\tau_{k,j}}})}-\dfrac{1}{f_{\tau_{k,j}}}$\;
    $\boldsymbol\nu_k \leftarrow \boldsymbol\Omega_k\sum_{i=1}^n\varphi_{i,k}\mathbf X_i^T(\mathbf y_i-\mathbf W_i\boldsymbol\mu_i)$\;
    $a_k \leftarrow a_0+\frac12\sum_{i=1}^n\varphi_{i,k}(m_i+\ell)$\;
    $b_k \leftarrow b_0+\frac12\sum_{i=1}^n\varphi_{i,k}
    \big[\|\mathbf y_i-\mathbf W_i\boldsymbol\mu_i\|_2^2+\operatorname{tr}(\mathbf W_i^T\mathbf W_i\boldsymbol\Sigma_i)
    +r_k\{\boldsymbol\mu_i^T\mathbf S_k\boldsymbol\mu_i+\operatorname{tr}(\mathbf S_k\boldsymbol\Sigma_i)\}\big]$
    $\hphantom{b_k \leftarrow{}}-\frac12\big[\sum_{i=1}^n\varphi_{i,k}\mathbf X_i^T(\mathbf y_i-\mathbf W_i\boldsymbol\mu_i)\big]^T\boldsymbol\nu_k$\;
}
\For{$k=1,\ldots,K$}{
    \For{$j=1,\ldots,D-4$}{
        $c_{\tau_k} \leftarrow 2\,\dfrac{g_0+D-4}{h_k}$\;
        $f_{\tau_{k,j}} \leftarrow \dfrac{a_k}{b_k}(\boldsymbol\nu_k)^2_{j+4}+(\boldsymbol\Omega_k)_{j+4,j+4}$\;
    }
}
\For{$k=1,\ldots,K$}{
    $h_k \leftarrow h_0+\sum_{j=1}^{D-4} \dfrac{\sqrt{f_{\tau_{k,j}}}\,K_{3/2}(\sqrt{c_{\tau_k}f_{\tau_{k,j}}})}{\sqrt{c_{\tau_k}}\,K_{1/2}(\sqrt{c_{\tau_k}f_{\tau_{k,j}}})}$\;
}
\For{$k=1,\ldots,K$}{
    $\mathbf S_k \leftarrow \big(\mathbf S_{0}^{-1} + \sum_{i=1}^n\varphi_{i,k}\frac{a_k}{b_k}(\boldsymbol\mu_i\boldsymbol\mu_i^T+\boldsymbol\Sigma_i)\big)^{-1}$\;
    $r_k \leftarrow r_{0} + \sum_{i=1}^n\varphi_{i,k}$\;
}
\For{$i=1,\ldots,n$}{
    $\boldsymbol\Sigma_i \leftarrow \big[\sum_{k=1}^K\varphi_{i,k}\frac{a_k}{b_k}(\mathbf W_i^T\mathbf W_i+r_k\mathbf S_k)\big]^{-1}$\;
    $\boldsymbol\mu_i \leftarrow \boldsymbol\Sigma_i\big[\sum_{k=1}^K\varphi_{i,k}\frac{a_k}{b_k}(\mathbf W_i^T\mathbf y_i-\mathbf W_i^T\mathbf X_i\boldsymbol\nu_k)\big]$\;
}
\For{$i=1,\ldots,n$}{
    \For{$k=1,\ldots,K$}{
        $\varphi_{i,k} \propto \exp\Big\{
        -\frac12\big(\frac{a_k}{b_k}\|\mathbf y_i-\mathbf X_i\boldsymbol\nu_k-\mathbf W_i\boldsymbol\mu_i\|_2^2
        +\operatorname{tr}(\mathbf X_i^T\mathbf X_i\boldsymbol\Omega_k)
        +\frac{a_k}{b_k}\operatorname{tr}(\mathbf W_i^T\mathbf W_i\boldsymbol\Sigma_i)\big)$
        $\hphantom{\varphi_{i,k} \propto{}}-\frac12\frac{a_k}{b_k}r_k\big(\boldsymbol\mu_i^T\mathbf S_k\boldsymbol\mu_i+\operatorname{tr}(\mathbf S_k\boldsymbol\Sigma_i)\big)
        +\big(\Psi(\gamma_{k,1})-\Psi(\gamma_{k,1}+\gamma_{k,2})\big)$
        $\hphantom{\varphi_{i,k} \propto{}}+\frac{m_i+\ell}{2}\big(\Psi(a_k)-\log b_k\big)
        +\frac12\big(\textstyle\sum_{p=1}^\ell\Psi(\frac{r_k+1-p}{2})+\ell\log2+\log|\mathbf S_k|\big)$
        $\hphantom{\varphi_{i,k} \propto{}}+\textstyle\sum_{h<k}\big(\Psi(\gamma_{h,2})-\Psi(\gamma_{h,1}+\gamma_{h,2})\big)
        \Big\}$\;
    }
    Normalize $\{\varphi_{i,k}\}_{k=1}^K$ so that $\sum_{k=1}^K\varphi_{i,k}=1$\;
}
}
\end{algorithm}

\begin{algorithm}[t!]
\footnotesize
\SetAlgoSkip{}
\SetInd{0.5em}{0.5em}
\caption{CAVI algorithm for the OU model}
\label{alg:cavi_ou}
\KwIn{$\{\mathbf y_i,\mathbf X_i,\{t_{ij}\}_{j=1}^{m_i}\}_{i=1}^n$; hyperparameters
$\alpha,a_0,b_0,g_0,h_0,\rho,p_{0},q_{0}$; truncation level $K$}
\Repeat{\textnormal{the relative change in all variational parameters falls below a prespecified tolerance}}{
\For{$k=1,\ldots,K-1$}{
    $\tilde\gamma_{k,1} \leftarrow 1+\sum_{i=1}^n\tilde\varphi_{i,k}$\;
    $\tilde\gamma_{k,2} \leftarrow \alpha+\sum_{i=1}^n\sum_{h=k+1}^{K}\tilde\varphi_{i,h}$\;
}
\For{$k=1,\ldots,K$}{
    $\tilde{\boldsymbol\Omega}_k \leftarrow \Big(\operatorname{blkdiag}\big\{\rho^{-1}\mathbf I_4,\operatorname{diag}(\mathbb E_{\mathcal Q_{\mathrm{OU}}}[\tau_{k,1}^{-1}],\ldots,\mathbb E_{\mathcal Q_{\mathrm{OU}}}[\tau_{k,D-4}^{-1}])\big\}+\sum_{i=1}^n\tilde\varphi_{i,k}\Big[\sum_{j=1}^{m_i}\mathbb E_{\mathcal Q_{\mathrm{OU}}}[d_{ij,k}]\,\mathbf x_{ij}\mathbf x_{ij}^T$
    $\hphantom{\tilde{\boldsymbol\Omega}_k \leftarrow{}}+\sum_{j=2}^{m_i}\mathbb E_{\mathcal Q_{\mathrm{OU}}}[o_{ij,k}]\big(\mathbf x_{ij}\mathbf x_{i,j-1}^T+\mathbf x_{i,j-1}\mathbf x_{ij}^T\big)\Big]\Big)^{-1}$
    $\hphantom{\tilde{\boldsymbol\Omega}_k \leftarrow{}}\text{where } \mathbb E_{\mathcal Q_{\mathrm{OU}}}[\tau_{k,j}^{-1}] = \dfrac{\sqrt{\tilde c_{\tau_k}}\,K_{3/2}(\sqrt{\tilde c_{\tau_k}\tilde f_{\tau_{k,j}}})}{\sqrt{\tilde f_{\tau_{k,j}}}\,K_{1/2}(\sqrt{\tilde c_{\tau_k}\tilde f_{\tau_{k,j}}})}-\dfrac{1}{\tilde f_{\tau_{k,j}}}$\;
    $\tilde{\boldsymbol\nu}_k \leftarrow \tilde{\boldsymbol\Omega}_k\sum_{i=1}^n\tilde\varphi_{i,k}\Big[\sum_{j=1}^{m_i}\mathbb E_{\mathcal Q_{\mathrm{OU}}}[d_{ij,k}]\,y_{ij}\mathbf x_{ij}+\sum_{j=2}^{m_i}\mathbb E_{\mathcal Q_{\mathrm{OU}}}[o_{ij,k}]\big(y_{ij}\mathbf x_{i,j-1}+y_{i,j-1}\mathbf x_{ij}\big)\Big]$\;
    $\tilde a_k \leftarrow a_0+\frac12\sum_{i=1}^n m_i\tilde\varphi_{i,k}$\;
    $\tilde b_k \leftarrow b_0+\frac12\sum_{i=1}^n\tilde\varphi_{i,k}\Big[\sum_{j=1}^{m_i}\mathbb E_{\mathcal Q_{\mathrm{OU}}}[d_{ij,k}]\,y_{ij}^2+2\sum_{j=2}^{m_i}\mathbb E_{\mathcal Q_{\mathrm{OU}}}[o_{ij,k}]\,y_{ij}y_{i,j-1}\Big] -\frac12\tilde{\boldsymbol\nu}_k^T\tilde{\boldsymbol\Omega}_k^{-1}\tilde{\boldsymbol\nu}_k$\;
}
\For{$k=1,\ldots,K$}{
    \For{$j=1,\ldots,D-4$}{
        $\tilde c_{\tau_k} \leftarrow 2\,\dfrac{g_0+D-4}{\tilde h_k}$\;
        $\tilde f_{\tau_{k,j}} \leftarrow \dfrac{\tilde a_k}{\tilde b_k}(\tilde{\boldsymbol\nu}_k)^2_{j+4}+(\tilde{\boldsymbol\Omega}_k)_{j+4,j+4}$\;
    }
}
\For{$k=1,\ldots,K$}{
    $\tilde h_k \leftarrow h_0+\sum_{j=1}^{D-4} \dfrac{\sqrt{\tilde f_{\tau_{k,j}}}\,K_{3/2}(\sqrt{\tilde c_{\tau_k}\tilde f_{\tau_{k,j}}})}{\sqrt{\tilde c_{\tau_k}}\,K_{1/2}(\sqrt{\tilde c_{\tau_k}\tilde f_{\tau_{k,j}}})}$\;
}
\For{$k=1,\ldots,K$}{
    Solve $\ell_k'(\hat\zeta_k)=0$ numerically for $\hat\zeta_k$\;
    $\tilde r_k \leftarrow 1-\hat\zeta_k^2\ell_k''(\hat\zeta_k)$\;
    $\tilde s_k \leftarrow -\hat\zeta_k\ell_k''(\hat\zeta_k)$\;
}
\For{$i=1,\ldots,n$}{
    \For{$k=1,\ldots,K$}{
        $\tilde\varphi_{i,k} \propto \exp\Big\{
        -\frac12\sum_{j=2}^{m_i}\mathbb E_{\mathcal Q_{\mathrm{OU}}}[\log w_{ij,k}]
        +\frac{m_i}{2}\big(\Psi(\tilde a_k)-\log\tilde b_k\big)$
        $\hphantom{\tilde\varphi_{i,k} \propto{}}-\frac12\frac{\tilde a_k}{\tilde b_k}\Big[\sum_{j=1}^{m_i}\mathbb E_{\mathcal Q_{\mathrm{OU}}}[d_{ij,k}]\,r_{ij,k}^2+2\sum_{j=2}^{m_i}\mathbb E_{\mathcal Q_{\mathrm{OU}}}[o_{ij,k}]\,r_{ij,k}r_{i,j-1,k}\Big]$
        $\hphantom{\tilde\varphi_{i,k} \propto{}}-\frac12\Big[\sum_{j=1}^{m_i}\mathbb E_{\mathcal Q_{\mathrm{OU}}}[d_{ij,k}]\,\mathbf x_{ij}^T\tilde{\boldsymbol\Omega}_k\mathbf x_{ij}+2\sum_{j=2}^{m_i}\mathbb E_{\mathcal Q_{\mathrm{OU}}}[o_{ij,k}]\,\mathbf x_{ij}^T\tilde{\boldsymbol\Omega}_k\mathbf x_{i,j-1}\Big]$
        $\hphantom{\tilde\varphi_{i,k} \propto{}}+\big(\Psi(\tilde\gamma_{k,1})-\Psi(\tilde\gamma_{k,1}+\tilde\gamma_{k,2})\big)
        +\textstyle\sum_{h<k}\big(\Psi(\tilde\gamma_{h,2})-\Psi(\tilde\gamma_{h,1}+\tilde\gamma_{h,2})\big)
        \Big\}$\;
    }
    Normalize $\{\tilde\varphi_{i,k}\}_{k=1}^K$ so that $\sum_{k=1}^K\tilde\varphi_{i,k}=1$\;
}
}
\end{algorithm}

\section{Simulation study}
\label{sec:Simulation study}
This section evaluates the proposed method using synthetic datasets through two complementary simulation studies. Since functional clustering methods with covariance structures directly matching either of the proposed specifications are generally limited in the existing literature, we evaluate only the proposed two models under correctly specified covariance structures in Section~\ref{subsec:proposed}. In Section~\ref{subsec:comparison}, we then compare the proposed models with existing model-based functional clustering approaches under covariance misspecification. Specifically, the data are generated from error structures that do not coincide with the covariance assumptions of any of the methods under comparison, allowing us to assess how robustly each method performs when its assumed structure fails to hold.

\subsection{Evaluation of the proposed methods}
\label{subsec:proposed}
The simulation study is designed to examine both clustering accuracy and mean-curve estimation across a range of data characteristics. Clustering performance is evaluated using the adjusted rand index (ARI), while mean-curve estimation accuracy is evaluated using the average $L_2$-error, defined as $n^{-1}\sum_{i=1}^{n}\lVert \hat f_i-f_i\rVert_2$, where $\hat f_i$ denotes the estimated mean curve corresponding to the $i$th functional observation. To reduce the influence of prior specification, all models are fitted using weakly informative priors. Under the standardized parameterization described in Section~2.1, the prior mean of each $\beta_k$ is set to the zero vector. In addition, $\rho$ is fixed at $10^{10}$ to preserve an effectively unpenalized polynomial component after standardization. For the automatic Bayesian hyper-lasso procedure, we adopt independent weakly informative gamma priors for $\phi_k$ and $\lambda_k$, with $a_0=b_0=g_0=h_0=10^{-10}$, following \citet{spline_lasso}. The concentration parameter of the DP is fixed at $\alpha=1$, following the recommendation of \citet{alpha_one}. The covariance-related hyperparameters are specified according to the assumed dependence structure. For the RE model, the precision matrix $\mathbf Q_k$ is assigned a Wishart prior with $\mathbf S_{0}=10^{10}\mathbf I_l$ and $r_{0}=l$, where $l$ denotes the dimension of the random effect. For the OU model, the dependence parameter $\tilde{\zeta}_k$ follows a gamma prior with $p_{0}=20$ and $q_{0}=1$, which provides a mildly informative prior on the standardized time scale while ensuring numerical stability during variational optimization. The maximum truncation level is fixed at $K=30$ throughout.

For each unit $i$, the number of observations, $m_i$, is independently generated from a zero-truncated Poisson distribution with intensity $\lambda$, denoted by $\mathrm{ZTP}(\lambda)$. Conditional on $m_i$, the observation times $t_{ij}$ are independently sampled from a uniform distribution on $[0,1]$ and then arranged in increasing order.
The simulation study considers five experimental factors: the mean-function scenario (Scenario~A or Scenario~B), the number of units $n\in \{100,300\}$, the ZTP intensity $\lambda\in \{3,10,30,50\}$, the noise standard deviation (SD) $\phi_k^{-1/2}\in \{0.1,0.3\}$, and the number of candidate knots in ${30,100}$. Their full factorial combination yields 64 simulation settings. 

\begin{figure}[t]
    \centering
    \includegraphics[width=0.99\textwidth]{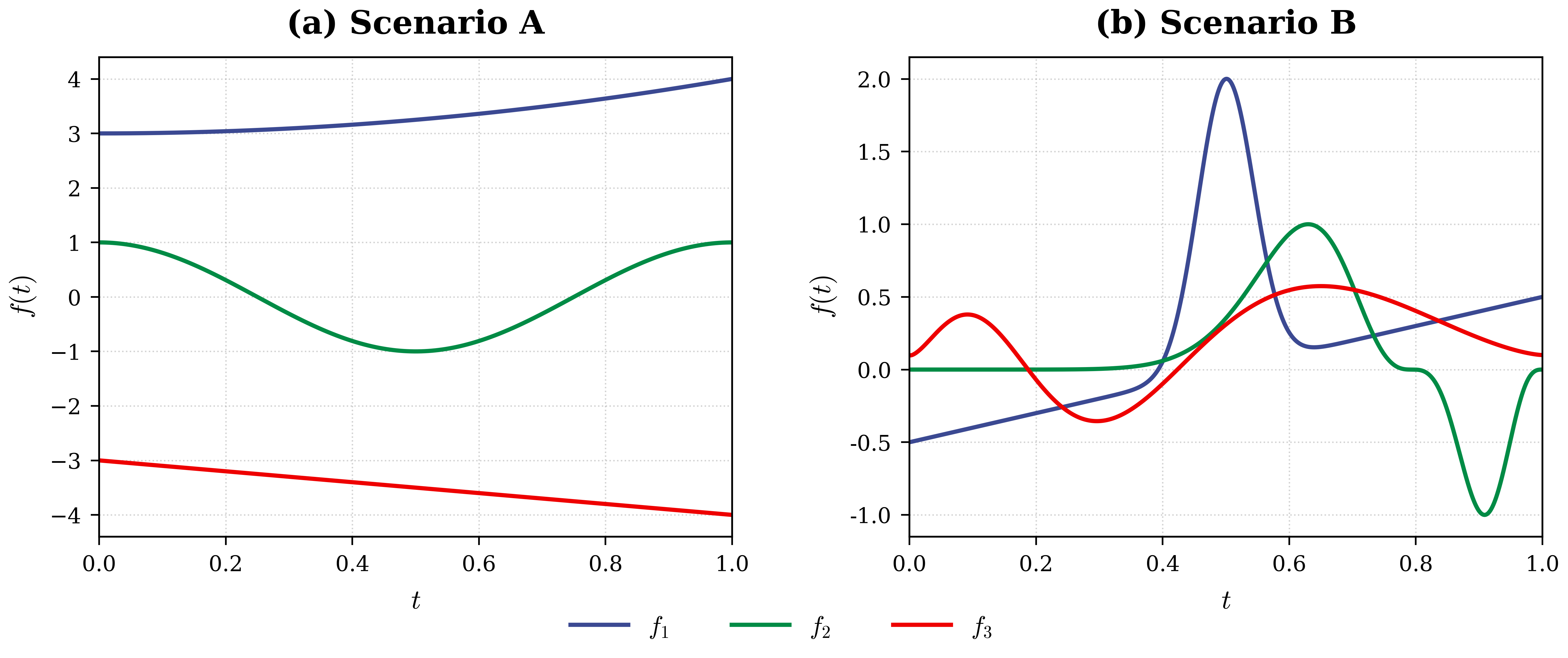}
    \caption{True cluster-specific mean functions $f_1, f_2, f_3$ used in the
    simulation study. Panel (a) shows the well-separated mean curves of
    Scenario~A, and panel (b) shows the more complex, overlapping mean
    curves of Scenario~B.}
    \label{fig:scenario_plot}
\end{figure}

Scenarios~A and B each consist of three cluster-specific mean functions $f_1$, $f_2$, and $f_3$. For each unit, the cluster assignment is independently generated from a categorical distribution with probabilities $(0.5,0.3,0.2)$ to represent an imbalanced clustering structure. The functions are defined for $t\in[0,1]$, and the corresponding mean curves are displayed in Figure~\ref{fig:scenario_plot}. Scenario~A consists of smooth and well-separated mean curves and provides a relatively simple setting for clustering and mean-curve estimation. Its mean functions are
\begin{align*}
f_1(t) &= 3+t^2,\\
f_2(t) &= \cos(2\pi t),\\
f_3(t) &= -3-t.
\end{align*}
Scenario~B consists of more complex mean curves with localized variation and substantial overlap between clusters. It therefore provides a more challenging setting and allows us to examine whether the proposed shrinkage procedure can recover both global and local features of the mean functions. Its mean functions are
\begin{align*}
f_1(t)
&=
t+2\exp\left\{-(16(t-0.5))^2\right\}-0.5,\\
f_2(t)
&=
\sin^3(2\pi t^3),\\
f_3(t)
&=
\sqrt{t(1-t)}
\sin\!\left(
\frac{2\pi(1+2^{-3/5})}{t+2^{-3/5}}
\right)
+0.1.
\end{align*}
The four values of $\lambda$ are chosen to represent observation patterns ranging from very sparse to dense. Because observation times are generated irregularly rather than observed on a common grid, the sparse settings remain applicable even when only a few measurements are available for each unit. These settings also allow us to assess how observation density interacts with the shapes of the mean functions. The well-separated curves in Scenario~A can be distinguished using relatively few observations, whereas the overlapping curves and localized features in Scenario~B generally require denser observations.
The two candidate-knot settings are included to assess the sensitivity of the proposed method to the initial number of knots. The setting with 30 candidate knots provides a moderately rich spline representation, whereas the setting with 100 candidate knots represents a substantially larger candidate set. If the shrinkage prior effectively suppresses unnecessary spline terms, the estimation results should remain stable once the candidate set is sufficiently rich. For both settings, the interior knots are placed at equally spaced locations over the interval $[0,1]$. 
For the RE model, the cluster-specific precision matrices $\mathbf Q_k$
in \eqref{eqn:re_random_effect}, corresponding to
$f_1$, $f_2$, and $f_3$, are, respectively,
\[
\begin{pmatrix}
1 & 1\\
1 & 1.5
\end{pmatrix},
\qquad
\begin{pmatrix}
3 & -2.3\\
-2.3 & 2.5
\end{pmatrix},
\qquad
\begin{pmatrix}
0.33 & 0.19\\
0.19 & 1.89
\end{pmatrix}.
\]
For the OU model, the cluster-specific decay parameters $\tilde\zeta_k$
in \eqref{eqn:ou_markov} are set to
16, 37, 27.
As discussed in Section~\ref{subsubsec:multistart}, the RE model is fitted using 30 different starting values, whereas the OU model is fitted using a single starting value. Every simulation setting is replicated over 30 independent random seeds. The VI algorithm is terminated when the maximum relative change in the variational parameters between successive iterations falls below $10^{-3}$.

\begin{table}[p]
\centering
\caption{
Simulation results under the RE covariance structure.
Each block reports, from top to bottom, the mean (SD) of the ARI, the mean
(SD) of the average $L_2$-error, and the mean (SD) of the computation time
in seconds, over 30 replications.
}
\label{tab:sim_results_RE}
\begingroup
\renewcommand{\baselinestretch}{1}\selectfont
\setlength{\tabcolsep}{3pt}
\renewcommand{\arraystretch}{1.0}
\scriptsize
\begin{tabular}{cccl
  r@{ }l
  r@{ }l
  r@{ }l
  r@{ }l}
\toprule
& & & &
\multicolumn{4}{c}{$\phi_k^{-1/2}=0.1$} &
\multicolumn{4}{c}{$\phi_k^{-1/2}=0.3$} \\
\cmidrule(lr){5-8}\cmidrule(lr){9-12}
& & $m_i$ & Metric &
\multicolumn{2}{c}{30 knots} &
\multicolumn{2}{c}{100 knots} &
\multicolumn{2}{c}{30 knots} &
\multicolumn{2}{c}{100 knots} \\
\midrule
\multirow[c]{24}{*}{Scenario A}
& \multirow[c]{12}{*}{$n=100$}
& \multirow[c]{3}{*}{ZTP(3)}
& ARI         & 0.998  & (0.013) & 1.000  & (0.000) & 1.000  & (0.000) & 1.000  & (0.000) \\
& & & $L_2$-error & 0.040  & (0.017) & 0.036  & (0.011) & 0.074  & (0.018) & 0.077  & (0.018) \\
& & & Time (s)    & 26.0   & (9.6)   & 121.1  & (48.1)  & 42.4   & (12.5)  & 225.7  & (64.0)  \\
\cmidrule(lr){3-12}
& & \multirow[c]{3}{*}{ZTP(10)}
& ARI         & 1.000  & (0.000) & 1.000  & (0.000) & 1.000  & (0.000) & 1.000  & (0.000) \\
& & & $L_2$-error & 0.026  & (0.014) & 0.027  & (0.013) & 0.045  & (0.013) & 0.046  & (0.012) \\
& & & Time (s)    & 24.2   & (5.8)   & 166.7  & (51.2)  & 27.4   & (18.1)  & 195.0  & (101.2) \\
\cmidrule(lr){3-12}
& & \multirow[c]{3}{*}{ZTP(30)}
& ARI         & 1.000  & (0.000) & 1.000  & (0.000) & 1.000  & (0.000) & 1.000  & (0.000) \\
& & & $L_2$-error & 0.031  & (0.029) & 0.034  & (0.041) & 0.031  & (0.011) & 0.033  & (0.010) \\
& & & Time (s)    & 58.3   & (25.6)  & 406.4  & (178.2) & 48.1   & (20.8)  & 430.2  & (204.9) \\
\cmidrule(lr){3-12}
& & \multirow[c]{3}{*}{ZTP(50)}
& ARI         & 1.000  & (0.000) & 1.000  & (0.000) & 1.000  & (0.000) & 1.000  & (0.000) \\
& & & $L_2$-error & 0.026  & (0.014) & 0.032  & (0.016) & 0.030  & (0.012) & 0.031  & (0.012) \\
& & & Time (s)    & 102.4  & (54.0)  & 633.3  & (314.9) & 89.3   & (45.7)  & 672.9  & (337.5) \\
\cmidrule(lr){2-12}
& \multirow[c]{12}{*}{$n=300$}
& \multirow[c]{3}{*}{ZTP(3)}
& ARI         & 1.000  & (0.000) & 1.000  & (0.000) & 1.000  & (0.000) & 1.000  & (0.000) \\
& & & $L_2$-error & 0.021  & (0.006) & 0.021  & (0.006) & 0.042  & (0.008) & 0.044  & (0.007) \\
& & & Time (s)    & 49.7   & (15.5)  & 279.7  & (88.0)  & 68.1   & (39.9)  & 365.5  & (146.4) \\
\cmidrule(lr){3-12}
& & \multirow[c]{3}{*}{ZTP(10)}
& ARI         & 1.000  & (0.000) & 1.000  & (0.000) & 1.000  & (0.000) & 1.000  & (0.000) \\
& & & $L_2$-error & 0.016  & (0.005) & 0.020  & (0.011) & 0.028  & (0.005) & 0.029  & (0.005) \\
& & & Time (s)    & 88.9   & (25.6)  & 507.0  & (165.3) & 85.4   & (28.4)  & 543.0  & (190.4) \\
\cmidrule(lr){3-12}
& & \multirow[c]{3}{*}{ZTP(30)}
& ARI         & 1.000  & (0.000) & 1.000  & (0.000) & 1.000  & (0.000) & 1.000  & (0.000) \\
& & & $L_2$-error & 0.015  & (0.008) & 0.022  & (0.010) & 0.020  & (0.005) & 0.021  & (0.005) \\
& & & Time (s)    & 249.1  & (151.7) & 1152.7 & (525.6) & 248.7  & (141.8) & 1373.4 & (633.0) \\
\cmidrule(lr){3-12}
& & \multirow[c]{3}{*}{ZTP(50)}
& ARI         & 1.000  & (0.000) & 1.000  & (0.000) & 1.000  & (0.000) & 1.000  & (0.000) \\
& & & $L_2$-error & 0.024  & (0.017) & 0.046  & (0.025) & 0.016  & (0.006) & 0.020  & (0.009) \\
& & & Time (s)    & 462.3  & (249.4) & 1841.5 & (744.6) & 494.1  & (251.7) & 2053.1 & (950.3) \\
\midrule
\multirow[c]{24}{*}{Scenario B}
& \multirow[c]{12}{*}{$n=100$}
& \multirow[c]{3}{*}{ZTP(3)}
& ARI         & 0.155  & (0.221) & 0.181  & (0.204) & 0.000  & (0.000) & 0.000  & (0.000) \\
& & & $L_2$-error & 0.289  & (0.059) & 0.284  & (0.050) & 0.329  & (0.013) & 0.329  & (0.013) \\
& & & Time (s)    & 81.4   & (22.8)  & 405.8  & (104.2) & 74.3   & (23.5)  & 377.4  & (126.4) \\
\cmidrule(lr){3-12}
& & \multirow[c]{3}{*}{ZTP(10)}
& ARI         & 0.857  & (0.158) & 0.893  & (0.145) & 0.581  & (0.149) & 0.573  & (0.144) \\
& & & $L_2$-error & 0.078  & (0.050) & 0.077  & (0.048) & 0.165  & (0.039) & 0.167  & (0.038) \\
& & & Time (s)    & 88.6   & (20.2)  & 513.8  & (134.1) & 107.5  & (23.0)  & 592.5  & (144.9) \\
\cmidrule(lr){3-12}
& & \multirow[c]{3}{*}{ZTP(30)}
& ARI         & 0.999  & (0.003) & 0.995  & (0.014) & 0.991  & (0.014) & 0.992  & (0.014) \\
& & & $L_2$-error & 0.027  & (0.011) & 0.027  & (0.012) & 0.054  & (0.014) & 0.061  & (0.013) \\
& & & Time (s)    & 117.2  & (48.2)  & 854.6  & (397.2) & 177.5  & (83.3)  & 1109.9 & (561.9) \\
\cmidrule(lr){3-12}
& & \multirow[c]{3}{*}{ZTP(50)}
& ARI         & 1.000  & (0.002) & 1.000  & (0.000) & 0.996  & (0.020) & 0.993  & (0.025) \\
& & & $L_2$-error & 0.027  & (0.011) & 0.025  & (0.011) & 0.043  & (0.011) & 0.054  & (0.012) \\
& & & Time (s)    & 187.6  & (99.4)  & 1404.0 & (695.5) & 228.0  & (123.8) & 1423.6 & (764.8) \\
\cmidrule(lr){2-12}
& \multirow[c]{12}{*}{$n=300$}
& \multirow[c]{3}{*}{ZTP(3)}
& ARI         & 0.657  & (0.116) & 0.645  & (0.121) & 0.044  & (0.101) & 0.049  & (0.101) \\
& & & $L_2$-error & 0.108  & (0.030) & 0.122  & (0.037) & 0.309  & (0.035) & 0.310  & (0.034) \\
& & & Time (s)    & 141.4  & (39.7)  & 894.2  & (207.1) & 147.6  & (42.1)  & 934.0  & (302.5) \\
\cmidrule(lr){3-12}
& & \multirow[c]{3}{*}{ZTP(10)}
& ARI         & 0.982  & (0.015) & 0.984  & (0.013) & 0.829  & (0.067) & 0.813  & (0.094) \\
& & & $L_2$-error & 0.024  & (0.005) & 0.024  & (0.005) & 0.081  & (0.018) & 0.089  & (0.024) \\
& & & Time (s)    & 164.3  & (43.8)  & 1250.5 & (461.5) & 193.5  & (88.2)  & 1526.8 & (737.9) \\
\cmidrule(lr){3-12}
& & \multirow[c]{3}{*}{ZTP(30)}
& ARI         & 0.999  & (0.002) & 1.000  & (0.000) & 0.994  & (0.007) & 0.994  & (0.009) \\
& & & $L_2$-error & 0.021  & (0.008) & 0.018  & (0.006) & 0.029  & (0.005) & 0.031  & (0.007) \\
& & & Time (s)    & 387.6  & (218.9) & 2682.7 & (1285.8) & 540.5  & (304.6) & 2666.7 & (1309.7) \\
\cmidrule(lr){3-12}
& & \multirow[c]{3}{*}{ZTP(50)}
& ARI         & 1.000  & (0.000) & 1.000  & (0.000) & 1.000  & (0.002) & 0.999  & (0.003) \\
& & & $L_2$-error & 0.029  & (0.012) & 0.023  & (0.010) & 0.023  & (0.004) & 0.023  & (0.004) \\
& & & Time (s)    & 620.2  & (336.7) & 3679.7 & (1733.7) & 849.9  & (448.1) & 3902.3 & (1857.4) \\
\bottomrule
\end{tabular}
\endgroup
\end{table}

\begin{table}[p]
\centering
\caption{
Simulation results under the OU covariance structure.
Each block reports, from top to bottom, the mean (SD) of the ARI, the mean
(SD) of the average $L_2$-error, and the mean (SD) of the computation time
in seconds, over 30 replications.
}
\label{tab:sim_results_OU}
\begingroup
\renewcommand{\baselinestretch}{1}\selectfont
\setlength{\tabcolsep}{3pt}
\renewcommand{\arraystretch}{1.0}
\scriptsize
\begin{tabular}{cccl
  r@{ }l
  r@{ }l
  r@{ }l
  r@{ }l}
\toprule
& & & &
\multicolumn{4}{c}{$\phi_k^{-1/2}=0.1$} &
\multicolumn{4}{c}{$\phi_k^{-1/2}=0.3$} \\
\cmidrule(lr){5-8}\cmidrule(lr){9-12}
& & $m_i$ & Metric &
\multicolumn{2}{c}{30 knots} &
\multicolumn{2}{c}{100 knots} &
\multicolumn{2}{c}{30 knots} &
\multicolumn{2}{c}{100 knots} \\
\midrule
\multirow[c]{24}{*}{Scenario A}
& \multirow[c]{12}{*}{$n=100$}
& \multirow[c]{3}{*}{ZTP(3)}
& ARI         & 0.710  & (0.014) & 0.702  & (0.019) & 0.659  & (0.171) & 0.671  & (0.167) \\
& & & $L_2$-error & 0.431  & (0.070) & 0.451  & (0.072) & 0.704  & (0.589) & 0.584  & (0.430) \\
& & & Time (s)    & 13.4   & (20.7)  & 51.1   & (60.2)  & 16.0   & (40.7)  & 6.6    & (5.4)   \\
\cmidrule(lr){3-12}
& & \multirow[c]{3}{*}{ZTP(10)}
& ARI         & 1.000  & (0.000) & 1.000  & (0.000) & 0.993  & (0.016) & 0.994  & (0.015) \\
& & & $L_2$-error & 0.015  & (0.003) & 0.015  & (0.003) & 0.046  & (0.013) & 0.048  & (0.013) \\
& & & Time (s)    & 6.7    & (5.5)   & 6.0    & (3.6)   & 4.4    & (8.9)   & 7.8    & (7.9)   \\
\cmidrule(lr){3-12}
& & \multirow[c]{3}{*}{ZTP(30)}
& ARI         & 1.000  & (0.000) & 1.000  & (0.000) & 0.993  & (0.033) & 0.998  & (0.009) \\
& & & $L_2$-error & 0.011  & (0.002) & 0.011  & (0.001) & 0.033  & (0.006) & 0.034  & (0.004) \\
& & & Time (s)    & 37.6   & (38.4)  & 50.2   & (61.4)  & 14.4   & (15.2)  & 36.8   & (44.0)  \\
\cmidrule(lr){3-12}
& & \multirow[c]{3}{*}{ZTP(50)}
& ARI         & 1.000  & (0.000) & 1.000  & (0.000) & 0.999  & (0.003) & 0.999  & (0.003) \\
& & & $L_2$-error & 0.010  & (0.002) & 0.010  & (0.002) & 0.029  & (0.007) & 0.031  & (0.006) \\
& & & Time (s)    & 72.9   & (85.8)  & 122.9  & (138.0) & 27.8   & (29.9)  & 81.4   & (92.4)  \\
\cmidrule(lr){2-12}
& \multirow[c]{12}{*}{$n=300$}
& \multirow[c]{3}{*}{ZTP(3)}
& ARI         & 0.771  & (0.020) & 0.759  & (0.032) & 0.533  & (0.292) & 0.559  & (0.252) \\
& & & $L_2$-error & 0.363  & (0.029) & 0.362  & (0.028) & 1.179  & (0.823) & 1.047  & (0.772) \\
& & & Time (s)    & 28.2   & (23.3)  & 108.1  & (108.8) & 29.6   & (84.6)  & 22.4   & (31.4)  \\
\cmidrule(lr){3-12}
& & \multirow[c]{3}{*}{ZTP(10)}
& ARI         & 0.998  & (0.004) & 0.996  & (0.013) & 0.994  & (0.006) & 0.992  & (0.010) \\
& & & $L_2$-error & 0.009  & (0.003) & 0.010  & (0.003) & 0.027  & (0.005) & 0.029  & (0.006) \\
& & & Time (s)    & 27.8   & (28.0)  & 35.0   & (37.0)  & 22.3   & (26.1)  & 51.0   & (74.0)  \\
\cmidrule(lr){3-12}
& & \multirow[c]{3}{*}{ZTP(30)}
& ARI         & 1.000  & (0.000) & 1.000  & (0.000) & 0.998  & (0.005) & 0.997  & (0.006) \\
& & & $L_2$-error & 0.006  & (0.001) & 0.006  & (0.001) & 0.019  & (0.003) & 0.020  & (0.004) \\
& & & Time (s)    & 155.2  & (189.6) & 195.8  & (182.1) & 71.6   & (68.0)  & 153.7  & (161.7) \\
\cmidrule(lr){3-12}
& & \multirow[c]{3}{*}{ZTP(50)}
& ARI         & 1.000  & (0.000) & 1.000  & (0.000) & 0.999  & (0.002) & 0.999  & (0.003) \\
& & & $L_2$-error & 0.006  & (0.001) & 0.006  & (0.001) & 0.018  & (0.002) & 0.019  & (0.003) \\
& & & Time (s)    & 211.3  & (125.7) & 357.5  & (183.2) & 141.2  & (173.8) & 309.5  & (341.5) \\
\midrule
\multirow[c]{24}{*}{Scenario B}
& \multirow[c]{12}{*}{$n=100$}
& \multirow[c]{3}{*}{ZTP(3)}
& ARI         & 0.503  & (0.149) & 0.503  & (0.154) & 0.090  & (0.137) & 0.079  & (0.135) \\
& & & $L_2$-error & 0.478  & (1.118) & 0.401  & (0.636) & 0.374  & (0.053) & 0.403  & (0.086) \\
& & & Time (s)    & 24.2   & (34.6)  & 153.3  & (131.7) & 19.2   & (14.5)  & 134.2  & (182.1) \\
\cmidrule(lr){3-12}
& & \multirow[c]{3}{*}{ZTP(10)}
& ARI         & 0.993  & (0.014) & 0.985  & (0.038) & 0.806  & (0.083) & 0.835  & (0.081) \\
& & & $L_2$-error & 0.042  & (0.007) & 0.055  & (0.012) & 0.122  & (0.021) & 0.116  & (0.017) \\
& & & Time (s)    & 12.4   & (11.7)  & 36.6   & (34.2)  & 15.6   & (13.2)  & 131.4  & (128.6) \\
\cmidrule(lr){3-12}
& & \multirow[c]{3}{*}{ZTP(30)}
& ARI         & 0.971  & (0.047) & 0.975  & (0.067) & 0.958  & (0.071) & 0.959  & (0.087) \\
& & & $L_2$-error & 0.046  & (0.029) & 0.042  & (0.024) & 0.074  & (0.019) & 0.072  & (0.012) \\
& & & Time (s)    & 32.4   & (36.2)  & 117.1  & (115.5) & 42.3   & (57.6)  & 161.9  & (178.4) \\
\cmidrule(lr){3-12}
& & \multirow[c]{3}{*}{ZTP(50)}
& ARI         & 0.994  & (0.031) & 0.994  & (0.031) & 0.972  & (0.058) & 0.987  & (0.023) \\
& & & $L_2$-error & 0.031  & (0.011) & 0.031  & (0.012) & 0.068  & (0.017) & 0.067  & (0.009) \\
& & & Time (s)    & 57.2   & (67.9)  & 230.4  & (253.8) & 58.3   & (59.2)  & 214.2  & (224.6) \\
\cmidrule(lr){2-12}
& \multirow[c]{12}{*}{$n=300$}
& \multirow[c]{3}{*}{ZTP(3)}
& ARI         & 0.685  & (0.035) & 0.672  & (0.041) & 0.372  & (0.047) & 0.370  & (0.046) \\
& & & $L_2$-error & 0.124  & (0.010) & 0.136  & (0.016) & 0.236  & (0.012) & 0.237  & (0.012) \\
& & & Time (s)    & 19.0   & (19.8)  & 63.1   & (58.3)  & 24.4   & (19.3)  & 177.3  & (126.3) \\
\cmidrule(lr){3-12}
& & \multirow[c]{3}{*}{ZTP(10)}
& ARI         & 0.994  & (0.012) & 0.996  & (0.005) & 0.863  & (0.033) & 0.861  & (0.033) \\
& & & $L_2$-error & 0.034  & (0.004) & 0.033  & (0.002) & 0.083  & (0.007) & 0.087  & (0.007) \\
& & & Time (s)    & 28.9   & (28.3)  & 132.7  & (137.5) & 44.9   & (47.8)  & 162.6  & (147.8) \\
\cmidrule(lr){3-12}
& & \multirow[c]{3}{*}{ZTP(30)}
& ARI         & 1.000  & (0.000) & 1.000  & (0.000) & 0.983  & (0.012) & 0.981  & (0.012) \\
& & & $L_2$-error & 0.028  & (0.002) & 0.028  & (0.002) & 0.042  & (0.003) & 0.055  & (0.004) \\
& & & Time (s)    & 89.9   & (90.8)  & 275.0  & (284.0) & 221.6  & (241.9) & 348.1  & (365.0) \\
\cmidrule(lr){3-12}
& & \multirow[c]{3}{*}{ZTP(50)}
& ARI         & 1.000  & (0.000) & 1.000  & (0.000) & 0.995  & (0.007) & 0.995  & (0.008) \\
& & & $L_2$-error & 0.027  & (0.002) & 0.026  & (0.002) & 0.039  & (0.003) & 0.049  & (0.006) \\
& & & Time (s)    & 152.8  & (144.8) & 412.1  & (426.6) & 340.1  & (384.1) & 511.3  & (529.6) \\
\bottomrule
\end{tabular}
\endgroup
\end{table}

Tables~\ref{tab:sim_results_RE} and \ref{tab:sim_results_OU} report the mean and SD of the ARI, average $L_2$-error, and computation time in seconds over 30 independent replications under the RE and OU covariance structures, respectively. Overall, the proposed methods achieve high clustering accuracy and accurate mean-curve estimation across a wide range of settings. Their performance follows the expected pattern, with accuracy generally improving as $n$ and $\lambda$ increase and the noise level decreases.
An important aspect of the simulation study is the sensitivity of the proposed methods to the number of candidate knots. Overall, increasing the number of candidate knots from 30 to 100 leaves accuracy results largely unchanged under both covariance structures. This stability indicates that the shrinkage prior effectively suppresses redundant basis terms, thereby capturing both global and local features of the functions without overfitting.

Performance differs across the two mean-function scenarios. In Scenario~A, the RE model yields median ARI values ranging from 0.998 to 1.000 across all simulation settings. Under the OU covariance structure, clustering is also nearly perfect when $\lambda\geq10$. Scenario~B is more challenging, particularly when $\lambda=3$, where only a few irregularly sampled observations are available for each subject. Under such sparse sampling, lower clustering accuracy is expected because the observed time points may not adequately capture the regions needed to distinguish the clusters. Regarding computation time, the OU model exhibits greater variability in several settings. This variability arises from the variational update for $\tilde{\zeta}_k$. As detailed in the Appendix, one of three numerical approximation schemes is selected according to the current variational parameters, and these schemes differ in computational cost.

\subsection{Comparison with existing methods under misspecification}
\label{subsec:comparison}
We compare the proposed models with three model-based functional clustering methods: \textit{funHDDC} \citep{Bouveyron}, \textit{E-FDMP} \citep{Rigon}, and the model proposed by \citet{Xian} that includes a subject-specific random effect.
The \textit{funHDDC} method represents each cluster using a cluster-specific low-dimensional functional subspace and estimates the subspace and mixture parameters through the expectation-maximization algorithm. It considers six parsimonious covariance models obtained by imposing different constraints on the cluster-specific covariance parameters. \textit{E-FDMP} uses an enriched Dirichlet process mixture and functional constraints to incorporate prior information about curve shapes, with posterior inference performed using VI. \citet{Xian} proposed a functional clustering method that jointly performs smoothing and clustering using VI, with models that either exclude or include a subject-specific random effect. We use the model including the random effect and refer to it as \textit{Xian-RE}. The implementation of \textit{funHDDC} is available in the R package \texttt{funHDDC}, while the implementations of \textit{E-FDMP} and \textit{Xian-RE} are available in the GitHub repositories provided by their authors.

We consider two misspecified error structures. The first is a latent smooth process (LSP),
\[
\epsilon_i(t_{ij})
=
\mathbf z(t_{ij})^T \mathbf a_i
+
e_i(t_{ij}),
\]
where
$\mathbf z(t)
=
\left(
\sin\left(\frac{2\pi\omega t}{T}\right),
\cos\left(\frac{2\pi\omega t}{T}\right)
\right)^T,$
$\mathbf a_i
\sim
\text{N}\left(\mathbf 0,\tau^2\mathbf I_2\right),$
and
$e_i(t_{ij})
\sim
\text{N}\left(0,\sigma^2\right),$
with $\omega=0.8$ and $\tau=0.14$. The second is a moving-average process of order 2, denoted by MA(2),
\[
\epsilon_i(t_{ij})
=
\sum_{r=0}^{2}
\theta_r e_i(t_{i,j-r}),
\]
where
$(\theta_0,\theta_1,\theta_2)
=
(1,0.8,0.3),$
and
$e_i(t_{ij})
\sim
\text{N}\left(
0,
\sigma^2/{\sum_{r=0}^{2}\theta_r^2}
\right).$
Unlike the simulation study in Section~\ref{subsec:proposed}, where the number of observations $m_i$
may vary across units, we consider a common observation grid in this
comparison because the competing methods require all functional
observations to be measured at the same time points. Accordingly, each unit is observed at the same $m$ equally spaced time points, and we consider the relatively dense settings $m\in\{30,50\}$ to ensure sufficient coverage of the functional domain. For each error structure, we combine the two values of $m$ with the two mean-function scenarios
introduced in Section~\ref{subsec:proposed} (Scenario~A and Scenario~B),
two sample sizes $n\in\{100,300\}$, and two noise levels $\sigma\in\{0.1,0.3\}$,
yielding 16 simulation settings. An unbalanced clustering structure is generated in the same manner as in Section~\ref{subsec:proposed}, with cluster probabilities $(0.5, 0.3, 0.2)$.

For the proposed models and \textit{E-FDMP}, in which the number of clusters is determined adaptively, the truncation level is fixed at 16. This upper bound corresponds to the largest candidate range over which \textit{funHDDC} can be fitted across all simulation settings and is sufficiently larger than the true number of clusters. In contrast, \textit{funHDDC} and \textit{Xian-RE}, which require the number of clusters to be specified in advance, are fitted with candidate numbers of clusters ranging from 2 to 16, and the final models are selected using the BIC and ELBO, respectively.
Since the original studies provide little guidance on basis selection, we set the number of cubic B-spline basis functions to $\min(m/4, 40)$, adapting the standard heuristic that balances flexibility against overfitting \citep{ruppert2002selecting}. Accordingly, we used 8 basis functions for $m=30$ and 12 basis functions for $m=50$.
For \textit{Xian-RE}, we use weakly informative proper priors for the corresponding hyperparameters, since the original paper does not fully specify the informative prior settings.
Following the fitting strategies adopted in the original studies,
multiple random initializations are also used for \textit{funHDDC},
\textit{E-FDMP}, and \textit{Xian-RE}. To ensure comparable treatment
across methods, 30 initializations are used for these methods and RE.
In contrast, OU is fitted using a single initialization, as it showed
stable performance with a single start in Section~\ref{subsec:proposed}. The remaining settings for the competing methods follow their original simulation studies. For our models, any unspecified settings for the proposed models are set as in Section~\ref{subsec:proposed}. 

\begin{figure}[t!]
    \centering
    \includegraphics[width=1\textwidth]{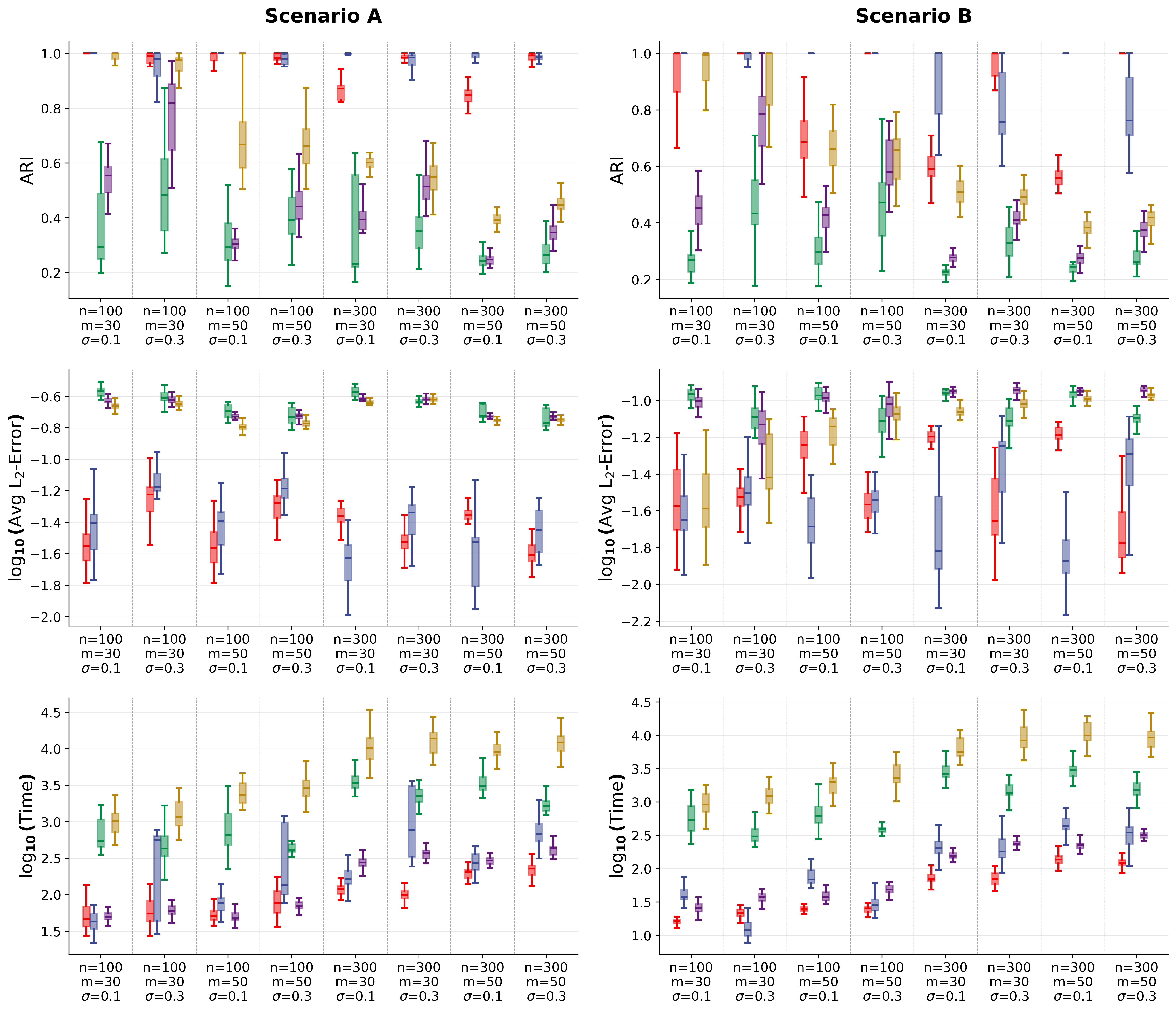}
    
    \includegraphics[width=0.5\textwidth]{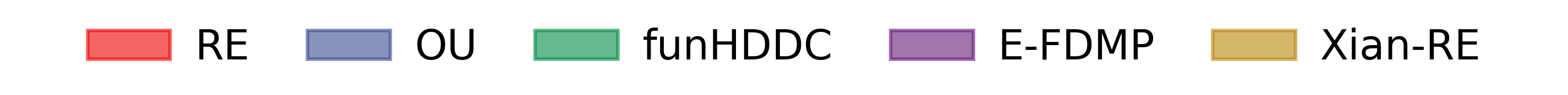}
    
    \caption{Comparison of clustering and estimation performance across sixteen simulation settings under the LSP error structure. The left and right columns represent Scenario A and Scenario B, respectively, each consisting of eight simulation settings. From top to bottom, the panels show the ARI, the common logarithm of the average $L_2$-error, and the common logarithm of the computation time (in seconds), reported as boxplots across 30 replications.}
    \label{fig:comparison_LSP}
\end{figure} 

\begin{figure}[t!]
    \centering
    \includegraphics[width=1\textwidth]{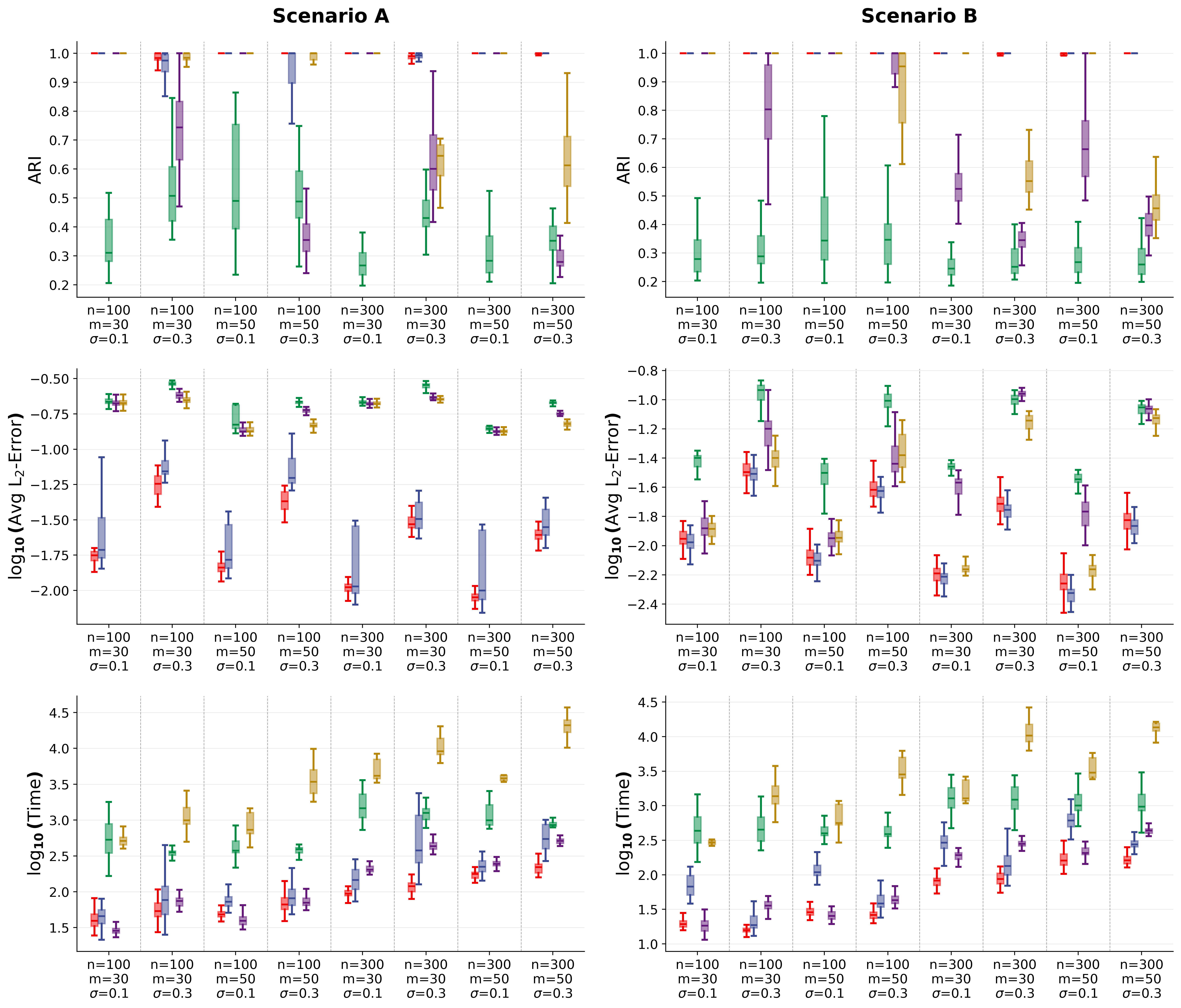}
    
    \includegraphics[width=0.5\textwidth]{plots/boxplot_legend.png}
    
    \caption{Comparison of clustering and estimation performance across sixteen simulation settings under the MA(2) error structure. The left and right columns represent Scenario A and Scenario B, respectively, each consisting of eight simulation settings. From top to bottom, the panels show the ARI, the common logarithm of the average $L_2$-error, and the common logarithm of the computation time (in seconds), reported as boxplots across 30 replications.}
    \label{fig:comparison_MA}
\end{figure} 

Figures~\ref{fig:comparison_LSP} and \ref{fig:comparison_MA} report the boxplots of the ARI, log average $L_2$-error, and log computation time across 30 independent replications for the five methods under the LSP and MA(2) error structures, respectively, across the 16 settings.
Across almost all simulation settings, OU and RE achieve the highest ARI and the lowest average $L_2$-error under both error structures. 
In contrast, \textit{funHDDC} consistently performs poorly under both error structures. It records the lowest ARI and the largest average $L_2$-error in most simulation settings. 
Under the LSP error structure, the ARI of E-FDMP consistently increases as $\sigma$ increases from 0.1 to 0.3, holding all other simulation conditions fixed. 
A possible explanation is that a larger value of $\sigma$ makes the overall error process closer to an unstructured noise model compatible with the modeling assumptions of E-FDMP. 
These results suggest that the performance of E-FDMP depends strongly on the interaction between the error structure and the simulation setting rather than exhibiting uniform robustness.
The performance of \textit{Xian-RE} is strongly affected by the noise level. Under MA(2), its median ARI decreases as $\sigma$ increases from 0.1 to 0.3, while the average running time increases. In contrast, OU and RE maintain stable ARI values and computation times across different noise levels.

There are only a few simulation settings in which OU or RE does not attain the best performance in terms of ARI or average $L_2$-error; in these cases, the performance gap is small and is accompanied by a substantial reduction in computation time.
Overall, OU and RE consistently achieve high clustering accuracy and low estimation error under both misspecified error structures. These results demonstrate that the proposed methods provide a favorable balance between robustness, estimation accuracy, and computational efficiency under model misspecification.

\section{Application}
\label{sec:Application}
We further evaluated the proposed method using three real-world functional datasets: the Berkeley growth dataset, the Canadian weather dataset, and the Italy power demand dataset. For each dataset, an appropriate covariance structure was selected based on the characteristics of the data, and the resulting clustering patterns were analyzed and interpreted.

\subsection{Berkeley growth data}
The Berkeley growth dataset is a longitudinal dataset consisting of repeated height measurements of children and adolescents \citep{growth_data}. The dataset contains 93 subjects, each observed from ages 1 to 18, with a total of 31 measurements recorded at intervals ranging from 0.25 to 1 year. Although the subjects share a common overall growth trend, individual-specific deviations are evident in both growth rate and growth timing. To capture such subject-level heterogeneity, we employed the RE model. Raw data of growth curves indicates substantial variability across individuals in their baseline height, growth rate, and overall curvature of the growth trajectory. Based on these characteristics, we chose the RE structure with an intercept, a slope, and a quadratic term. Furthermore, the growth trajectories are relatively smooth and contain a moderate number of observations per subject, making a large number of knots unnecessary; therefore, the default setting of 30 knots was used throughout the analysis.

\begin{figure}[t]
    \centering
    \includegraphics[width=0.55\textwidth]{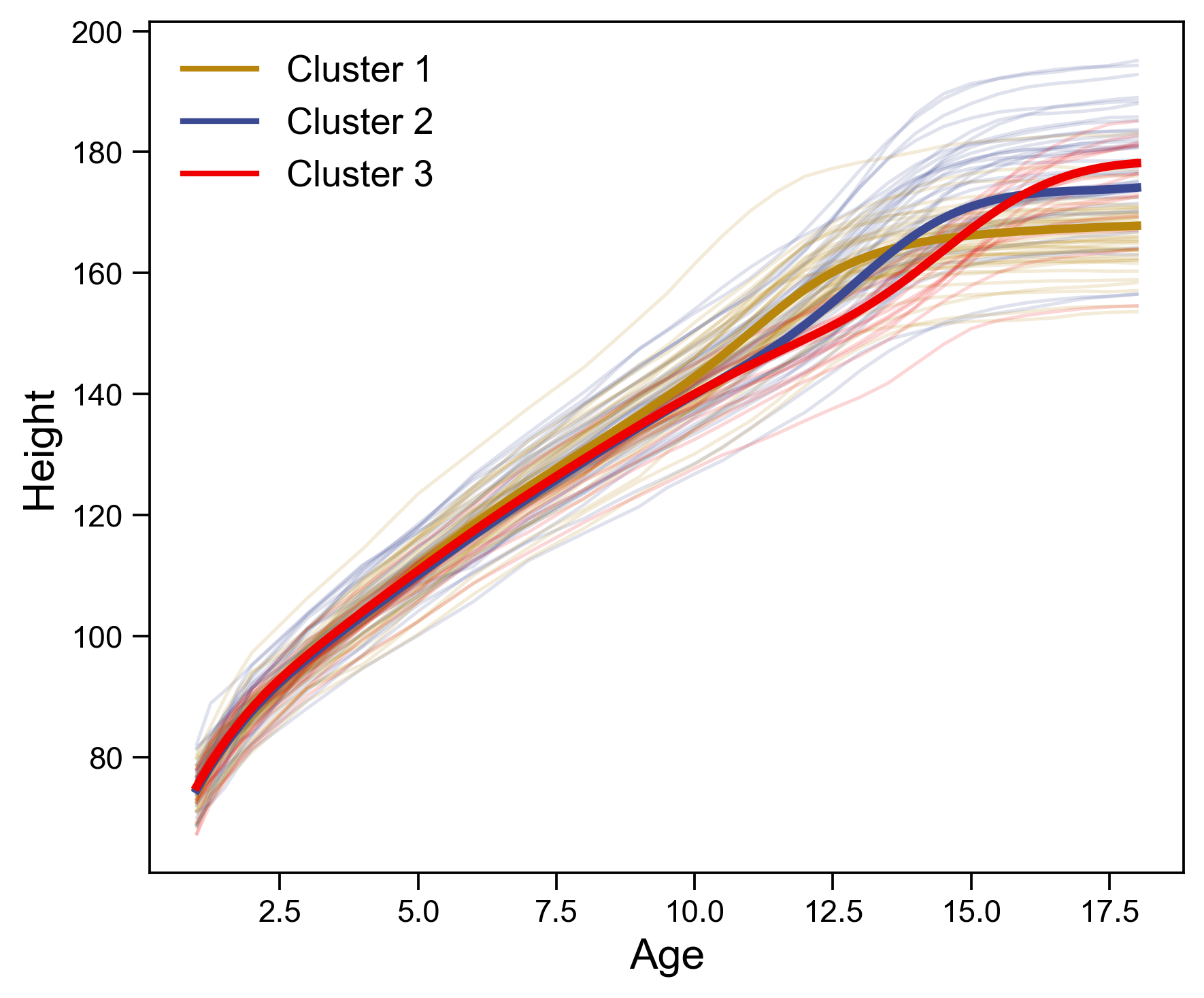}
    \caption{Clustering results for the Berkeley growth data obtained from the RE model. Bold curves represent the estimated mean curves for each cluster, while thin faded lines indicate the raw observed trajectories.}
    \label{fig:growth_plot}
\end{figure}

As shown in Figure~\ref{fig:growth_plot}, three clusters were identified, each exhibiting a distinct growth pattern. Cluster 1 showed a relatively early slowdown in growth, beginning around ages 12–14, and tended to have a smaller final height than the other clusters. Cluster 2 and generally attained a greater final height and maintained its growth trajectory for a longer period than Cluster 1, with the growth rate gradually slowing around ages 14–16. Cluster 3 showed steady growth through the latest ages. 
Interestingly, although the Berkeley growth dataset contains gender information, the discovered cluster structure does not appear to be determined solely by gender. For example, Cluster 2 contains both 27 males and 11 females. This result suggests that the proposed model is capable of identifying clusters based on individual growth patterns rather than merely separating subjects according to average height levels or gender.

\subsection{Canadian weather data}
The Canadian weather dataset contains daily average temperature measurements collected at 35 weather stations across Canada \citep{weather_data}. Each station is observed at the same 365 daily time points over one year. A distinctive feature of the data is that summer temperatures are relatively similar across regions, whereas winter temperatures vary substantially. Based on these characteristics, we adopted the RE model. Because nearby temperature curves often have similar shapes but differ in their overall levels, we use only a random intercept in the random-effects structure. Regarding the number of knots, the temperature curves exhibit relatively simple and smooth annual patterns. We therefore use 30 candidate knots, as in the Berkeley growth data.

\begin{figure}[t!]
    \centering
    \begin{minipage}[c]{0.75\textwidth}
        \centering
        \begin{subfigure}[t]{\linewidth}
            \centering
            \includegraphics[width=0.9\linewidth]{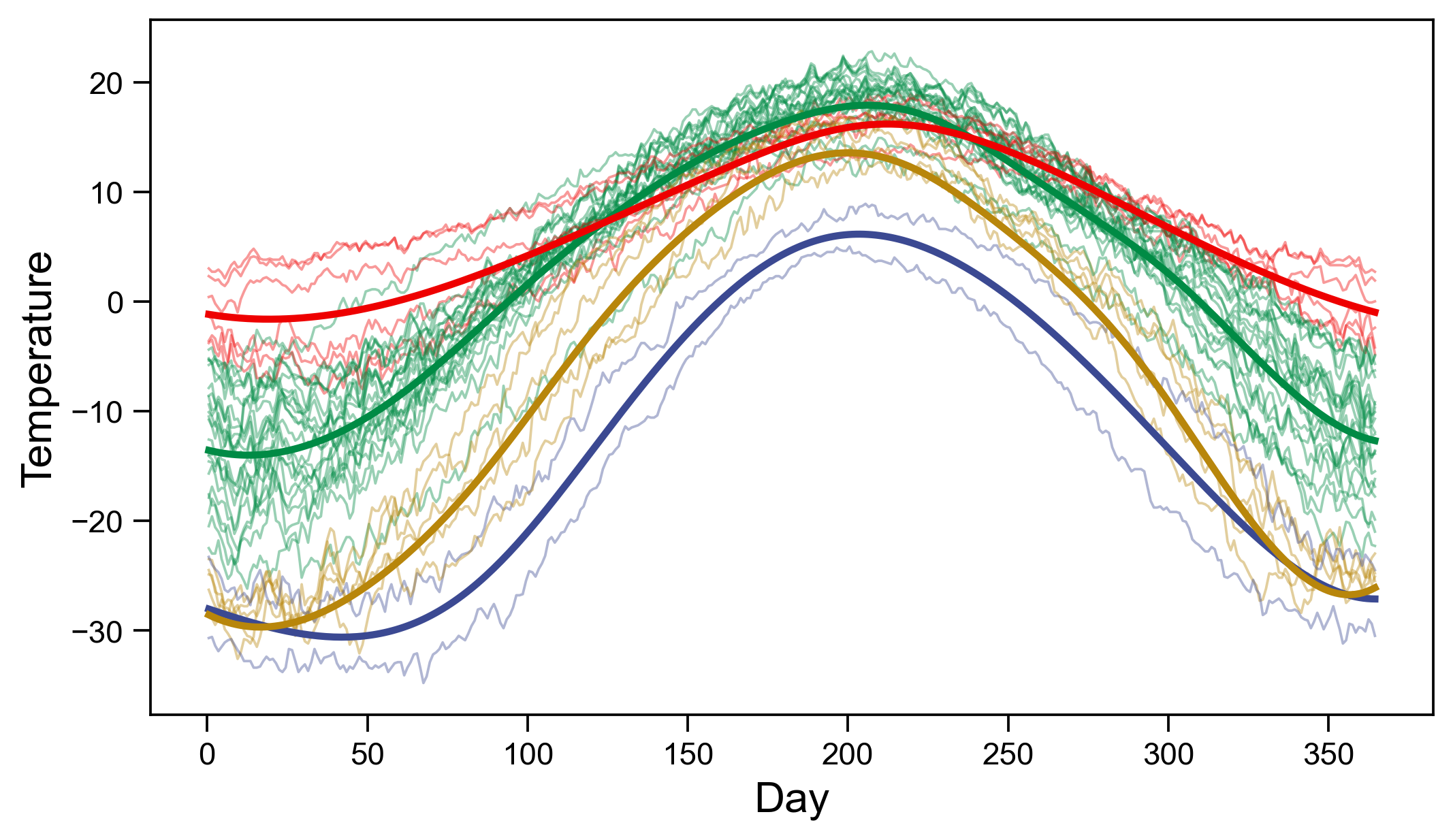}
            \label{fig:weather_plot}
        \end{subfigure}
        
        \vspace{0.3em}

        \begin{subfigure}[t]{\linewidth}
            \centering
            \hspace{2.5em}
            \includegraphics[width=0.83\linewidth]{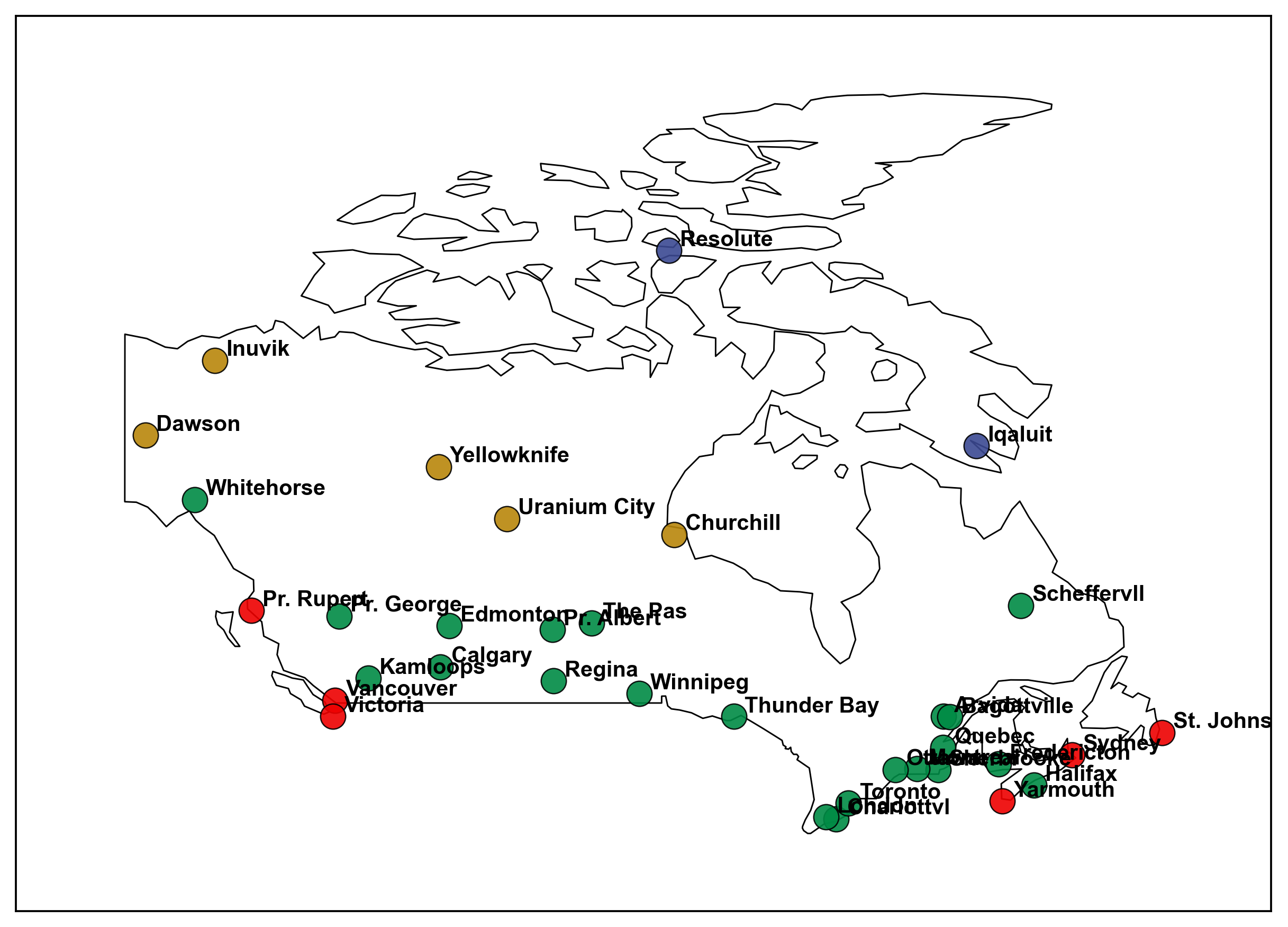}
            \label{fig:weather_map}
        \end{subfigure}
    \end{minipage}%
    \hspace{-2.5em}
    \begin{minipage}[c]{0.15\textwidth}
        \centering
        \raisebox{5.5em}{\includegraphics[width=\linewidth]{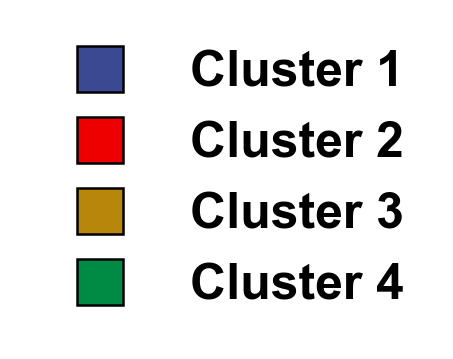}}
    \end{minipage}

    \caption{Clustering results for the Canadian weather data obtained from the RE model. In the top panel, bold curves represent the estimated mean curves for each cluster, while thin faded lines indicate the raw observed trajectories.}
    \label{fig:weather_results}
\end{figure}

Figure~\ref{fig:weather_results} presents the clustering results and their visualization on the map of Canada. The estimated clusters exhibit strong geographical coherence, with neighboring regions tending to be assigned to the same cluster. To further illustrate the characteristics of the identified climate clusters, we briefly describe several representative clusters. Cluster 1 primarily consists of Arctic stations and is characterized by the largest seasonal amplitude among all clusters. Clusters 2 corresponds mainly to the Pacific and Atlantic coastal regions. Due to the influence of the oceans, both clusters exhibit considerably smaller seasonal amplitudes than inland regions. Cluster 3 consists of northern inland stations, including Inuvik, Yellowknife, and Churchill. Although winter temperatures in this cluster are nearly as low as those observed in Cluster 1, summer temperatures increase substantially. As a result, Cluster 3 exhibits an even larger seasonal amplitude than the Arctic cluster and can be interpreted as representing a typical northern continental climate regime.

\subsection{Italy power demand data}
The Italy power demand dataset consists of 1,096 daily electricity demand trajectories, each observed at 24 hourly time points \citep{italy_data}. The data contain two seasonal classes corresponding to April--September and October--March. The trajectories exhibit differences in the timing and rate of local increases and decreases rather than simple between-unit differences in intercept or slope. We therefore employed the OU model rather than the RE model. For the spline basis, 30 candidate knots were used.

\begin{figure}[t]
    \centering
    \includegraphics[width=0.99\textwidth]{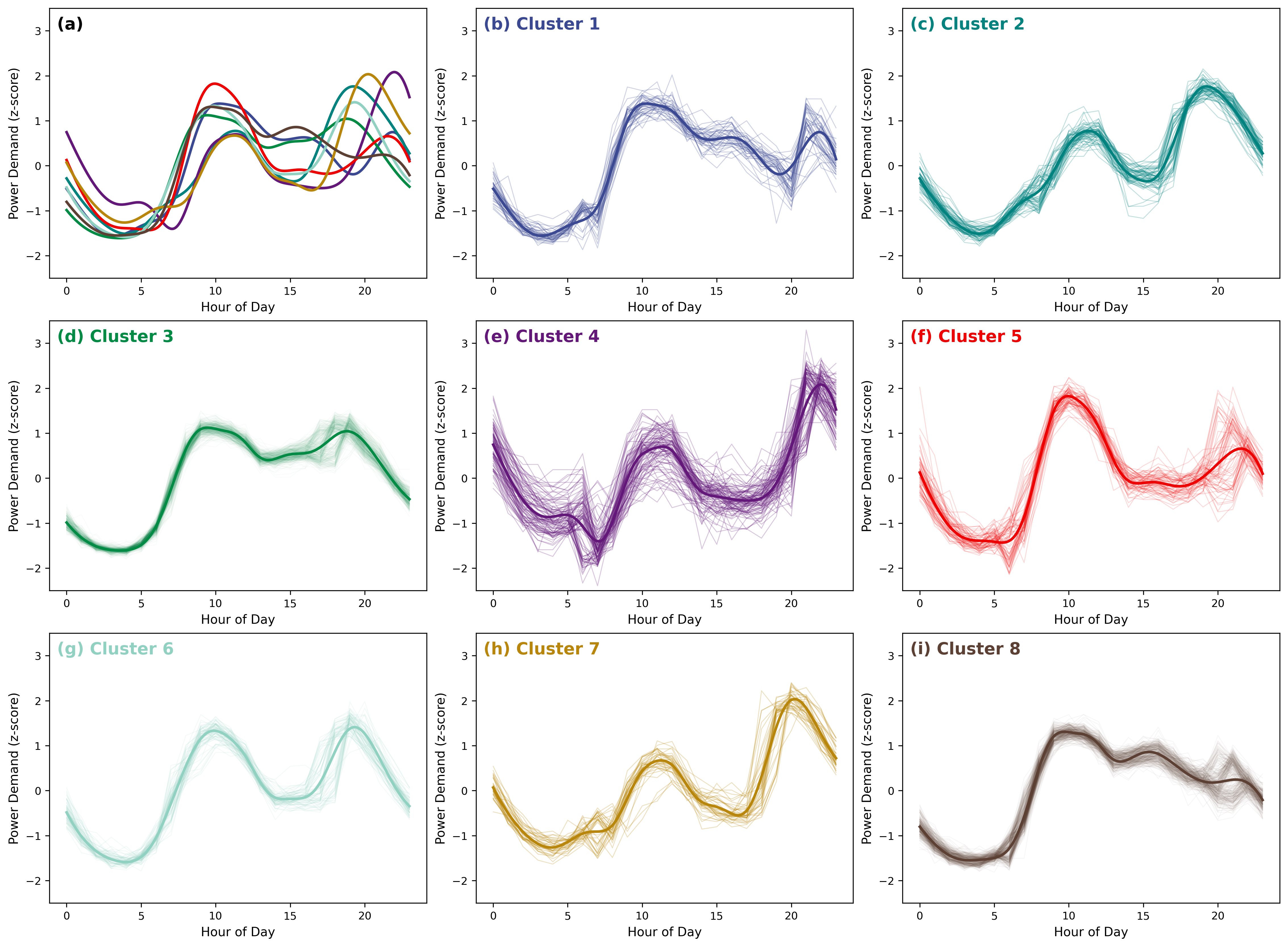}
    \caption{Clustering results for the Italy power demand data obtained from the OU model. Panel (a) shows the estimated mean curves for all eight clusters overlaid, with each cluster shown in its own color (consistent across all panels). Panels (b)--(i) show each cluster individually, corresponding to Clusters 1--8: bold curves represent the estimated mean curve for that cluster, while thin faded lines indicate the raw observed power demand trajectories assigned to it.}
    \label{fig:italy_plot}
\end{figure}

The OU model identified eight clusters with distinct daily demand patterns, as shown in Figure~\ref{fig:italy_plot}. To briefly highlight a few representative clusters, Clusters 2, 4, and 7 exhibit relatively late first peaks around 11 a.m.--12 p.m. and pronounced evening peaks. These patterns may be interpreted as being consistent with non-working-day demand patterns in Italy, where morning activity tends to be delayed and evening demand becomes more prominent. In contrast, Clusters 3 and 8, the two largest clusters, show a rapid increase in demand around 7--9 a.m. followed by sustained daytime activity. These patterns may be interpreted as typical weekday demand profiles. The two clusters differ in their evening behavior: Cluster 3 exhibits a distinct second peak, whereas Cluster 8 shows a weaker evening peak. More than 97\% of the observations in Clusters 3 and 8 correspond to October--March and April--September, respectively. This difference may partly reflect greater heating-related electricity demand during the colder months, together with the relatively limited use of air conditioning during the period represented by the data \citep{italy_data}.

\section{Discussion}
\label{sec:discussion}
In this paper, we proposed a Bayesian functional clustering framework that jointly performs clustering and cluster-specific mean curve estimation. The proposed framework combines a DP mixture with hyper-lasso shrinkage priors on spline bases. The DP mixture allows the number of occupied clusters to be inferred from the data. The hyper-lasso prior provides data-adaptive regularization of the spline coefficients and reduces the need to carefully tune the number of candidate knots in advance. To account for within-curve dependence, we considered RE and OU covariance structures. For both models, we developed CAVI algorithms that replace sampling-based posterior computation with deterministic optimization. The resulting framework therefore addresses cluster-number selection, smoothness control, within-curve dependence, and computational efficiency within a unified Bayesian model.

Several limitations remain. First, the current framework does not incorporate subject-level covariates. Scalar or functional covariates could be incorporated into the mean structure or cluster allocation mechanism. Second, the proposed inference relies on mean-field variational approximations, which may not fully capture posterior dependence. More flexible variational approximations could address this limitation, although at a higher computational cost.

\section*{Acknowledgment}

This research was supported by the National Research Foundation of Korea (NRF) grant funded by the Korean government (MSIT) (2022R1C1C1006735).

\appendix
\section*{Appendix A: Variational inference for the RE Model}
\label{app:re}

Under the mean-field assumption, the variational posterior measure factorizes as
\[
\mathcal Q_{\mathrm{RE}}(\boldsymbol\theta_{\mathrm{RE}})
=
\prod_{k=1}^{K-1}\mathcal Q_k^{1}(v_k)
\prod_{k=1}^{K}
\left[
\mathcal Q_k^{2}(\boldsymbol\beta_k,\phi_k)
\prod_{j=1}^{D-4}\mathcal Q_{k,j}^{3}(\tau_{k,j})
\mathcal Q_k^{4}(\lambda_k)
\mathcal Q_k^{5}(\mathbf Q_k)
\right]
\prod_{i=1}^{n}
\left[
\mathcal Q_i^{6}(\boldsymbol\xi_i)
\mathcal Q_i^{7}(z_i)
\right],
\]
where
\[
\boldsymbol\theta_{\mathrm{RE}}
=
\left(
v_{1:K-1},
\boldsymbol\beta_{1:K},
\phi_{1:K},
\boldsymbol\tau_{1:K},
\lambda_{1:K},
\mathbf Q_{1:K},
\boldsymbol\xi_{1:n},
z_{1:n}
\right).
\]
The corresponding density is denoted by
$q_{\mathrm{RE}}(\boldsymbol\theta_{\mathrm{RE}})$.
Each factor and its associated variational parameters are summarized in
Table~\ref{tab:re_notation}. Throughout this appendix, $D$ denotes the
dimension of the spline coefficient vector $\boldsymbol\beta_k$ in
\eqref{regression_spline}, and $\ell$ denotes the dimension of the
random effects vector $\boldsymbol\xi_i$, so that
$\mathbf Q_k\in\mathbb R^{\ell\times\ell}$.
The update of each variational factor and the resulting parameter expressions
are derived in Appendix~A.1, whereas the corresponding ELBO contributions are
derived in Appendix~A.2.
\begin{table}[h]
\centering
\caption{Variational factors and notation for the RE model.}
\label{tab:re_notation}
\begin{tabular}{ll}
\hline
Factor & Variational distribution \\
\hline
$v_k$ &
$\operatorname{Beta}(\gamma_{k,1},\gamma_{k,2})$ \\
$(\boldsymbol\beta_k,\phi_k)$ &
$\operatorname{NG}(\boldsymbol\nu_k,\boldsymbol\Omega_k,a_k,b_k)$ \\
$\tau_{k,j}$ &
$\operatorname{GIG}(1/2,c_{\tau_k},f_{\tau_{k,j}})$ \\
$\lambda_k$ &
$\operatorname{Gamma}(g,h_k)$ \\
$\mathbf Q_k$ &
$\operatorname{Wishart}(\mathbf S_k,r_k)$ \\
$\boldsymbol\xi_i$ &
$\text{N}(\boldsymbol\mu_i,\boldsymbol\Sigma_i)$ \\
$z_i$ &
$\operatorname{Discrete}(\varphi_{i,1},\ldots,\varphi_{i,K})$ \\
\hline
\end{tabular}
\end{table}

\subsection*{A.1 Updating rules}
\label{app:re:update}

The coordinate-ascent updates are derived under the mean-field assumption.
For the $l$th parameter block, the general CAVI update is
\[
q_l(\boldsymbol\theta_l)
\propto
\exp\left\{
\mathbb E_{-\boldsymbol\theta_l}
\left[
\log p(\boldsymbol\theta_{\mathrm{RE}},\mathbf y)
\right]
\right\},
\]
where $\mathbb E_{-\boldsymbol\theta_l}$ denotes expectation with respect to
all variational factors except $\mathcal Q_l$. The joint density of $\mathbf y$ and
$\boldsymbol\theta_{\mathrm{RE}}$ required for these updates is proportional to
\begin{align*}
p(\mathbf y,\boldsymbol\theta_{\mathrm{RE}})
&\propto
\prod_{i=1}^{n}
\prod_{k=1}^{K}
\left[
p(\mathbf y_i\mid z_i=k)\,
p(\boldsymbol\xi_i\mid z_i=k)\,
p(z_i=k\mid\mathbf v)
\right]
\\
&\quad\times
\prod_{k=1}^{K}
\left[
p(\boldsymbol\beta_k\mid\phi_k,\boldsymbol\tau_k)\,
p(\phi_k)\,
p(\boldsymbol\tau_k\mid\lambda_k)\,
p(\lambda_k)\,
p(\mathbf Q_k)
\right]
\prod_{k=1}^{K-1}p(v_k)
\\
&\propto
\prod_{i=1}^{n}
\prod_{k=1}^{K}
\left[
\phi_k^{m_i/2}
\exp\left\{
-\frac{\phi_k}{2}
\left\|
\mathbf y_i-\mathbf X_i\boldsymbol\beta_k
-\mathbf W_i\boldsymbol\xi_i
\right\|_2^2
\right\}
\right]^{\mathbbm 1(z_i=k)}
\\
&\quad\times
\prod_{i=1}^{n}
\prod_{k=1}^{K}
\left[
\phi_k^{\ell/2}
|\mathbf Q_k|^{1/2}
\exp\left\{
-\frac{\phi_k}{2}
\boldsymbol\xi_i^T\mathbf Q_k\boldsymbol\xi_i
\right\}
\right]^{\mathbbm 1(z_i=k)} \times
\prod_{i=1}^{n}
\prod_{k=1}^{K}
v_k^{\mathbbm 1(z_i=k)}
(1-v_k)^{\mathbbm 1(z_i>k)}
\\
&\quad\times
\prod_{k=1}^{K}
\left[
\phi_k^{D/2}
\left(
\prod_{j=1}^{D-4}\tau_{k,j}
\right)^{-1/2}
\exp\left\{
-\frac{\phi_k}{2}
\boldsymbol\beta_k^T
\mathbf C_{\boldsymbol\tau_k}^{-1}
\boldsymbol\beta_k
\right\}
\right]
\\
&\quad\times
\prod_{k=1}^{K}
\left[
\phi_k^{a_0-1}e^{-b_0\phi_k}
\prod_{j=1}^{D-4}
\left\{
\lambda_k e^{-\lambda_k\tau_{k,j}}
\right\}
\lambda_k^{g_0-1}e^{-h_0\lambda_k}
\right]
\\
&\quad\times
\prod_{k=1}^{K-1}
(1-v_k)^{\alpha-1}
\prod_{k=1}^{K}
\left[
|\mathbf Q_k|^{(r_{0}-\ell-1)/2}
\exp\left\{
-\frac{1}{2}
\operatorname{tr}
\left(
\mathbf S_{0}^{-1}\mathbf Q_k
\right)
\right\}
\right].
\end{align*}
Taking the logarithm and retaining only terms involving model parameters gives
\begin{align*}
\log p(\mathbf y,\boldsymbol\theta_{\mathrm{RE}})
&\propto
\sum_{i=1}^{n}
\sum_{k=1}^{K}
\mathbbm 1(z_i=k)
\Bigg[
-\frac{\phi_k}{2}
\left\|
\mathbf y_i-\mathbf X_i\boldsymbol\beta_k
-\mathbf W_i\boldsymbol\xi_i
\right\|_2^2
-\frac{\phi_k}{2}
\boldsymbol\xi_i^T\mathbf Q_k\boldsymbol\xi_i
+\log v_k
\Bigg]
\\
&\quad+
\sum_{k=1}^{K}
\Bigg[
-\frac{1}{2}
\sum_{j=1}^{D-4}\log\tau_{k,j}
-\frac{\phi_k}{2}
\boldsymbol\beta_k^T
\mathbf C_{\boldsymbol\tau_k}^{-1}
\boldsymbol\beta_k
-b_0\phi_k
-\lambda_k\sum_{j=1}^{D-4}\tau_{k,j}
-h_0\lambda_k
\Bigg]
\\
&\quad+
\sum_{k=1}^{K}
\left[
\frac{D}{2}+a_0-1
+\frac{1}{2}
\sum_{i=1}^{n}
(m_i+\ell)\mathbbm 1(z_i=k)
\right]
\log\phi_k
\\
&\quad+
\sum_{k=1}^{K}
(D-4+g_0-1)\log\lambda_k +
\sum_{k=1}^{K-1}
\left(
\alpha-1+\sum_{i=1}^{n}\mathbbm 1(z_i>k)
\right)
\log(1-v_k)
\\
&\quad+
\sum_{k=1}^{K}
\Bigg[
\frac{
\sum_{i=1}^{n}\mathbbm 1(z_i=k)
+r_{0}-\ell-1
}{2}
\log|\mathbf Q_k|
-\frac{1}{2}
\operatorname{tr}
\left(
\mathbf S_{0}^{-1}\mathbf Q_k
\right)
\Bigg].
\end{align*}

\subsubsection*{A.1.1 Updating $v_k$}

After collecting the terms in the log-joint density that depend on $v_k$
and taking the expectation with respect to all other variational factors,
we obtain
\begin{align*}
q_k^{1}(v_k)
&\propto
\exp\Bigg\{
\mathbb E_{\mathcal Q_{\mathrm{RE}},-v_k}
\Bigg[
\sum_{i=1}^{n}\mathbbm 1(z_i=k)\log v_k
+
\left(
\sum_{i=1}^{n}\mathbbm 1(z_i>k)
+\alpha-1
\right)
\log(1-v_k)
\Bigg]
\Bigg\}
\\
&\propto
v_k^{\sum_{i=1}^{n}\varphi_{i,k}}
(1-v_k)^{
\alpha-1+
\sum_{i=1}^{n}\sum_{h=k+1}^{K}\varphi_{i,h}
}.
\end{align*}
This is the kernel of a Beta distribution. Therefore,
\[
\mathcal Q_k^{1}(v_k)
=
\operatorname{Beta}(\gamma_{k,1},\gamma_{k,2}),
\qquad k=1,\ldots,K-1,
\]
where
\begin{align*}
\gamma_{k,1}
&=
1+\sum_{i=1}^{n}\varphi_{i,k},
\\
\gamma_{k,2}
&=
\alpha
+\sum_{i=1}^{n}\sum_{h=k+1}^{K}\varphi_{i,h}.
\end{align*}
The two moments of $v_k$ used in subsequent updates are
\begin{align}
\mathbb E_{\mathcal Q_{\mathrm{RE}}}[\log v_k]
&=
\Psi(\gamma_{k,1})
-\Psi(\gamma_{k,1}+\gamma_{k,2}),
\nonumber\\
\mathbb E_{\mathcal Q_{\mathrm{RE}}}[\log(1-v_k)]
&=
\Psi(\gamma_{k,2})
-\Psi(\gamma_{k,1}+\gamma_{k,2}).
\label{eq:vk_moments}
\end{align}

\subsubsection*{A.1.2 Updating $\boldsymbol\beta_k$ and $\phi_k$}
\label{subsubsec:beta_re_update}

After collecting the terms in the log-joint density that involve
$\boldsymbol\beta_k$ and $\phi_k$ and taking the expectation with respect
to all other variational factors, we obtain
\begin{align*}
q_k^{2}(\boldsymbol\beta_k,\phi_k)
&\propto
\exp\Bigg\{
\mathbb E_{\mathcal Q_{\mathrm{RE}},
-(\boldsymbol\beta_k,\phi_k)}
\Bigg[
\sum_{i=1}^{n}
\mathbbm 1(z_i=k)
\Bigg\{
-\frac{\phi_k}{2}
\left\|
\mathbf y_i-\mathbf X_i\boldsymbol\beta_k
-\mathbf W_i\boldsymbol\xi_i
\right\|_2^2
-\frac{\phi_k}{2}
\boldsymbol\xi_i^T\mathbf Q_k\boldsymbol\xi_i
\Bigg\}
\\
&\quad
-\frac{\phi_k}{2}
\boldsymbol\beta_k^T
\mathbf C_{\boldsymbol\tau_k}^{-1}
\boldsymbol\beta_k
-b_0\phi_k
+
\left\{
\frac{D}{2}+a_0-1
+\frac{1}{2}
\sum_{i=1}^{n}
(m_i+\ell)\mathbbm 1(z_i=k)
\right\}
\log\phi_k
\Bigg]
\Bigg\}.
\end{align*}
Here, the residual quadratic form can be written as
\begin{align}
\left\|
\mathbf y_i-\mathbf X_i\boldsymbol\beta_k
-\mathbf W_i\boldsymbol\xi_i
\right\|_2^2
&=
\mathbf y_i^T\mathbf y_i
-2\boldsymbol\beta_k^T\mathbf X_i^T\mathbf y_i
-2\boldsymbol\xi_i^T\mathbf W_i^T\mathbf y_i
+\boldsymbol\beta_k^T\mathbf X_i^T\mathbf X_i\boldsymbol\beta_k
\nonumber\\
&\quad
+2\boldsymbol\beta_k^T
\mathbf X_i^T\mathbf W_i\boldsymbol\xi_i
+\boldsymbol\xi_i^T
\mathbf W_i^T\mathbf W_i\boldsymbol\xi_i.
\label{eq:re_residual_expansion}
\end{align}
Taking the expectation with respect to $\boldsymbol\xi_i$ gives
\begin{align*}
\mathbb E_{\mathcal Q_{\mathrm{RE}},-\boldsymbol\beta_k}
\left[
\left\|
\mathbf y_i-\mathbf X_i\boldsymbol\beta_k
-\mathbf W_i\boldsymbol\xi_i
\right\|_2^2
\right]
&=
\boldsymbol\beta_k^T
\mathbf X_i^T\mathbf X_i
\boldsymbol\beta_k
-
2\boldsymbol\beta_k^T
\mathbf X_i^T
\left(
\mathbf y_i-\mathbf W_i\boldsymbol\mu_i
\right)
+
\left\|
\mathbf y_i-\mathbf W_i\boldsymbol\mu_i
\right\|_2^2
\\
& \quad+
\operatorname{tr}
\left(
\mathbf W_i^T\mathbf W_i\boldsymbol\Sigma_i
\right).
\end{align*}
Similarly, using
$\mathbb E_{\mathcal Q_{\mathrm{RE}}}[\mathbf Q_k]
=r_k\mathbf S_k$ and
$\mathbb E_{\mathcal Q_{\mathrm{RE}}}
[\boldsymbol\xi_i\boldsymbol\xi_i^T]
=
\boldsymbol\mu_i\boldsymbol\mu_i^T+\boldsymbol\Sigma_i$,
the expectation of the RE quadratic form is
\begin{align*}
\mathbb E_{\mathcal Q_{\mathrm{RE}}}
\left[
\boldsymbol\xi_i^T\mathbf Q_k\boldsymbol\xi_i
\right]
&=
r_k
\operatorname{tr}
\left(
\mathbf S_k\boldsymbol\Sigma_i
\right)
+
r_k
\boldsymbol\mu_i^T
\mathbf S_k\boldsymbol\mu_i.
\end{align*}
Substituting these expectations into the variational update gives
\begin{align*}
q_k^{2}(\boldsymbol\beta_k,\phi_k)
&\propto
\phi_k^{
D/2+a_0-1+
\frac{1}{2}
\sum_{i=1}^{n}
(m_i+\ell)\varphi_{i,k}
}
\\
&\quad\times
\exp\Bigg\{
-\phi_k
\Bigg[
b_0
+
\frac{1}{2}
\sum_{i=1}^{n}
\varphi_{i,k}
\Bigg\{
\left\|
\mathbf y_i-\mathbf W_i\boldsymbol\mu_i
\right\|_2^2
+
\operatorname{tr}
\left(
\mathbf W_i^T\mathbf W_i\boldsymbol\Sigma_i
\right)
\\
&\quad
+
r_k
\left[
\boldsymbol\mu_i^T\mathbf S_k\boldsymbol\mu_i
+
\operatorname{tr}
\left(
\mathbf S_k\boldsymbol\Sigma_i
\right)
\right]
\Bigg\}
\Bigg]
\Bigg\}
\\
&\quad\times
\exp\Bigg\{
-\frac{\phi_k}{2}
\Bigg[
\boldsymbol\beta_k^T
\left\{
\mathbb E_{\mathcal Q_{\mathrm{RE}}}
\left[
\mathbf C_{\boldsymbol\tau_k}^{-1}
\right]
+
\sum_{i=1}^{n}
\varphi_{i,k}\mathbf X_i^T\mathbf X_i
\right\}
\boldsymbol\beta_k
\\
&\quad
-
2\boldsymbol\beta_k^T
\sum_{i=1}^{n}
\varphi_{i,k}
\mathbf X_i^T
\left(
\mathbf y_i-\mathbf W_i\boldsymbol\mu_i
\right)
\Bigg]
\Bigg\}.
\end{align*}
For notational convenience, define
\begin{align*}
\mathbf A_{1k}^{-1}
&=
\mathbb E_{\mathcal Q_{\mathrm{RE}}}
\left[
\mathbf C_{\boldsymbol\tau_k}^{-1}
\right]
+
\sum_{i=1}^{n}
\varphi_{i,k}
\mathbf X_i^T\mathbf X_i,
\\
\mathbf B_{1k}
&=
\sum_{i=1}^{n}
\varphi_{i,k}
\mathbf X_i^T
\left(
\mathbf y_i-\mathbf W_i\boldsymbol\mu_i
\right).
\end{align*}
Completing the square with respect to $\boldsymbol\beta_k$ yields
\begin{align*}
q_k^{2}(\boldsymbol\beta_k,\phi_k)
&\propto
\phi_k^{
a_0-1+
\frac{1}{2}
\sum_{i=1}^{n}
(m_i+\ell)\varphi_{i,k}
}
\\
&\quad\times
\exp\Bigg\{
-\phi_k
\Bigg[
b_0
+
\frac{1}{2}
\sum_{i=1}^{n}
\varphi_{i,k}
\Bigg\{
\left\|
\mathbf y_i-\mathbf W_i\boldsymbol\mu_i
\right\|_2^2
+
\operatorname{tr}
\left(
\mathbf W_i^T\mathbf W_i\boldsymbol\Sigma_i
\right)
\\
&\quad
+
r_k
\left[
\boldsymbol\mu_i^T\mathbf S_k\boldsymbol\mu_i
+
\operatorname{tr}
\left(
\mathbf S_k\boldsymbol\Sigma_i
\right)
\right]
\Bigg\}
-
\frac{1}{2}
\mathbf B_{1k}^T\mathbf A_{1k}\mathbf B_{1k}
\Bigg]
\Bigg\}
\\
&\quad\times
\left|
\phi_k^{-1}\mathbf A_{1k}
\right|^{-1/2}
\exp\Bigg\{
-\frac{1}{2}
\left(
\boldsymbol\beta_k-\mathbf A_{1k}\mathbf B_{1k}
\right)^T
\left(
\phi_k^{-1}\mathbf A_{1k}
\right)^{-1}
\left(
\boldsymbol\beta_k-\mathbf A_{1k}\mathbf B_{1k}
\right)
\Bigg\}.
\end{align*}
Therefore,
\[
\mathcal Q_k^{2}(\boldsymbol\beta_k,\phi_k)
=
\operatorname{NG}
\left(
\boldsymbol\nu_k,
\boldsymbol\Omega_k,
a_k,
b_k
\right),
\]
where
\begin{align*}
\boldsymbol\Omega_k
&=
\left(
\mathbb E_{\mathcal Q_{\mathrm{RE}}}
\left[
\mathbf C_{\boldsymbol\tau_k}^{-1}
\right]
+
\sum_{i=1}^{n}
\varphi_{i,k}\mathbf X_i^T\mathbf X_i
\right)^{-1},
\\
\boldsymbol\nu_k
&=
\boldsymbol\Omega_k
\sum_{i=1}^{n}
\varphi_{i,k}
\mathbf X_i^T
\left(
\mathbf y_i-\mathbf W_i\boldsymbol\mu_i
\right),\\
a_k
&=
a_0
+
\frac{1}{2}
\sum_{i=1}^{n}
(m_i+\ell)\varphi_{i,k},
\\
b_k
&=
b_0
+
\frac{1}{2}
\sum_{i=1}^{n}
\varphi_{i,k}
\Bigg[
\left\|
\mathbf y_i-\mathbf W_i\boldsymbol\mu_i
\right\|_2^2
+
\operatorname{tr}
\left(
\mathbf W_i^T\mathbf W_i\boldsymbol\Sigma_i
\right)
\\
&\quad
+
r_k
\left\{
\boldsymbol\mu_i^T\mathbf S_k\boldsymbol\mu_i
+
\operatorname{tr}
\left(
\mathbf S_k\boldsymbol\Sigma_i
\right)
\right\}
\Bigg]
-
\frac{1}{2}
\left[
\sum_{i=1}^{n}
\varphi_{i,k}
\mathbf X_i^T
\left(
\mathbf y_i-\mathbf W_i\boldsymbol\mu_i
\right)
\right]^T
\boldsymbol\nu_k.
\end{align*}
Here,
\begin{align}
\mathbb E_{\mathcal Q_{\mathrm{RE}}}
\left[
\mathbf C_{\boldsymbol\tau_k}^{-1}
\right]
&=
\operatorname{blockdiag}
\left\{
\rho^{-1}\mathbf I_4,\,
\operatorname{diag}
\left(
\mathbb E_{\mathcal Q_{\mathrm{RE}}}[\tau_{k,1}^{-1}],
\ldots,
\mathbb E_{\mathcal Q_{\mathrm{RE}}}[\tau_{k,D-4}^{-1}]
\right)
\right\}.
\label{eq:Ctau_expectation}
\end{align}
The moments used in the remaining updates are
\[
\mathbb E_{\mathcal Q_{\mathrm{RE}}}[\boldsymbol\beta_k]
=
\boldsymbol\nu_k,
\qquad
\mathbb E_{\mathcal Q_{\mathrm{RE}}}[\phi_k]
=
\frac{a_k}{b_k}.
\]

\subsubsection*{A.1.3 Updating $\tau_{k,j}$}

After collecting the terms in the log-joint density that involve
$\tau_{k,j}$ and taking the expectation with respect to all other
variational factors, we obtain
\begin{align*}
q_{k,j}^{3}(\tau_{k,j})
&\propto
\exp\Bigg\{
\mathbb E_{\mathcal Q_{\mathrm{RE}},-\tau_{k,j}}
\left[
-\frac{1}{2}\log\tau_{k,j}
-\frac{\phi_k}{2\tau_{k,j}}
(\boldsymbol\beta_k)_{j+4}^{2}
-\lambda_k\tau_{k,j}
\right]
\Bigg\}
\\
&\propto
\tau_{k,j}^{-1/2}
\exp\left\{
-\frac{1}{2}
\left(
c_{\tau_k}\tau_{k,j}
+
\frac{f_{\tau_{k,j}}}{\tau_{k,j}}
\right)
\right\},
\end{align*}
where
\begin{align*}
c_{\tau_k}
&=
2\,\mathbb E_{\mathcal Q_{\mathrm{RE}}}[\lambda_k],
\\
f_{\tau_{k,j}}
&=
\frac{a_k}{b_k}(\boldsymbol\nu_k)^2_{j+4}
+
(\boldsymbol\Omega_k)_{j+4,j+4}.
\end{align*}
This is the kernel of a generalized inverse Gaussian distribution. Therefore,
\[
\mathcal Q_{k,j}^{3}(\tau_{k,j})
=
\operatorname{GIG}
\left(
\frac12,c_{\tau_k},f_{\tau_{k,j}}
\right).
\]
The update of
$\mathbb E_{\mathcal Q_{\mathrm{RE}}}[\lambda_k]$
is derived in the next subsection. The moments of $\tau_{k,j}$ required in
the remaining updates follow from the properties of the generalized inverse
Gaussian distribution \citep{expectation_tau}:
\begin{align}
\mathbb E_{\mathcal Q_{\mathrm{RE}}}[\tau_{k,j}]
&=
\frac{
\sqrt{f_{\tau_{k,j}}}\,
K_{3/2}(\sqrt{c_{\tau_k}f_{\tau_{k,j}}})
}{
\sqrt{c_{\tau_k}}\,
K_{1/2}(\sqrt{c_{\tau_k}f_{\tau_{k,j}}})
},
\label{eq:tau_moment}
\\
\mathbb E_{\mathcal Q_{\mathrm{RE}}}[\tau_{k,j}^{-1}]
&=
\frac{
\sqrt{c_{\tau_k}}\,
K_{3/2}(\sqrt{c_{\tau_k}f_{\tau_{k,j}}})
}{
\sqrt{f_{\tau_{k,j}}}\,
K_{1/2}(\sqrt{c_{\tau_k}f_{\tau_{k,j}}})
}
-
\frac{1}{f_{\tau_{k,j}}},
\label{eq:tau_inv_moment}
\\
\mathbb E_{\mathcal Q_{\mathrm{RE}}}[\log\tau_{k,j}]
&=
\frac12
\log\left(
\frac{f_{\tau_{k,j}}}{c_{\tau_k}}
\right)
+
\left.
\frac{\partial}{\partial p}
\log
K_p\left(
\sqrt{c_{\tau_k}f_{\tau_{k,j}}}
\right)
\right|_{p=1/2}.
\label{eq:tau_log_moment}
\end{align}
Here, $K_p(\cdot)$ denotes the modified Bessel function of the second kind
of order $p$.

\subsubsection*{A.1.4 Updating $\lambda_k$}

After collecting the terms in the log-joint density that involve
$\lambda_k$ and taking the expectation with respect to all other
variational factors, we obtain
\begin{align*}
q_k^{4}(\lambda_k)
&\propto
\exp\Bigg\{
\mathbb E_{\mathcal Q_{\mathrm{RE}},-\lambda_k}
\left[
(g_0+D-4-1)\log\lambda_k
-\lambda_k\sum_{j=1}^{D-4}\tau_{k,j}
-h_0\lambda_k
\right]
\Bigg\}
\\
&\propto
\lambda_k^{g_0+D-4-1}
\exp\left\{
-\lambda_k
\left(
h_0
+
\sum_{j=1}^{D-4}
\mathbb E_{\mathcal Q_{\mathrm{RE}}}[\tau_{k,j}]
\right)
\right\}.
\end{align*}
This is the kernel of a Gamma distribution. Therefore,
\[
\mathcal Q_k^{4}(\lambda_k)
=
\operatorname{Gamma}(g_0+D-4,h_k),
\]
where
\begin{align*}
h_k
&=
h_0
+
\sum_{j=1}^{D-4}
\frac{
\sqrt{f_{\tau_{k,j}}}\,
K_{3/2}(\sqrt{c_{\tau_k}f_{\tau_{k,j}}})
}{
\sqrt{c_{\tau_k}}\,
K_{1/2}(\sqrt{c_{\tau_k}f_{\tau_{k,j}}})
}.
\end{align*}
The moments of $\lambda_k$ used in the remaining updates are
\[
\mathbb E_{\mathcal Q_{\mathrm{RE}}}[\lambda_k]
=
\frac{g_0+D-4}{h_k},
\qquad
\mathbb E_{\mathcal Q_{\mathrm{RE}}}[\log\lambda_k]
=
\Psi(g_0+D-4)-\log h_k.
\]

\subsubsection*{A.1.5 Updating $\mathbf Q_k$}

After collecting the terms in the log-joint density that involve
$\mathbf Q_k$ and taking the expectation with respect to all other
variational factors, we obtain
\begin{align*}
q_k^{5}(\mathbf Q_k)
&\propto
\exp\Bigg\{
\mathbb E_{\mathcal Q_{\mathrm{RE}},-\mathbf Q_k}
\Bigg[
-\frac12
\sum_{i=1}^{n}
\mathbbm 1(z_i=k)
\phi_k
\boldsymbol\xi_i^T\mathbf Q_k\boldsymbol\xi_i
+
\frac{
\sum_{i=1}^{n}\mathbbm 1(z_i=k)
+r_0-\ell-1
}{2}
\log|\mathbf Q_k|
\\
&\quad
-\frac12
\operatorname{tr}
\left(
\mathbf S_0^{-1}\mathbf Q_k
\right)
\Bigg]
\Bigg\}.
\end{align*}
Using
$\mathbb E_{\mathcal Q_{\mathrm{RE}}}[\phi_k]=a_k/b_k$ and
$\mathbb E_{\mathcal Q_{\mathrm{RE}}}
[\boldsymbol\xi_i\boldsymbol\xi_i^T]
=
\boldsymbol\mu_i\boldsymbol\mu_i^T+\boldsymbol\Sigma_i$,
we have
\[
\mathbb E_{\mathcal Q_{\mathrm{RE}},-\mathbf Q_k}
\left[
\phi_k
\boldsymbol\xi_i^T\mathbf Q_k\boldsymbol\xi_i
\right]
=
\frac{a_k}{b_k}
\operatorname{tr}
\left[
\left(
\boldsymbol\mu_i\boldsymbol\mu_i^T+\boldsymbol\Sigma_i
\right)
\mathbf Q_k
\right].
\]
Hence,
\begin{align*}
q_k^{5}(\mathbf Q_k)
&\propto
|\mathbf Q_k|^{
\left(
\sum_{i=1}^{n}\varphi_{i,k}
+r_0-\ell-1
\right)/2
}
\times
\exp\Bigg\{
-\frac12
\operatorname{tr}
\Bigg[
\Bigg\{
\mathbf S_0^{-1}
+
\sum_{i=1}^{n}
\varphi_{i,k}
\frac{a_k}{b_k}
\left(
\boldsymbol\mu_i\boldsymbol\mu_i^T+\boldsymbol\Sigma_i
\right)
\Bigg\}
\mathbf Q_k
\Bigg]
\Bigg\}.
\end{align*}
This is the kernel of a Wishart distribution. Therefore,
\[
\mathcal Q_k^{5}(\mathbf Q_k)
=
\operatorname{Wishart}(\mathbf S_k,r_k),
\]
where
\begin{align*}
\mathbf S_k
&=
\left[
\mathbf S_0^{-1}
+
\sum_{i=1}^{n}
\varphi_{i,k}
\frac{a_k}{b_k}
\left(
\boldsymbol\mu_i\boldsymbol\mu_i^T+\boldsymbol\Sigma_i
\right)
\right]^{-1},
\\
r_k
&=
r_0
+
\sum_{i=1}^{n}\varphi_{i,k}.
\end{align*}
Using the standard moments of the Wishart distribution
\citep{wishart_moments}, the quantities required in the remaining updates are
\begin{align*}
\mathbb E_{\mathcal Q_{\mathrm{RE}}}[\mathbf Q_k]
&=
r_k\mathbf S_k,
\nonumber\\
\mathbb E_{\mathcal Q_{\mathrm{RE}}}[\log|\mathbf Q_k|]
&=
\sum_{j=1}^{\ell}
\Psi\left(
\frac{r_k+1-j}{2}
\right)
+
\ell\log 2
+
\log|\mathbf S_k|.
\end{align*}

\subsubsection*{A.1.6 Updating $\boldsymbol\xi_i$}

After collecting the terms in the log-joint density that involve
$\boldsymbol\xi_i$ and taking the expectation with respect to all other
variational factors, we obtain
\begin{align*}
q_i^{6}(\boldsymbol\xi_i)
&\propto
\exp\Bigg\{
\mathbb E_{\mathcal Q_{\mathrm{RE}},-\boldsymbol\xi_i}
\Bigg[
\sum_{k=1}^{K}
\mathbbm 1(z_i=k)
\Bigg\{
-\frac{\phi_k}{2}
\left\|
\mathbf y_i-\mathbf X_i\boldsymbol\beta_k
-\mathbf W_i\boldsymbol\xi_i
\right\|_2^2
-\frac{\phi_k}{2}
\boldsymbol\xi_i^T\mathbf Q_k\boldsymbol\xi_i
\Bigg\}
\Bigg]
\Bigg\}.
\end{align*}
For the residual term, we use the expansion in
\eqref{eq:re_residual_expansion} and take the expectation with respect to
$(\boldsymbol\beta_k,\phi_k)$ while retaining its dependence on
$\boldsymbol\xi_i$. Therefore,
\begin{align*}
&\mathbb E_{\mathcal Q_{\mathrm{RE}},-\boldsymbol\xi_i}
\left[
\phi_k
\left\|
\mathbf y_i-\mathbf X_i\boldsymbol\beta_k
-\mathbf W_i\boldsymbol\xi_i
\right\|_2^2
\right]=
\frac{a_k}{b_k}
\left\|
\mathbf y_i-\mathbf X_i\boldsymbol\nu_k
-\mathbf W_i\boldsymbol\xi_i
\right\|_2^2
+
\operatorname{tr}
\left(
\mathbf X_i^T\mathbf X_i\boldsymbol\Omega_k
\right).
\end{align*}
For the RE quadratic term,
\begin{align*}
\mathbb E_{\mathcal Q_{\mathrm{RE}},-\boldsymbol\xi_i}
\left[
\phi_k
\boldsymbol\xi_i^T\mathbf Q_k\boldsymbol\xi_i
\right]
&=
\frac{a_k}{b_k}
r_k
\boldsymbol\xi_i^T\mathbf S_k\boldsymbol\xi_i.
\end{align*}
Substituting these expectations and retaining only terms that depend on
$\boldsymbol\xi_i$ gives
\begin{align*}
q_i^{6}(\boldsymbol\xi_i)
&\propto
\exp\Bigg\{
-\frac12
\boldsymbol\xi_i^T
\left[
\sum_{k=1}^{K}
\varphi_{i,k}
\frac{a_k}{b_k}
\left(
\mathbf W_i^T\mathbf W_i+r_k\mathbf S_k
\right)
\right]
\boldsymbol\xi_i
+
\boldsymbol\xi_i^T
\left[
\sum_{k=1}^{K}
\varphi_{i,k}
\frac{a_k}{b_k}
\left(
\mathbf W_i^T\mathbf y_i
-\mathbf W_i^T\mathbf X_i\boldsymbol\nu_k
\right)
\right]
\Bigg\}.
\end{align*}
As in the update of $(\boldsymbol\beta_k,\phi_k)$ in
Section~A.1.2, define
\begin{align*}
\mathbf A_{2i}^{-1}
&=
\sum_{k=1}^{K}
\varphi_{i,k}
\frac{a_k}{b_k}
\left(
\mathbf W_i^T\mathbf W_i+r_k\mathbf S_k
\right),
\\
\mathbf B_{2i}
&=
\sum_{k=1}^{K}
\varphi_{i,k}
\frac{a_k}{b_k}
\left(
\mathbf W_i^T\mathbf y_i
-\mathbf W_i^T\mathbf X_i\boldsymbol\nu_k
\right).
\end{align*}
Then, completing the square with respect to $\boldsymbol\xi_i$ gives
\begin{align*}
q_i^{6}(\boldsymbol\xi_i)
&\propto
\exp\left\{
-\frac12
\left(
\boldsymbol\xi_i-\mathbf A_{2i}\mathbf B_{2i}
\right)^T
\mathbf A_{2i}^{-1}
\left(
\boldsymbol\xi_i-\mathbf A_{2i}\mathbf B_{2i}
\right)
\right\}.
\end{align*}
Therefore,
\[
\mathcal Q_i^{6}(\boldsymbol\xi_i)
=
\text{N}(\boldsymbol\mu_i,\boldsymbol\Sigma_i),
\]
where
\begin{align*}
\boldsymbol\Sigma_i
&=
\left[
\sum_{k=1}^{K}
\varphi_{i,k}
\frac{a_k}{b_k}
\left(
\mathbf W_i^T\mathbf W_i+r_k\mathbf S_k
\right)
\right]^{-1},
\\
\boldsymbol\mu_i
&=
\boldsymbol\Sigma_i
\left[
\sum_{k=1}^{K}
\varphi_{i,k}
\frac{a_k}{b_k}
\left(
\mathbf W_i^T\mathbf y_i
-\mathbf W_i^T\mathbf X_i\boldsymbol\nu_k
\right)
\right].
\end{align*}

\subsubsection*{A.1.7 Updating $z_i$}

After collecting the terms in the log-joint density that involve
$z_i$ and taking the expectation with respect to all other variational
factors, we obtain, for $k=1,\ldots,K$,
\begin{align*}
q_i^{7}(z_i=k)
&\propto
\exp\Bigg\{
\mathbb E_{\mathcal Q_{\mathrm{RE}},-z_i}
\Bigg[
-\frac{\phi_k}{2}
\left\|
\mathbf y_i-\mathbf X_i\boldsymbol\beta_k
-\mathbf W_i\boldsymbol\xi_i
\right\|_2^2
-\frac{\phi_k}{2}
\boldsymbol\xi_i^T\mathbf Q_k\boldsymbol\xi_i+
\frac{m_i+\ell}{2}\log\phi_k
\\
&\quad
+
\frac12\log|\mathbf Q_k|
+
\log\pi_k(\mathbf v)
\Bigg]
\Bigg\}.
\end{align*}
For the residual term, we again use
\eqref{eq:re_residual_expansion}. Taking the expectation with respect to
$(\boldsymbol\beta_k,\phi_k)$ and $\boldsymbol\xi_i$ gives
\begin{align*}
\mathbb E_{\mathcal Q_{\mathrm{RE}},-z_i}
\left[
\phi_k
\left\|
\mathbf y_i-\mathbf X_i\boldsymbol\beta_k
-\mathbf W_i\boldsymbol\xi_i
\right\|_2^2
\right]
&=
\frac{a_k}{b_k}
\left\|
\mathbf y_i-\mathbf X_i\boldsymbol\nu_k
-\mathbf W_i\boldsymbol\mu_i
\right\|_2^2
+
\operatorname{tr}
\left(
\mathbf X_i^T\mathbf X_i\boldsymbol\Omega_k
\right)
\\
&\quad
+
\frac{a_k}{b_k}
\operatorname{tr}
\left(
\mathbf W_i^T\mathbf W_i\boldsymbol\Sigma_i
\right).
\end{align*}
Similarly, the expectation of the RE quadratic term is
\begin{align*}
\mathbb E_{\mathcal Q_{\mathrm{RE}},-z_i}
\left[
\phi_k
\boldsymbol\xi_i^T\mathbf Q_k\boldsymbol\xi_i
\right]
&=
\frac{a_k}{b_k}
r_k
\left\{
\boldsymbol\mu_i^T\mathbf S_k\boldsymbol\mu_i
+
\operatorname{tr}
\left(
\mathbf S_k\boldsymbol\Sigma_i
\right)
\right\}.
\end{align*}
Substituting these expectations gives
\begin{align*}
q_i^{7}(z_i=k)
&\propto
\exp\Bigg\{
-\frac12
\Bigg[
\frac{a_k}{b_k}
\left\|
\mathbf y_i-\mathbf X_i\boldsymbol\nu_k
-\mathbf W_i\boldsymbol\mu_i
\right\|_2^2
+
\operatorname{tr}
\left(
\mathbf X_i^T\mathbf X_i\boldsymbol\Omega_k
\right)
\\
&\quad
+
\frac{a_k}{b_k}
\operatorname{tr}
\left(
\mathbf W_i^T\mathbf W_i\boldsymbol\Sigma_i
\right)
\Bigg]
-
\frac12
\frac{a_k}{b_k}
r_k
\left\{
\boldsymbol\mu_i^T\mathbf S_k\boldsymbol\mu_i
+
\operatorname{tr}
\left(
\mathbf S_k\boldsymbol\Sigma_i
\right)
\right\}
\\
&\quad
+
\frac{m_i+\ell}{2}
\left\{
\Psi(a_k)-\log b_k
\right\}
+
\frac12
\Bigg[
\sum_{j=1}^{\ell}
\Psi\left(
\frac{r_k+1-j}{2}
\right)
+
\ell\log 2
+
\log|\mathbf S_k|
\Bigg]
\\
&\quad
+
\big(\Psi(\gamma_{k,1})-\Psi(\gamma_{k,1}+\gamma_{k,2})\big)
+
\textstyle\sum_{h<k}\big(\Psi(\gamma_{h,2})-\Psi(\gamma_{h,1}+\gamma_{h,2})\big)
\Bigg\}
\end{align*}
Therefore,
\[
\mathcal Q_i^{7}(z_i)
=
\operatorname{Discrete}
(\varphi_{i,1},\ldots,\varphi_{i,K}),
\]
where
\begin{align*}
\varphi_{i,k}
&\propto
\exp\Bigg\{
-\frac12
\Bigg[
\frac{a_k}{b_k}
\left\|
\mathbf y_i-\mathbf X_i\boldsymbol\nu_k
-\mathbf W_i\boldsymbol\mu_i
\right\|_2^2
+
\operatorname{tr}
\left(
\mathbf X_i^T\mathbf X_i\boldsymbol\Omega_k
\right)
\\
&\quad
+
\frac{a_k}{b_k}
\operatorname{tr}
\left(
\mathbf W_i^T\mathbf W_i\boldsymbol\Sigma_i
\right)
\Bigg]
-
\frac12
\frac{a_k}{b_k}
r_k
\left\{
\boldsymbol\mu_i^T\mathbf S_k\boldsymbol\mu_i
+
\operatorname{tr}
\left(
\mathbf S_k\boldsymbol\Sigma_i
\right)
\right\}
\\
&\quad
+
\frac{m_i+\ell}{2}
\left\{
\Psi(a_k)-\log b_k
\right\}
+
\frac12
\Bigg[
\sum_{j=1}^{\ell}
\Psi\left(
\frac{r_k+1-j}{2}
\right)
+
\ell\log 2
+
\log|\mathbf S_k|
\Bigg]
\\
&\quad
+
\big(\Psi(\gamma_{k,1})-\Psi(\gamma_{k,1}+\gamma_{k,2})\big)
+
\textstyle\sum_{h<k}\big(\Psi(\gamma_{h,2})-\Psi(\gamma_{h,1}+\gamma_{h,2})\big)
\Bigg\},
\end{align*}
normalized so that
$\sum_{k=1}^{K}\varphi_{i,k}=1$.

\subsection*{A.2 Evidence lower bound}
\label{app:re:elbo}
The ELBO
$\mathcal L(\mathcal Q)
=
\mathbb E_{\mathcal Q}[\log p(\boldsymbol\theta,\mathbf y)]
-
\mathbb E_{\mathcal Q}[\log q(\boldsymbol\theta)]$
decomposes under the mean-field factorization as
\begin{align*}
\mathcal L_{\mathrm{RE}}
&=
\sum_{k=1}^{K-1}
\left(
\mathbb E_{\mathcal Q_{\mathrm{RE}}}[\log p(v_k)]
-
\mathbb E_{\mathcal Q_{\mathrm{RE}}}[\log q_k^1(v_k)]
\right)
\\
&\quad+
\sum_{k=1}^{K}
\left(
\mathbb E_{\mathcal Q_{\mathrm{RE}}}
[\log p(\boldsymbol\beta_k,\phi_k\mid\boldsymbol\tau_k)]
-
\mathbb E_{\mathcal Q_{\mathrm{RE}}}
[\log q_k^2(\boldsymbol\beta_k,\phi_k)]
\right)
\\
&\quad+
\sum_{k=1}^{K}\sum_{j=1}^{D-4}
\left(
\mathbb E_{\mathcal Q_{\mathrm{RE}}}
[\log p(\tau_{k,j}\mid\lambda_k)]
-
\mathbb E_{\mathcal Q_{\mathrm{RE}}}
[\log q_{k,j}^3(\tau_{k,j})]
\right)
\\
&\quad+
\sum_{k=1}^{K}
\left(
\mathbb E_{\mathcal Q_{\mathrm{RE}}}[\log p(\lambda_k)]
-
\mathbb E_{\mathcal Q_{\mathrm{RE}}}[\log q_k^4(\lambda_k)]
\right)
\\
&\quad+
\sum_{k=1}^{K}
\left(
\mathbb E_{\mathcal Q_{\mathrm{RE}}}[\log p(\mathbf Q_k)]
-
\mathbb E_{\mathcal Q_{\mathrm{RE}}}[\log q_k^5(\mathbf Q_k)]
\right)
\\
&\quad+
\sum_{i=1}^{n}
\left(
\mathbb E_{\mathcal Q_{\mathrm{RE}}}
[\log p(\boldsymbol\xi_i\mid z_i,\boldsymbol\phi,\mathbf Q)]
-
\mathbb E_{\mathcal Q_{\mathrm{RE}}}
[\log q_i^6(\boldsymbol\xi_i)]
\right)
\\
&\quad+
\sum_{i=1}^{n}
\left(
\mathbb E_{\mathcal Q_{\mathrm{RE}}}
[\log p(z_i\mid\mathbf v)]
-
\mathbb E_{\mathcal Q_{\mathrm{RE}}}
[\log q_i^7(z_i)]
\right)
\\
&\quad+
\sum_{i=1}^{n}
\mathbb E_{\mathcal Q_{\mathrm{RE}}}
[\log p(\mathbf y_i\mid z_i,\boldsymbol\beta,\boldsymbol\phi,\boldsymbol\xi_i)].
\end{align*}
We derive each of these eight contributions below.

\subsubsection*{A.2.1 Bound for $v_k$}

The two expectations required for the ELBO contribution of $v_k$ are given by
\begin{align*}
\mathbb E_{\mathcal Q_{\mathrm{RE}}}[\log p(v_k)]
&=
\log\Gamma(\alpha+1)-\log\Gamma(\alpha)
+(\alpha-1)\left(\Psi(\gamma_{k,2})-\Psi(\gamma_{k,1}+\gamma_{k,2})\right),
\\
\mathbb E_{\mathcal Q_{\mathrm{RE}}}[\log \mathcal Q_k^1(v_k)]
&=
\log\Gamma(\gamma_{k,1}+\gamma_{k,2})-\log\Gamma(\gamma_{k,1})-\log\Gamma(\gamma_{k,2})
\\
&\quad
+(\gamma_{k,1}-1)\left(\Psi(\gamma_{k,1})-\Psi(\gamma_{k,1}+\gamma_{k,2})\right)
\\
&\quad
+(\gamma_{k,2}-1)\left(\Psi(\gamma_{k,2})-\Psi(\gamma_{k,1}+\gamma_{k,2})\right).
\end{align*}
Therefore, the ELBO contribution associated with $v_k$ is
\begin{align*}
\mathbb E_{\mathcal Q_{\mathrm{RE}}}[\log p(v_k)]
-\mathbb E_{\mathcal Q_{\mathrm{RE}}}[\log \mathcal Q_k^1(v_k)]
&=
\log\Gamma(\alpha+1)-\log\Gamma(\alpha)
-\log\Gamma(\gamma_{k,1}+\gamma_{k,2})
\\
&\quad
+\log\Gamma(\gamma_{k,1})+\log\Gamma(\gamma_{k,2})
\\
&\quad
+(1-\gamma_{k,1})\left(\Psi(\gamma_{k,1})-\Psi(\gamma_{k,1}+\gamma_{k,2})\right)
\\
&\quad
+(\alpha-\gamma_{k,2})\left(\Psi(\gamma_{k,2})-\Psi(\gamma_{k,1}+\gamma_{k,2})\right).
\end{align*}

\subsubsection*{A.2.2 Bound for $\boldsymbol\beta_k,\phi_k$}

The two expectations required for the ELBO contribution of $\boldsymbol\beta_k,\phi_k$ are given by
\begin{align*}
\mathbb E_{\mathcal Q_{\mathrm{RE}}}[\log p(\boldsymbol\beta_k,\phi_k)]
&=
a_0\log b_0-\log\Gamma(a_0)-\frac D2\log(2\pi)-\frac12
\mathbb E_{\mathcal Q_{\mathrm{RE}}}
[\log|\mathbf C_{\boldsymbol\tau_k}|]
\\
&\quad
+\left(a_0+\frac D2-1\right)\left(\Psi(a_k)-\log b_k\right)
-b_0\frac{a_k}{b_k}
\\
&\quad
-\frac12
\left[
\frac{a_k}{b_k}\boldsymbol\nu_k^T
\mathbb E_{\mathcal Q_{\mathrm{RE}}}[\mathbf C_{\boldsymbol\tau_k}^{-1}]
\boldsymbol\nu_k
+\operatorname{tr}\!\left(
\mathbb E_{\mathcal Q_{\mathrm{RE}}}[\mathbf C_{\boldsymbol\tau_k}^{-1}]
\boldsymbol\Omega_k
\right)
\right],
\\
\mathbb E_{\mathcal Q_{\mathrm{RE}}}[\log \mathcal Q_k^2(\boldsymbol\beta_k,\phi_k)]
&=
a_k\log b_k-\log\Gamma(a_k)-\frac D2\log(2\pi)-\frac12\log|\boldsymbol\Omega_k|
\\
&\quad
+\left(a_k+\frac D2-1\right)\left(\Psi(a_k)-\log b_k\right)-\frac D2-a_k,
\end{align*}
where
\[
\mathbb E_{\mathcal Q_{\mathrm{RE}}}
[\log|\mathbf C_{\boldsymbol\tau_k}|]
=
4\log\rho
+
\sum_{j=1}^{D-4}
\mathbb E_{\mathcal Q_{\mathrm{RE}}}[\log\tau_{k,j}],
\]
and
$\mathbb E_{\mathcal Q_{\mathrm{RE}}}
[\mathbf C_{\boldsymbol\tau_k}^{-1}]$
is given in \eqref{eq:Ctau_expectation}. Therefore, the ELBO contribution
associated with $\boldsymbol\beta_k,\phi_k$ is
\begin{align*}
\mathbb E_{\mathcal Q_{\mathrm{RE}}}[\log p(\boldsymbol\beta_k,\phi_k)]
-\mathbb E_{\mathcal Q_{\mathrm{RE}}}[\log \mathcal Q_k^2(\boldsymbol\beta_k,\phi_k)]
&=
a_0\log b_0-a_k\log b_k
-\log\Gamma(a_0)+\log\Gamma(a_k)
\\
&\quad
-\frac12\log|\mathbf C_{\boldsymbol\tau_k}|+\frac12\log|\boldsymbol\Omega_k|
\\
&\quad
+(a_0-a_k)\Psi(a_k)+a_k-b_0\frac{a_k}{b_k}
\\
&\quad
-\frac12
\left(
\frac{a_k}{b_k}\boldsymbol\nu_k^T
\mathbb E_{\mathcal Q_{\mathrm{RE}}}[\mathbf C_{\boldsymbol\tau_k}^{-1}]
\boldsymbol\nu_k
+\operatorname{tr}\!\left(
\mathbb E_{\mathcal Q_{\mathrm{RE}}}[\mathbf C_{\boldsymbol\tau_k}^{-1}]
\boldsymbol\Omega_k
\right)
\right).
\end{align*}

\subsubsection*{A.2.3 Bound for $\tau_{k,j}$}

The two expectations required for the ELBO contribution of $\tau_{k,j}$ are
\begin{align*}
\mathbb E_{\mathcal Q_{\mathrm{RE}}}
[\log p(\tau_{k,j}\mid\lambda_k)]
&=
\Psi(g_0+D-4)-\log h_k
-
\frac{g_0+D-4}{h_k}
\mathbb E_{\mathcal Q_{\mathrm{RE}}}[\tau_{k,j}],
\\
\mathbb E_{\mathcal Q_{\mathrm{RE}}}
[\log q_{k,j}^3(\tau_{k,j})]
&=
\frac14
\log\left(
\frac{c_{\tau_k}}{f_{\tau_{k,j}}}
\right)
-
\log\left[
2K_{1/2}
\left(
\sqrt{c_{\tau_k}f_{\tau_{k,j}}}
\right)
\right]
\\
&\quad
-\frac12
\mathbb E_{\mathcal Q_{\mathrm{RE}}}[\log\tau_{k,j}]
\\
&\quad
-\frac12
\left[
c_{\tau_k}
\mathbb E_{\mathcal Q_{\mathrm{RE}}}[\tau_{k,j}]
+
f_{\tau_{k,j}}
\mathbb E_{\mathcal Q_{\mathrm{RE}}}[\tau_{k,j}^{-1}]
\right].
\end{align*}
The required moments of $\tau_{k,j}$ are given in
\eqref{eq:tau_moment}--\eqref{eq:tau_log_moment}. Therefore,
\begin{align*}
\mathbb E_{\mathcal Q_{\mathrm{RE}}}
[\log p(\tau_{k,j}\mid\lambda_k)]
-
\mathbb E_{\mathcal Q_{\mathrm{RE}}}
[\log q_{k,j}^3(\tau_{k,j})]
&=
\Psi(g_0+D-4)-\log h_k
-
\frac{g_0+D-4}{h_k}
\mathbb E_{\mathcal Q_{\mathrm{RE}}}[\tau_{k,j}]
\\
&\quad
-\frac14
\log\left(
\frac{c_{\tau_k}}{f_{\tau_{k,j}}}
\right)
+
\log\left[
2K_{1/2}
\left(
\sqrt{c_{\tau_k}f_{\tau_{k,j}}}
\right)
\right]
\\
&\quad
+\frac12
\mathbb E_{\mathcal Q_{\mathrm{RE}}}[\log\tau_{k,j}]
\\
&\quad
+\frac12
\left[
c_{\tau_k}
\mathbb E_{\mathcal Q_{\mathrm{RE}}}[\tau_{k,j}]
+
f_{\tau_{k,j}}
\mathbb E_{\mathcal Q_{\mathrm{RE}}}[\tau_{k,j}^{-1}]
\right].
\end{align*}

\subsubsection*{A.2.4 Bound for $\lambda_k$}

The two expectations required for the ELBO contribution of $\lambda_k$ are
\begin{align*}
\mathbb E_{\mathcal Q_{\mathrm{RE}}}[\log p(\lambda_k)]
&=
g_0\log h_0-\log\Gamma(g_0)
\\
&\quad
+(g_0-1)
\left[
\Psi(g_0+D-4)-\log h_k
\right]
-h_0\frac{g_0+D-4}{h_k},
\\
\mathbb E_{\mathcal Q_{\mathrm{RE}}}[\log q_k^4(\lambda_k)]
&=
(g_0+D-4)\log h_k
-\log\Gamma(g_0+D-4)
\\
&\quad
+(g_0+D-5)
\left[
\Psi(g_0+D-4)-\log h_k
\right]
-(g_0+D-4).
\end{align*}
Therefore, the ELBO contribution associated with
$\lambda_k$ is
\begin{align*}
\mathbb E_{\mathcal Q_{\mathrm{RE}}}[\log p(\lambda_k)]
-
\mathbb E_{\mathcal Q_{\mathrm{RE}}}[\log q_k^4(\lambda_k)]
&=
g_0(\log h_0 - \log h_k)
-
(D-4)\Psi(g_0+D-4)
\\
&\quad
+\log\Gamma(g_0+D-4)
- \log\Gamma(g_0) + g_0+D-4 
\\
&\quad
- h_0\frac{g_0+D-4}{h_k}.
\end{align*}

\subsubsection*{A.2.5 Bound for $\mathbf Q_k$}
The two expectations required for the ELBO contribution of $\mathbf Q_k$ are given by
\begin{align*}
\mathbb E_{\mathcal Q_{\mathrm{RE}}}[\log p(\mathbf Q_k)]
&=
\frac{r_{0}-\ell-1}{2}\mathbb E_{\mathcal Q_{\mathrm{RE}}}[\log|\mathbf Q_k|]
-\frac12\operatorname{tr}\!\left(\mathbf S_{0}^{-1}\mathbb E_{\mathcal Q_{\mathrm{RE}}}[\mathbf Q_k]\right)
\\
&\quad
-\frac{r_{0}\ell}{2}\log2-\frac{r_{0}}{2}\log|\mathbf S_{0}|
-\log\Gamma_\ell\!\left(\frac{r_{0}}{2}\right),
\\
\mathbb E_{\mathcal Q_{\mathrm{RE}}}[\log \mathcal Q_k^5(\mathbf Q_k)]
&=
\frac{r_k-\ell-1}{2}\mathbb E_{\mathcal Q_{\mathrm{RE}}}[\log|\mathbf Q_k|]
-\frac12\operatorname{tr}\!\left(\mathbf S_k^{-1}\mathbb E_{\mathcal Q_{\mathrm{RE}}}[\mathbf Q_k]\right)
\\
&\quad
-\frac{r_k\ell}{2}\log2-\frac{r_k}{2}\log|\mathbf S_k|
-\log\Gamma_\ell\!\left(\frac{r_k}{2}\right),
\end{align*}
where $\Gamma_{\ell}(\cdot)$ denotes the multivariate gamma function,
$\Gamma_{\ell}(a)=\pi^{\ell(\ell-1)/4}\prod_{j=1}^{\ell}\Gamma\big(a+\frac{1-j}{2}\big)$,
which appears in the normalizing constant of the Wishart density
\citep{wishart_moments}. Using the moments derived in Appendix~A.1.5,
$\mathbb E_{\mathcal Q_{\mathrm{RE}}}[\mathbf Q_k]$, $
\mathbb E_{\mathcal Q_{\mathrm{RE}}}[\log|\mathbf Q_k|]$
the ELBO contribution associated with $\mathbf Q_k$ simplifies to
$\mathbf Q_k$ is
\begin{align*}
\mathbb E_{\mathcal Q_{\mathrm{RE}}}[\log p(\mathbf Q_k)]
-\mathbb E_{\mathcal Q_{\mathrm{RE}}}[\log \mathcal Q_k^5(\mathbf Q_k)]
&=
\frac{r_{0}-r_k}{2}\sum_{j=1}^{\ell}\Psi\!\left(\frac{r_k+1-j}{2}\right)
+\frac{r_{0}}{2}\left(\log|\mathbf S_k|-\log|\mathbf S_{0}|\right)
\\
&\quad
+\log\Gamma_\ell\!\left(\frac{r_k}{2}\right)-\log\Gamma_\ell\!\left(\frac{r_{0}}{2}\right)
+\frac{r_k\ell}{2}-\frac{r_k}{2}\operatorname{tr}\!\left(\mathbf S_{0}^{-1}\mathbf S_k\right).
\end{align*}

\subsubsection*{A.2.6 Bound for $\boldsymbol\xi_i$}

The two expectations required for the ELBO contribution of $\boldsymbol\xi_i$ are given by
\begin{align*}
\mathbb E_{\mathcal Q_{\mathrm{RE}}}[\log p(\boldsymbol\xi_i)]
&=
\sum_{k=1}^K\varphi_{i,k}
\Bigg[
-\frac\ell2\log2\pi
+\frac\ell2\left(\Psi(a_k)-\log b_k\right)
\\
&\quad
+\frac12
\left(
\sum_{j=1}^\ell\Psi\!\left(\frac{r_k+1-j}{2}\right)+\ell\log2+\log|\mathbf S_k|
\right)
-\frac12\frac{a_k}{b_k}r_k
\left(
\operatorname{tr}(\mathbf S_k\boldsymbol\Sigma_i)+\boldsymbol\mu_i^T\mathbf S_k\boldsymbol\mu_i
\right)
\Bigg],
\\
\mathbb E_{\mathcal Q_{\mathrm{RE}}}[\log \mathcal Q_i^6(\boldsymbol\xi_i)]
&=
-\frac\ell2\log2\pi-\frac12\log|\boldsymbol\Sigma_i|-\frac\ell2.
\end{align*}
Therefore, the ELBO contribution associated with $\boldsymbol\xi_i$ is
\begin{align*}
\mathbb E_{\mathcal Q_{\mathrm{RE}}}[\log p(\boldsymbol\xi_i)]
-\mathbb E_{\mathcal Q_{\mathrm{RE}}}[\log \mathcal Q_i^6(\boldsymbol\xi_i)]
&=
\sum_{k=1}^K\varphi_{i,k}
\Bigg[
\frac\ell2\left(\Psi(a_k)-\log b_k\right)
\\
& \quad
+\frac12
\left(
\sum_{j=1}^\ell\Psi\!\left(\frac{r_k+1-j}{2}\right)+\ell\log2+\log|\mathbf S_k|
\right)
\\
& \quad
-\frac12\frac{a_k}{b_k}r_k
\left(
\operatorname{tr}(\mathbf S_k\boldsymbol\Sigma_i)+\boldsymbol\mu_i^T\mathbf S_k\boldsymbol\mu_i
\right)
\Bigg]
+\frac12\log|\boldsymbol\Sigma_i|+\frac\ell2.
\end{align*}

\subsubsection*{A.2.7 Bound for $z_i$}

The two expectations required for the ELBO contribution of $z_i$ are given by
\begin{align*}
\mathbb E_{\mathcal Q_{\mathrm{RE}}}[\log p(z_i\mid\mathbf v)]
&=
\sum_{k=1}^{K-1}
\varphi_{i,k}
\left[
\Psi(\gamma_{k,1})
-\Psi(\gamma_{k,1}+\gamma_{k,2})
\right]
\\
&\quad
+
\sum_{k=1}^{K-1}
\left(
\sum_{h=k+1}^{K}\varphi_{i,h}
\right)
\left[
\Psi(\gamma_{k,2})
-\Psi(\gamma_{k,1}+\gamma_{k,2})
\right],
\\
\mathbb E_{\mathcal Q_{\mathrm{RE}}}[\log q_i^7(z_i)]
&=
\sum_{k=1}^{K}
\varphi_{i,k}\log\varphi_{i,k}.
\end{align*}
Therefore, the ELBO contribution associated with $z_i$ is
\begin{align*}
\mathbb E_{\mathcal Q_{\mathrm{RE}}}[\log p(z_i)]
-\mathbb E_{\mathcal Q_{\mathrm{RE}}}[\log \mathcal Q_i^7(z_i)]
&=
\sum_{k=1}^{K-1}\varphi_{i,k}\left(\Psi(\gamma_{k,1})-\Psi(\gamma_{k,1}+\gamma_{k,2})\right)
\\
&\quad
+\sum_{k=1}^{K-1}\left(\sum_{h>k}\varphi_{i,h}\right)\left(\Psi(\gamma_{k,2})-\Psi(\gamma_{k,1}+\gamma_{k,2})\right)
\\
&\quad
-\sum_{k=1}^K\varphi_{i,k}\log\varphi_{i,k}.
\end{align*}

\subsubsection*{A.2.8 Bound for $\mathbf y_i$}

Since $\mathbf y_i$ is observed, this term has no matching variational
factor, so the ELBO contribution is simply
$\mathbb E_{\mathcal Q_{\mathrm{RE}}}[\log p(\mathbf y_i\mid z_i)]$, given by
\begin{align*}
\mathbb E_{\mathcal Q_{\mathrm{RE}}}[\log p(\mathbf y_i\mid z_i)]
&=
\sum_{k=1}^K\varphi_{i,k}
\Bigg[
-\frac{m_i}{2}\log(2\pi)+\frac{m_i}{2}\left(\Psi(a_k)-\log b_k\right)
\\
& \quad
-\frac12
\left\{
\operatorname{tr}(\mathbf X_i^T\mathbf X_i\boldsymbol\Omega_k)
+\frac{a_k}{b_k}\operatorname{tr}(\mathbf W_i^T\mathbf W_i\boldsymbol\Sigma_i)
+\frac{a_k}{b_k}\left\|\mathbf y_i-\mathbf X_i\boldsymbol\nu_k-\mathbf W_i\boldsymbol\mu_i\right\|_2^2
\right\}
\Bigg].
\end{align*}
Substituting each of the eight expressions above into the decomposition given at the start of this section yields the complete ELBO:
\begin{align*}
\mathcal L_{\mathrm{RE}}
&=\sum_{k=1}^{K-1}\Bigg[
\log\Gamma(\alpha+1) - \log\Gamma(\alpha)
-\log\Gamma(\gamma_{k,1}+\gamma_{k,2})+\log\Gamma(\gamma_{k,1})+\log\Gamma(\gamma_{k,2})\\
&\quad
+ (1-\gamma_{k,1})\Big(\Psi(\gamma_{k,1})-\Psi(\gamma_{k,1}+\gamma_{k,2})\Big)
+(\alpha-\gamma_{k,2})\Big(\Psi(\gamma_{k,2})-\Psi(\gamma_{k,1}+\gamma_{k,2})\Big)
\Bigg]\\[1.5ex]
&\quad+\sum_{k=1}^{K}\Bigg[
a_0\log b_0-a_k\log b_k-\log\Gamma(a_0)+\log\Gamma(a_k)
-\frac{1}{2}\log|\mathbf C_{\boldsymbol\tau_k}|
+\frac{1}{2}\log|\mathbf{\Omega}_{k}|\\
&\quad
+(a_0-a_k)\Big(\Psi(a_k)-\log b_k\Big)
+\frac D2 +a_k-b_0\frac{a_k}{b_k}\\
&\quad
-\frac{1}{2}\Bigg(
\frac{a_k}{b_k}\,\boldsymbol\nu_{k}^T
\mathbb E_{\mathcal Q}[\mathbf C_{\boldsymbol\tau_k}^{-1}]
\,\boldsymbol\nu_{k}
+\mathrm{tr}\!\Big(\mathbb E_{\mathcal Q}[\mathbf C_{\boldsymbol\tau_k}^{-1}]
\,\mathbf{\Omega}_{k}\Big)
\Bigg)
\Bigg]\\[1.5ex] 
&\quad+\sum_{k=1}^{K}\sum_{j=1}^{D-4}\Bigg[\Psi(g_0+D-4)-\log h_k
-
\frac{g_0+D-4}{h_k}
\mathbb E_{\mathcal Q_{\mathrm{RE}}}[\tau_{k,j}]
-\frac14
\log\left(
\frac{c_{\tau_k}}{f_{\tau_{k,j}}}
\right)\\
&\quad
+
\log\left[
2K_{1/2}
\left(
\sqrt{c_{\tau_k}f_{\tau_{k,j}}}
\right)
\right]
+\frac12
\mathbb E_{\mathcal Q_{\mathrm{RE}}}[\log\tau_{k,j}]
+\frac12
\left[
c_{\tau_k}
\mathbb E_{\mathcal Q_{\mathrm{RE}}}[\tau_{k,j}]
+
f_{\tau_{k,j}}
\mathbb E_{\mathcal Q_{\mathrm{RE}}}[\tau_{k,j}^{-1}]
\right]
\Bigg]\\[1.5ex]
&\quad+\sum_{k=1}^{K}\Bigg[
g_0(\log h_0 - \log h_k)
- (D-4)\Psi(g_0+D-4)+\log\Gamma(g_0+D-4)
- \log\Gamma(g_0) \\
& \quad + g_0+D-4  - h_0\frac{g_0+D-4}{h_k}
\Bigg]\\[1.5ex]
&\quad+\sum_{k=1}^{K}\Bigg[
\frac{r_{0}-r_k}{2}\sum_{j=1}^{\ell}\Psi\!\left(\frac{r_k+1-j}{2}\right)
+\frac{r_{0}}{2}\Big(\log|\mathbf S_k|-\log|\mathbf S_{0}|\Big)\\
&\quad
+\log\Gamma_{\ell}\!\Big(\frac{r_k}{2}\Big)
-\log\Gamma_{\ell}\!\Big(\frac{r_{0}}{2}\Big)
+\frac{r_k\ell}{2}
-\frac{r_k}{2}\mathrm{tr}\!\big(\mathbf S_{0}^{-1}\mathbf S_k\big)
\Bigg]\\[1.5ex]
&\quad+\sum_{i=1}^{n}\Bigg\{
\sum_{k=1}^K\varphi_{i,k}
\Bigg[
\frac\ell2\left(\Psi(a_k)-\log b_k\right)
+\frac12
\left(
\sum_{j=1}^\ell\Psi\!\left(\frac{r_k+1-j}{2}\right)+\ell\log2+\log|\mathbf S_k|
\right)
\\
& \quad
-\frac12\frac{a_k}{b_k}r_k
\left(
\operatorname{tr}(\mathbf S_k\boldsymbol\Sigma_i)+\boldsymbol\mu_i^T\mathbf S_k\boldsymbol\mu_i
\right)
\Bigg]
+\frac12\log|\boldsymbol\Sigma_i|+\frac\ell2
\Bigg\}\\[1.5ex]
&\quad+\sum_{i=1}^{n}\Bigg\{\sum_{k=1}^{K-1} \varphi_{i,k}\Big(\Psi(\gamma_{k,1})-\Psi(\gamma_{k,1}+\gamma_{k,2})\Big)
+\sum_{k=1}^{K-1}\Big(\sum_{j>k}\varphi_{i,j}\Big)\Big(\Psi(\gamma_{k,2})-\Psi(\gamma_{k,1}+\gamma_{k,2})\Big)\\[1.5ex]
&\quad
-\sum_{k=1}^K \varphi_{i,k}\log \varphi_{i,k}\Bigg\}+\sum_{i=1}^{n}\sum_{k=1}^K \varphi_{i,k}
\Bigg[
-\frac{m_i}{2}\log(2\pi)
+\frac{m_i}{2}\big(\Psi(a_k)-\log b_k\big)\\
&\quad
-\frac{1}{2}\Big\{
\mathrm{tr}(\mathbf X_i^{\!T}\mathbf X_i\,\mathbf{\Omega}_{k})
+\frac{a_k}{b_k}\,
\mathrm{tr}(\mathbf W_i^{\!T}\mathbf W_i\,\boldsymbol\Sigma_i)
+\frac{a_k}{b_k}
\big\|
\mathbf y_i-\mathbf X_i\boldsymbol\nu_{k}
-\mathbf W_i\boldsymbol\mu_i
\big\|_2^2
\Big\}
\Bigg].
\end{align*}

\section*{Appendix B: Variational inference for the OU model}
\label{app:ou}

Under the mean-field assumption, the variational posterior measure factorizes as
\[
\mathcal Q_{\mathrm{OU}}(\boldsymbol\theta_{\mathrm{OU}})
=
\prod_{k=1}^{K-1}\tilde{\mathcal Q}_k^{1}(v_k)
\prod_{k=1}^{K}
\left[
\tilde{\mathcal Q}_k^{2}(\boldsymbol\beta_k,\phi_k)
\prod_{j=1}^{D-4}\tilde{\mathcal Q}_{k,j}^{3}(\tau_{k,j})
\tilde{\mathcal Q}_k^{4}(\lambda_k)
\tilde{\mathcal Q}_k^{5}(\tilde\zeta_k)
\right]
\prod_{i=1}^{n}
\tilde{\mathcal Q}_i^{6}(z_i),
\]
where
\[
\boldsymbol\theta_{\mathrm{OU}}
=
\left(
v_{1:K-1},
\boldsymbol\beta_{1:K},
\phi_{1:K},
\boldsymbol\tau_{1:K},
\lambda_{1:K},
\tilde\zeta_{1:K},
z_{1:n}
\right).
\]
The corresponding density is denoted by
$q_{\mathrm{OU}}(\boldsymbol\theta_{\mathrm{OU}})$.
Each factor and its associated variational parameters are summarized in
Table~\ref{tab:ou_notation}. Throughout this appendix, $D$ denotes the
dimension of the spline coefficient vector $\boldsymbol\beta_k$ in
\eqref{regression_spline}.

\begin{table}[h]
\centering
\caption{Variational factors and notation for the OU model.}
\label{tab:ou_notation}
\begin{tabular}{ll}
\hline
Factor & Variational distribution \\
\hline
$v_k$ &
$\operatorname{Beta}(\tilde\gamma_{k,1},\tilde\gamma_{k,2})$ \\
$(\boldsymbol\beta_k,\phi_k)$ &
$\operatorname{NG}
(\tilde{\boldsymbol\nu}_k,\tilde{\boldsymbol\Omega}_k,
\tilde a_k,\tilde b_k)$ \\
$\tau_{k,j}$ &
$\operatorname{GIG}
(1/2,\tilde c_{\tau_k},\tilde f_{\tau_{k,j}})$ \\
$\lambda_k$ &
$\operatorname{Gamma}(g_0+D-4,\tilde h_k)$ \\
$\tilde\zeta_k$ &
$\operatorname{Gamma}(\tilde r_k,\tilde s_k)$ \\
$z_i$ &
$\operatorname{Discrete}
(\tilde\varphi_{i,1},\ldots,\tilde\varphi_{i,K})$ \\
\hline
\end{tabular}
\end{table}
The OU model is identical to the RE model except for its covariance structure.
Accordingly, the updates for $v_k$, $(\boldsymbol\beta_k,\phi_k)$,
$\tau_{k,j}$, and $\lambda_k$ follow the same general derivations as their
RE counterparts, with the whitened quantities
$\mathbf y_{i,k}^*$ and $\mathbf X_{i,k}^*$ replacing the raw data where
the likelihood is involved. The update of each variational factor and the
resulting parameter expressions are derived in Appendix~B.1, whereas the
corresponding ELBO contributions are derived in Appendix~B.2.

\subsection*{B.1 Updating rules}
\label{app:ou:update}

We apply a whitening transformation to the OU model. This simplifies the
derivations and makes the resulting expressions consistent with those of
the RE model. For unit $i$, component $k$, and observation index $j\geq2$,
let
\[
\Delta_{ij}=t_{ij}-t_{i,j-1}
\]
denote the time gap between consecutive observations. The OU process with
rate $\tilde\zeta_k>0$ induces the local autocorrelation
\[
\zeta_{ij,k}
=
\exp(-\tilde\zeta_k\Delta_{ij}),
\]
as given by the Markov representation of the OU process
\citep{makov_repre}. The corresponding whitening weight is defined as
\[
w_{ij,k}
=
\frac{1}{1-\zeta_{ij,k}^{2}},
\qquad
j=2,\ldots,m_i,
\]
following the Gaussian Markov representation
\citep{whitening}, with the conventions
$w_{i1,k}=w_{i,m_i+1,k}=1$ and
$\zeta_{i1,k}=\zeta_{i,m_i+1,k}=0$.
Consecutive observations $y_{i,j-1}$ and $y_{ij}$ are correlated through
$\zeta_{ij,k}$. Subtracting the predictable component
$\zeta_{ij,k}y_{i,j-1}$ from $y_{ij}$ and rescaling by
$w_{ij,k}^{1/2}$ removes this correlation and standardizes the residual
variance.

Let $\mathbf T_{ik}(\tilde\zeta_k)\in\mathbb R^{m_i\times m_i}$ denote the
lower-bidiagonal whitening matrix \citep{whitening}, defined by
\[
[\mathbf T_{ik}]_{jl}
=
\begin{cases}
1
& j=l=1,\\
w_{ij,k}^{1/2}
& j=l\geq2,\\
-\zeta_{ij,k}w_{ij,k}^{1/2}
& l=j-1,\;j\geq2,\\
0
& \text{otherwise}.
\end{cases}
\]
The whitened response vector and design matrix are defined as
\[
\mathbf y_{i,k}^*
=
\mathbf T_{ik}\mathbf y_i,
\qquad
\mathbf X_{i,k}^*
=
\mathbf T_{ik}\mathbf X_i.
\]
Element-wise,
\[
y_{ij,k}^*
=
\begin{cases}
y_{i1},
& j=1,\\[4pt]
w_{ij,k}^{1/2}
\left(
y_{ij}-\zeta_{ij,k}y_{i,j-1}
\right),
& j\geq2.
\end{cases}
\]
Let $\mathbf x_{ij}^T$ and $\mathbf x_{ij,k}^{*\top}$ denote the $j$th
rows of $\mathbf X_i$ and $\mathbf X_{i,k}^*$, respectively.

Under the OU model, the likelihood for unit $i$ assigned to component $k$
is obtained through the change of variables
$\mathbf y_i\mapsto\mathbf y_{i,k}^*$:
\[
p(\mathbf y_i
\mid z_i=k,\boldsymbol\beta_k,\phi_k,\tilde\zeta_k)
=
p(\mathbf y_{i,k}^*
\mid z_i=k,\boldsymbol\beta_k,\phi_k)
\left|
\det\mathbf T_{ik}(\tilde\zeta_k)
\right|,
\]
where
$\left|
\det\mathbf T_{ik}(\tilde\zeta_k)
\right|
=
\prod_{j=2}^{m_i}w_{ij,k}^{1/2}.$

The joint density of $\mathbf y$ and $\boldsymbol\theta_{\mathrm{OU}}$ is
proportional to
\begin{align*}
p(\mathbf y,\boldsymbol\theta_{\mathrm{OU}})
&\propto
\prod_{i=1}^{n}
\prod_{k=1}^{K}
\left[
p(\mathbf y_{i,k}^*
\mid z_i=k,\boldsymbol\beta_k,\phi_k)
\left|
\det\mathbf T_{ik}
\right|
\right]^{\mathbbm 1(z_i=k)}
\\
&\quad\times
\prod_{i=1}^{n}
\prod_{k=1}^{K-1}
v_k^{\mathbbm 1(z_i=k)}
(1-v_k)^{\mathbbm 1(z_i>k)}
\\
&\quad\times
\prod_{k=1}^{K}
\left[
p(\tilde\zeta_k)\,
p(\boldsymbol\beta_k\mid\phi_k,\boldsymbol\tau_k)\,
p(\phi_k)\,
p(\boldsymbol\tau_k\mid\lambda_k)\,
p(\lambda_k)
\right]
\prod_{k=1}^{K-1}p(v_k)
\\
&\propto
\prod_{i=1}^{n}
\prod_{k=1}^{K}
\left[
\phi_k^{m_i/2}
\exp\left\{
-\frac{\phi_k}{2}
\left\|
\mathbf y_{i,k}^*
-\mathbf X_{i,k}^*\boldsymbol\beta_k
\right\|_2^2
\right\}
\prod_{j=2}^{m_i}w_{ij,k}^{1/2}
\right]^{\mathbbm 1(z_i=k)}
\\
&\quad\times
\prod_{i=1}^{n}
\prod_{k=1}^{K-1}
v_k^{\mathbbm 1(z_i=k)}
(1-v_k)^{\mathbbm 1(z_i>k)}
\\
&\quad\times
\prod_{k=1}^{K}
\Bigg[
\tilde\zeta_k^{p_0-1}
e^{-q_0\tilde\zeta_k}
\phi_k^{D/2}
\left(
\prod_{j=1}^{D-4}\tau_{k,j}
\right)^{-1/2}
\exp\left\{
-\frac{\phi_k}{2}
\boldsymbol\beta_k^T
\mathbf C_{\boldsymbol\tau_k}^{-1}
\boldsymbol\beta_k
\right\}
\\
&\quad
\times
\phi_k^{a_0-1}e^{-b_0\phi_k}
\prod_{j=1}^{D-4}
\left\{
\lambda_k e^{-\lambda_k\tau_{k,j}}
\right\}
\lambda_k^{g_0-1}e^{-h_0\lambda_k}
\Bigg]
\prod_{k=1}^{K-1}
(1-v_k)^{\alpha-1}.
\end{align*}

Taking the logarithm and retaining only terms involving model parameters gives
\begin{align*}
\log p(\mathbf y,\boldsymbol\theta_{\mathrm{OU}})
&\propto
\sum_{i=1}^{n}
\sum_{k=1}^{K}
\mathbbm 1(z_i=k)
\Bigg[
-\frac{\phi_k}{2}
\left\|
\mathbf y_{i,k}^*
-\mathbf X_{i,k}^*\boldsymbol\beta_k
\right\|_2^2
+
\frac{m_i}{2}\log\phi_k
+
\frac12
\sum_{j=2}^{m_i}\log w_{ij,k}
\Bigg]
\\
&\quad+
\sum_{i=1}^{n}
\sum_{k=1}^{K-1}
\mathbbm 1(z_i=k)\log v_k
\\
&\quad+
\sum_{k=1}^{K-1}
\left(
\alpha-1
+
\sum_{i=1}^{n}\mathbbm 1(z_i>k)
\right)
\log(1-v_k)
\\
&\quad+
\sum_{k=1}^{K}
\left[
(p_0-1)\log\tilde\zeta_k
-q_0\tilde\zeta_k
\right]
\\
&\quad+
\sum_{k=1}^{K}
\Bigg[
-\frac12
\sum_{j=1}^{D-4}\log\tau_{k,j}
-\frac{\phi_k}{2}
\boldsymbol\beta_k^T
\mathbf C_{\boldsymbol\tau_k}^{-1}
\boldsymbol\beta_k
-b_0\phi_k
-\lambda_k\sum_{j=1}^{D-4}\tau_{k,j}
-h_0\lambda_k
\Bigg]
\\
&\quad+
\sum_{k=1}^{K}
\left[
\frac D2+a_0-1
+
\frac12
\sum_{i=1}^{n}
m_i\mathbbm 1(z_i=k)
\right]
\log\phi_k
\\
&\quad+
\sum_{k=1}^{K}
(D-4+g_0-1)\log\lambda_k.
\end{align*}

\subsubsection*{B.1.1 Updating $v_k$}

After collecting the terms in the log-joint density that involve $v_k$
and taking the expectation with respect to all other variational factors,
we obtain
\begin{align*}
\tilde q_k^{1}(v_k)
&\propto
\exp\Bigg\{
\mathbb E_{\mathcal Q_{\mathrm{OU}},-v_k}
\Bigg[
\sum_{i=1}^{n}
\mathbbm 1(z_i=k)\log v_k
+
\left(
\sum_{i=1}^{n}\mathbbm 1(z_i>k)
+\alpha-1
\right)
\log(1-v_k)
\Bigg]
\Bigg\}
\\
&\propto
v_k^{\sum_{i=1}^{n}\tilde\varphi_{i,k}}
(1-v_k)^{
\alpha-1+
\sum_{i=1}^{n}
\sum_{h=k+1}^{K}\tilde\varphi_{i,h}
}.
\end{align*}
This is the kernel of a Beta distribution. Therefore,
\[
\tilde{\mathcal Q}_k^{1}(v_k)
=
\operatorname{Beta}
(\tilde\gamma_{k,1},\tilde\gamma_{k,2}),
\qquad
k=1,\ldots,K-1,
\]
where
\begin{align*}
\tilde\gamma_{k,1}
&=
1+\sum_{i=1}^{n}\tilde\varphi_{i,k},
\\
\tilde\gamma_{k,2}
&=
\alpha
+
\sum_{i=1}^{n}
\sum_{h=k+1}^{K}\tilde\varphi_{i,h}.
\end{align*}
The two moments of $v_k$ used in subsequent updates are
\begin{align}
\mathbb E_{\mathcal Q_{\mathrm{OU}}}[\log v_k]
&=
\Psi(\tilde\gamma_{k,1})
-\Psi(\tilde\gamma_{k,1}+\tilde\gamma_{k,2}),
\nonumber\\
\mathbb E_{\mathcal Q_{\mathrm{OU}}}[\log(1-v_k)]
&=
\Psi(\tilde\gamma_{k,2})
-\Psi(\tilde\gamma_{k,1}+\tilde\gamma_{k,2}).
\label{eq:vk_moments_OU}
\end{align}

\subsubsection*{B.1.2 Updating $\boldsymbol\beta_k$ and $\phi_k$}

After collecting the terms in the log-joint density that involve
$\boldsymbol\beta_k$ and $\phi_k$ and taking the expectation with respect
to all other variational factors, we obtain
\begin{align*}
\tilde q_k^{2}(\boldsymbol\beta_k,\phi_k)
&\propto
\exp\Bigg\{
\mathbb E_{\mathcal Q_{\mathrm{OU}},
-(\boldsymbol\beta_k,\phi_k)}
\Bigg[
\sum_{i=1}^{n}
\mathbbm 1(z_i=k)
\left\{
-\frac{\phi_k}{2}
\left\|
\mathbf y_{i,k}^*
-\mathbf X_{i,k}^*\boldsymbol\beta_k
\right\|_2^2
\right\}
\\
&\quad
-\frac{\phi_k}{2}
\boldsymbol\beta_k^T
\mathbf C_{\boldsymbol\tau_k}^{-1}
\boldsymbol\beta_k
-b_0\phi_k
+
\left\{
\frac D2+a_0-1
+
\frac12
\sum_{i=1}^{n}
m_i\mathbbm 1(z_i=k)
\right\}
\log\phi_k
\Bigg]
\Bigg\}.
\end{align*}
Compared with the update of $(\boldsymbol\beta_k,\phi_k)$ in
Appendix~A.1.2, the RE contribution is absent and the raw data
$\mathbf y_i$ and $\mathbf X_i$ are replaced by the whitened quantities
$\mathbf y_{i,k}^*$ and $\mathbf X_{i,k}^*$.
Taking the required expectations gives
\begin{align*}
\tilde q_k^{2}(\boldsymbol\beta_k,\phi_k)
&\propto
\phi_k^{
D/2+a_0-1+
\frac12
\sum_{i=1}^{n}
m_i\tilde\varphi_{i,k}
}
\\
&\quad\times
\exp\Bigg\{
-\frac{\phi_k}{2}
\Bigg[
\boldsymbol\beta_k^T
\Bigg\{
\mathbb E_{\mathcal Q_{\mathrm{OU}}}
\left[
\mathbf C_{\boldsymbol\tau_k}^{-1}
\right]
+
\sum_{i=1}^{n}
\tilde\varphi_{i,k}
\mathbb E_{\mathcal Q_{\mathrm{OU}}}
\left[
\mathbf X_{i,k}^{*\top}\mathbf X_{i,k}^*
\right]
\Bigg\}
\boldsymbol\beta_k
\\
&\quad
-
2\boldsymbol\beta_k^T
\sum_{i=1}^{n}
\tilde\varphi_{i,k}
\mathbb E_{\mathcal Q_{\mathrm{OU}}}
\left[
\mathbf X_{i,k}^{*\top}\mathbf y_{i,k}^*
\right]
\\
&\quad
+
\sum_{i=1}^{n}
\tilde\varphi_{i,k}
\mathbb E_{\mathcal Q_{\mathrm{OU}}}
\left[
\mathbf y_{i,k}^{*\top}\mathbf y_{i,k}^*
\right]
+
2b_0
\Bigg]
\Bigg\}.
\end{align*}
As in the update of $(\boldsymbol\beta_k,\phi_k)$ in Appendix~A.1.2, define
\begin{align*}
\mathbf A_{3k}^{-1}
&=
\mathbb E_{\mathcal Q_{\mathrm{OU}}}
\left[
\mathbf C_{\boldsymbol\tau_k}^{-1}
\right]
+
\sum_{i=1}^{n}
\tilde\varphi_{i,k}
\mathbb E_{\mathcal Q_{\mathrm{OU}}}
\left[
\mathbf X_{i,k}^{*\top}\mathbf X_{i,k}^*
\right],
\\
\mathbf B_{3k}
&=
\sum_{i=1}^{n}
\tilde\varphi_{i,k}
\mathbb E_{\mathcal Q_{\mathrm{OU}}}
\left[
\mathbf X_{i,k}^{*\top}\mathbf y_{i,k}^*
\right].
\end{align*}
Then, completing the square with respect to $\boldsymbol\beta_k$ gives
\begin{align*}
\tilde q_k^{2}(\boldsymbol\beta_k,\phi_k)
&\propto
\phi_k^{
a_0-1+
\frac12
\sum_{i=1}^{n}
m_i\tilde\varphi_{i,k}
}
\\
&\quad\times
\exp\Bigg\{
-\phi_k
\Bigg[
b_0
+
\frac12
\sum_{i=1}^{n}
\tilde\varphi_{i,k}
\mathbb E_{\mathcal Q_{\mathrm{OU}}}
\left[
\mathbf y_{i,k}^{*\top}\mathbf y_{i,k}^*
\right]
-
\frac12
\mathbf B_{3k}^T
\mathbf A_{3k}
\mathbf B_{3k}
\Bigg]
\Bigg\}
\\
&\quad\times
\left|
\phi_k^{-1}\mathbf A_{3k}
\right|^{-1/2}
\exp\Bigg\{
-\frac12
\left(
\boldsymbol\beta_k-\mathbf A_{3k}\mathbf B_{3k}
\right)^T
\left(
\phi_k^{-1}\mathbf A_{3k}
\right)^{-1}
\left(
\boldsymbol\beta_k-\mathbf A_{3k}\mathbf B_{3k}
\right)
\Bigg\}.
\end{align*}
Therefore,
\[
\tilde{\mathcal Q}_k^{2}(\boldsymbol\beta_k,\phi_k)
=
\operatorname{NG}
\left(
\tilde{\boldsymbol\nu}_k,
\tilde{\boldsymbol\Omega}_k,
\tilde a_k,
\tilde b_k
\right),
\]
where
\begin{align*}
\tilde{\boldsymbol\Omega}_k
&=
\left(
\mathbb E_{\mathcal Q_{\mathrm{OU}}}
\left[
\mathbf C_{\boldsymbol\tau_k}^{-1}
\right]
+
\sum_{i=1}^{n}
\tilde\varphi_{i,k}
\mathbb E_{\mathcal Q_{\mathrm{OU}}}
\left[
\mathbf X_{i,k}^{*\top}\mathbf X_{i,k}^*
\right]
\right)^{-1},
\\
\tilde{\boldsymbol\nu}_k
&=
\tilde{\boldsymbol\Omega}_k
\sum_{i=1}^{n}
\tilde\varphi_{i,k}
\mathbb E_{\mathcal Q_{\mathrm{OU}}}
\left[
\mathbf X_{i,k}^{*\top}\mathbf y_{i,k}^*
\right],
\\
\tilde a_k
&=
a_0
+
\frac12
\sum_{i=1}^{n}
m_i\tilde\varphi_{i,k},
\\
\tilde b_k
&=
b_0
+
\frac12
\sum_{i=1}^{n}
\tilde\varphi_{i,k}
\mathbb E_{\mathcal Q_{\mathrm{OU}}}
\left[
\mathbf y_{i,k}^{*\top}\mathbf y_{i,k}^*
\right]
-
\frac12
\left[
\sum_{i=1}^{n}
\tilde\varphi_{i,k}
\mathbb E_{\mathcal Q_{\mathrm{OU}}}
\left[
\mathbf X_{i,k}^{*\top}\mathbf y_{i,k}^*
\right]
\right]^T
\tilde{\boldsymbol\nu}_k.
\end{align*}
Here,
\[
\mathbb E_{\mathcal Q_{\mathrm{OU}}}
\left[
\mathbf C_{\boldsymbol\tau_k}^{-1}
\right]
=
\operatorname{blockdiag}
\left\{
\rho^{-1}\mathbf I_4,\,
\operatorname{diag}
\left(
\mathbb E_{\mathcal Q_{\mathrm{OU}}}[\tau_{k,1}^{-1}],
\ldots,
\mathbb E_{\mathcal Q_{\mathrm{OU}}}[\tau_{k,D-4}^{-1}]
\right)
\right\},
\]
where
$\mathbb E_{\mathcal Q_{\mathrm{OU}}}[\tau_{k,j}^{-1}]$
is derived in the following subsection.
Since $\mathbf T_{ik}$ is the linear whitening map, the
three moments above all reduce to the same quadratic form
$\mathbf X_{i,k}^{*\top}\mathbf X_{i,k}^*=\mathbf X_i^T(\mathbf T_{ik}^T
\mathbf T_{ik})\mathbf X_i$, and analogously
$\mathbf X_{i,k}^{*\top}\mathbf y_{i,k}^*=\mathbf X_i^T(\mathbf T_{ik}^T
\mathbf T_{ik})\mathbf y_i$ and
$\mathbf y_{i,k}^{*\top}\mathbf y_{i,k}^*=\mathbf y_i^T(\mathbf T_{ik}^T
\mathbf T_{ik})\mathbf y_i$. Expanding $\mathbf T_{ik}^T\mathbf T_{ik}$
term by term and taking the expectation over $\tilde\zeta_k$ therefore gives
the three component-specific moments required for the update of
\begin{align*}
\mathbb E_{\mathcal Q_{\mathrm{OU}}}
\left[
\mathbf X_{i,k}^{*\top}\mathbf X_{i,k}^*
\right]
&=
\sum_{j=1}^{m_i}
\mathbb E_{\mathcal Q_{\mathrm{OU}}}[w_{ij,k}]
\mathbf x_{ij}\mathbf x_{ij}^T
+
\sum_{j=2}^{m_i}
\mathbb E_{\mathcal Q_{\mathrm{OU}}}
[w_{ij,k}\zeta_{ij,k}^2]
\mathbf x_{i,j-1}\mathbf x_{i,j-1}^T
\\
&\quad
-
\sum_{j=2}^{m_i}
\mathbb E_{\mathcal Q_{\mathrm{OU}}}
[w_{ij,k}\zeta_{ij,k}]
\left(
\mathbf x_{ij}\mathbf x_{i,j-1}^T
+
\mathbf x_{i,j-1}\mathbf x_{ij}^T
\right),
\\[1.5ex]
\mathbb E_{\mathcal Q_{\mathrm{OU}}}
\left[
\mathbf X_{i,k}^{*\top}\mathbf y_{i,k}^*
\right]
&=
\sum_{j=1}^{m_i}
\mathbb E_{\mathcal Q_{\mathrm{OU}}}[w_{ij,k}]
\mathbf x_{ij}y_{ij}
+
\sum_{j=2}^{m_i}
\mathbb E_{\mathcal Q_{\mathrm{OU}}}
[w_{ij,k}\zeta_{ij,k}^2]
\mathbf x_{i,j-1}y_{i,j-1}
\\
&\quad
-
\sum_{j=2}^{m_i}
\mathbb E_{\mathcal Q_{\mathrm{OU}}}
[w_{ij,k}\zeta_{ij,k}]
\left(
\mathbf x_{ij}y_{i,j-1}
+
\mathbf x_{i,j-1}y_{ij}
\right),
\\[1.5ex]
\mathbb E_{\mathcal Q_{\mathrm{OU}}}
\left[
\mathbf y_{i,k}^{*\top}\mathbf y_{i,k}^*
\right]
&=
\sum_{j=1}^{m_i}
\mathbb E_{\mathcal Q_{\mathrm{OU}}}[w_{ij,k}]
y_{ij}^2
+
\sum_{j=2}^{m_i}
\mathbb E_{\mathcal Q_{\mathrm{OU}}}
[w_{ij,k}\zeta_{ij,k}^2]
y_{i,j-1}^2
\\
&\quad
-
2\sum_{j=2}^{m_i}
\mathbb E_{\mathcal Q_{\mathrm{OU}}}
[w_{ij,k}\zeta_{ij,k}]
y_{ij}y_{i,j-1}.
\end{align*}
Here, $w_{i1,k}=1$, so that
$\mathbb E_{\mathcal Q_{\mathrm{OU}}}[w_{i1,k}]=1$.
All expectations above are taken with respect to the current variational
factor $\tilde{\mathcal Q}_k^5(\tilde\zeta_k)$ and are evaluated in
Appendix~B.1.5. The moments used in the remaining updates are $\mathbb E_{\mathcal Q_{\mathrm{OU}}}[\boldsymbol\beta_k]
=
\tilde{\boldsymbol\nu}_k$ and $
\mathbb E_{\mathcal Q_{\mathrm{OU}}}[\phi_k]
=
{\tilde a_k}/{\tilde b_k}$.

\subsubsection*{B.1.3 Updating $\tau_{k,j}$}

After collecting the terms in the log-joint density that involve
$\tau_{k,j}$ and taking the expectation with respect to all other
variational factors, we obtain
\begin{align*}
\tilde q_{k,j}^{3}(\tau_{k,j})
&\propto
\exp\Bigg\{
\mathbb E_{\mathcal Q_{\mathrm{OU}},-\tau_{k,j}}
\left[
-\frac{1}{2}\log\tau_{k,j}
-\frac{\phi_k}{2\tau_{k,j}}
(\boldsymbol\beta_k)_{j+4}^{2}
-\lambda_k\tau_{k,j}
\right]
\Bigg\}
\\
&\propto
\tau_{k,j}^{-1/2}
\exp\left\{
-\frac{1}{2}
\left(
\tilde c_{\tau_k}\tau_{k,j}
+
\frac{\tilde f_{\tau_{k,j}}}{\tau_{k,j}}
\right)
\right\},
\end{align*}
where
\begin{align*}
\tilde c_{\tau_k}
&=
2\,\mathbb E_{\mathcal Q_{\mathrm{OU}}}[\lambda_k],
\\
\tilde f_{\tau_{k,j}}
&=
\frac{\tilde a_k}{\tilde b_k}
(\tilde{\boldsymbol\nu}_k)^2_{j+4}
+
(\tilde{\boldsymbol\Omega}_k)_{j+4,j+4}.
\end{align*}
This is the kernel of a generalized inverse Gaussian distribution. Therefore,
\[
\tilde{\mathcal Q}_{k,j}^{3}(\tau_{k,j})
=
\operatorname{GIG}
\left(
\frac12,
\tilde c_{\tau_k},
\tilde f_{\tau_{k,j}}
\right).
\]
The update of
$\mathbb E_{\mathcal Q_{\mathrm{OU}}}[\lambda_k]$
is derived in the next subsection. The moments of $\tau_{k,j}$ required in
the remaining updates are
\begin{align}
\mathbb E_{\mathcal Q_{\mathrm{OU}}}[\tau_{k,j}]
&=
\frac{
\sqrt{\tilde f_{\tau_{k,j}}}\,
K_{3/2}
\left(
\sqrt{\tilde c_{\tau_k}\tilde f_{\tau_{k,j}}}
\right)
}{
\sqrt{\tilde c_{\tau_k}}\,
K_{1/2}
\left(
\sqrt{\tilde c_{\tau_k}\tilde f_{\tau_{k,j}}}
\right)
},
\label{eq:tau_moment_OU}
\\
\mathbb E_{\mathcal Q_{\mathrm{OU}}}[\tau_{k,j}^{-1}]
&=
\frac{
\sqrt{\tilde c_{\tau_k}}\,
K_{3/2}
\left(
\sqrt{\tilde c_{\tau_k}\tilde f_{\tau_{k,j}}}
\right)
}{
\sqrt{\tilde f_{\tau_{k,j}}}\,
K_{1/2}
\left(
\sqrt{\tilde c_{\tau_k}\tilde f_{\tau_{k,j}}}
\right)
}
-
\frac{1}{\tilde f_{\tau_{k,j}}},
\label{eq:tau_inv_moment_OU}
\\
\mathbb E_{\mathcal Q_{\mathrm{OU}}}[\log\tau_{k,j}]
&=
\frac12
\log\left(
\frac{\tilde f_{\tau_{k,j}}}{\tilde c_{\tau_k}}
\right)
+
\left.
\frac{\partial}{\partial p}
\log
K_p
\left(
\sqrt{\tilde c_{\tau_k}\tilde f_{\tau_{k,j}}}
\right)
\right|_{p=1/2}.
\label{eq:tau_log_moment_OU}
\end{align}

\subsubsection*{B.1.4 Updating $\lambda_k$}

After collecting the terms in the log-joint density that involve
$\lambda_k$ and taking the expectation with respect to all other
variational factors, we obtain
\begin{align*}
\tilde q_k^{4}(\lambda_k)
&\propto
\exp\Bigg\{
\mathbb E_{\mathcal Q_{\mathrm{OU}},-\lambda_k}
\left[
(g_0+D-4-1)\log\lambda_k
-\lambda_k\sum_{j=1}^{D-4}\tau_{k,j}
-h_0\lambda_k
\right]
\Bigg\}
\\
&\propto
\lambda_k^{g_0+D-4-1}
\exp\left\{
-\lambda_k
\left(
h_0
+
\sum_{j=1}^{D-4}
\mathbb E_{\mathcal Q_{\mathrm{OU}}}[\tau_{k,j}]
\right)
\right\}.
\end{align*}
This is the kernel of a Gamma distribution. Therefore,
\[
\tilde{\mathcal Q}_k^{4}(\lambda_k)
=
\operatorname{Gamma}
\left(
g_0+D-4,\,
\tilde h_k
\right),
\]
where
\begin{align*}
\tilde h_k
&=
h_0
+
\sum_{j=1}^{D-4}
\frac{
\sqrt{\tilde f_{\tau_{k,j}}}\,
K_{3/2}
\left(
\sqrt{\tilde c_{\tau_k}\tilde f_{\tau_{k,j}}}
\right)
}{
\sqrt{\tilde c_{\tau_k}}\,
K_{1/2}
\left(
\sqrt{\tilde c_{\tau_k}\tilde f_{\tau_{k,j}}}
\right)
}.
\end{align*}
The moments of $\lambda_k$ used in the remaining updates are
\[
\mathbb E_{\mathcal Q_{\mathrm{OU}}}[\lambda_k]
=
\frac{g_0+D-4}{\tilde h_k},
\qquad
\mathbb E_{\mathcal Q_{\mathrm{OU}}}[\log\lambda_k]
=
\Psi(g_0+D-4)-\log\tilde h_k.
\]

\subsubsection*{B.1.5 Updating $\tilde\zeta_k$}

\paragraph{Approximate update for $\tilde\zeta_k$.}

After collecting the terms in the log-joint density that involve
$\tilde\zeta_k$ and taking the expectation with respect to all other
variational factors, we obtain
\begin{align*}
\tilde q_k^{5,*}(\tilde\zeta_k)
&\propto
\exp\Bigg\{
\mathbb E_{\mathcal Q_{\mathrm{OU}},-\tilde\zeta_k}
\Bigg[
\sum_{i=1}^{n}
\mathbbm 1(z_i=k)
\Bigg\{
-\frac{\phi_k}{2}
\left\|
\mathbf y_{i,k}^*
-\mathbf X_{i,k}^*\boldsymbol\beta_k
\right\|_2^2
+
\frac12
\sum_{j=2}^{m_i}
\log w_{ij,k}
\Bigg\}
\\
&\quad
+
(p_0-1)\log\tilde\zeta_k
-q_0\tilde\zeta_k
\Bigg]
\Bigg\}.
\end{align*}
To evaluate the residual term $\mathbb E_{\mathcal Q_{\mathrm{OU}},-\tilde\zeta_k}
\big[
-\frac{\phi_k}{2}
\bigl\| \mathbf y_{i,k}^* - \mathbf X_{i,k}^*\boldsymbol\beta_k \bigr\|_2^2
\big]$, write
$\mathbf T_{ik}^T\mathbf T_{ik}$ as the tridiagonal matrix
\[
\mathbf T_{ik}^T\mathbf T_{ik}
=
\begin{pmatrix}
d_{i1,k} & o_{i2,k} &  &  \\
o_{i2,k} & d_{i2,k} & o_{i3,k} &  \\
 & o_{i3,k} & \ddots & \ddots \\
 & & \ddots & d_{i,m_i-1,k} & o_{i,m_i,k}\\
 & & & o_{i,m_i,k} & d_{i,m_i,k}
\end{pmatrix},
\]
where
\begin{align}
d_{ij,k}(\tilde\zeta_k)
&=
\begin{cases}
w_{i2,k},
& j=1,\\
w_{ij,k}+w_{i,j+1,k}-1,
& j=2,\ldots,m_i-1,\\
w_{im_i,k},
& j=m_i,
\end{cases}
\nonumber\\
o_{ij,k}(\tilde\zeta_k)
&=
-\zeta_{ij,k}w_{ij,k},
\qquad
j=2,\ldots,m_i.
\label{eq:d_o_def}
\end{align}
Taking the expectation of the residual term with respect to
$(\boldsymbol\beta_k,\phi_k)$ while holding $\tilde\zeta_k$ fixed gives
\begin{align}
\mathbb E_{\mathcal Q_{\mathrm{OU}},-\tilde\zeta_k}
\left[
-\frac{\phi_k}{2}
\left\|
\mathbf y_{i,k}^*
-\mathbf X_{i,k}^*\boldsymbol\beta_k
\right\|_2^2
\right]
\nonumber
&=
\mathbb E_{\mathcal Q_{\mathrm{OU}},-\tilde\zeta_k}
\left[
-\frac{\phi_k}{2}
(\mathbf y_i-\mathbf X_i\boldsymbol\beta_k)^T
(\mathbf T_{ik}^T\mathbf T_{ik})
(\mathbf y_i-\mathbf X_i\boldsymbol\beta_k)
\right]
\nonumber\\
&=
-\frac12
\frac{\tilde a_k}{\tilde b_k}
\Bigg[
\sum_{j=1}^{m_i}
d_{ij,k}
\left(
y_{ij}-\mathbf x_{ij}^T\tilde{\boldsymbol\nu}_k
\right)^2
\nonumber\\
&\quad
+
2\sum_{j=2}^{m_i}
o_{ij,k}
\left(
y_{ij}-\mathbf x_{ij}^T\tilde{\boldsymbol\nu}_k
\right)
\left(
y_{i,j-1}
-\mathbf x_{i,j-1}^T\tilde{\boldsymbol\nu}_k
\right)
\Bigg]
\nonumber\\
&\quad
-\frac12
\Bigg[
\sum_{j=1}^{m_i}
d_{ij,k}
\mathbf x_{ij}^T
\tilde{\boldsymbol\Omega}_k
\mathbf x_{ij}
+
2\sum_{j=2}^{m_i}
o_{ij,k}
\mathbf x_{ij}^T
\tilde{\boldsymbol\Omega}_k
\mathbf x_{i,j-1}
\Bigg].
\label{eq:ou_quadratic_expectation}
\end{align}
Substituting \eqref{eq:ou_quadratic_expectation} into the coordinate update
gives
\[
\tilde q_k^{5,*}(\tilde\zeta_k)
\propto
\exp\{\ell_k(\tilde\zeta_k)\},
\]
where
\begin{align}
\ell_k(\tilde\zeta_k)
&=
(p_0-1)\log\tilde\zeta_k
-q_0\tilde\zeta_k
+
\frac12
\sum_{i=1}^{n}
\tilde\varphi_{i,k}
\sum_{j=2}^{m_i}
\log w_{ij,k}
\nonumber\\
&\quad
-\frac12
\sum_{i=1}^{n}
\tilde\varphi_{i,k}
\Bigg[
\frac{\tilde a_k}{\tilde b_k}
\Bigg\{
\sum_{j=1}^{m_i}
d_{ij,k}
\left(
y_{ij}-\mathbf x_{ij}^T\tilde{\boldsymbol\nu}_k
\right)^2
\nonumber\\
&\quad
+
2\sum_{j=2}^{m_i}
o_{ij,k}
\left(
y_{ij}-\mathbf x_{ij}^T\tilde{\boldsymbol\nu}_k
\right)
\left(
y_{i,j-1}
-\mathbf x_{i,j-1}^T\tilde{\boldsymbol\nu}_k
\right)
\Bigg\}
\nonumber\\
&\quad
+
\sum_{j=1}^{m_i}
d_{ij,k}
\mathbf x_{ij}^T
\tilde{\boldsymbol\Omega}_k
\mathbf x_{ij}
+
2\sum_{j=2}^{m_i}
o_{ij,k}
\mathbf x_{ij}^T
\tilde{\boldsymbol\Omega}_k
\mathbf x_{i,j-1}
\Bigg].
\label{eq:ell_k}
\end{align}

The coordinate-optimal density
$\tilde q_k^{5,*}(\tilde\zeta_k)$ does not belong to a standard
distributional family. We therefore approximate it by a Gamma distribution
whose mode and local curvature agree with those of
$\ell_k(\tilde\zeta_k)$.
Let
\[
\hat\zeta_k
=
\arg\max_{\tilde\zeta_k>0}
\ell_k(\tilde\zeta_k).
\]
For a Gamma density
$\operatorname{Gamma}(r,s)$, the mode and second derivative at the mode are
\[
\frac{r-1}{s},
\qquad
-\frac{r-1}{x^2},
\]
respectively. Matching the Gamma mode to $\hat\zeta_k$ and its curvature to
$\ell_k''(\hat\zeta_k)$ and solving the resulting two equations yields
\[
\tilde r_k
=
1-\hat\zeta_k^2\ell_k''(\hat\zeta_k),
\qquad
\tilde s_k
=
-\hat\zeta_k\ell_k''(\hat\zeta_k).
\]
Therefore, we use the approximation
\[
\tilde{\mathcal Q}_k^5(\tilde\zeta_k)
\approx
\operatorname{Gamma}(\tilde r_k,\tilde s_k).
\]

To compute $\hat\zeta_k$, $\tilde r_k$, and $\tilde s_k$, we require the
first two derivatives of $\ell_k$. Differentiating
$\zeta_{ij,k}=\exp(-\tilde\zeta_k\Delta_{ij})$ gives
\[
\frac{d}{d\tilde\zeta_k}\zeta_{ij,k}
=
-\Delta_{ij}\zeta_{ij,k},
\qquad
\frac{d^2}{d\tilde\zeta_k^2}\zeta_{ij,k}
=
\Delta_{ij}^2\zeta_{ij,k}.
\]
Differentiating
$w_{ij,k}$ and
$o_{ij,k}$ gives
\begin{align*}
w'_{ij,k}
&=
-\frac{
2\Delta_{ij}\zeta_{ij,k}^2
}{
(1-\zeta_{ij,k}^2)^2
},
&
w''_{ij,k}
&=
\frac{
4\Delta_{ij}^2\zeta_{ij,k}^2
(1+\zeta_{ij,k}^2)
}{
(1-\zeta_{ij,k}^2)^3
},
\\
o'_{ij,k}
&=
\frac{
\Delta_{ij}\zeta_{ij,k}
(1+\zeta_{ij,k}^2)
}{
(1-\zeta_{ij,k}^2)^2
},
&
o''_{ij,k}
&=
-\frac{
\Delta_{ij}^2\zeta_{ij,k}
(1+6\zeta_{ij,k}^2+\zeta_{ij,k}^4)
}{
(1-\zeta_{ij,k}^2)^3
}.
\end{align*}
From \eqref{eq:d_o_def},
\begin{align*}
d'_{i1,k}
&=
w'_{i2,k},
\\
d'_{ij,k}
&=
w'_{ij,k}+w'_{i,j+1,k},
\qquad
j=2,\ldots,m_i-1,
\\
d'_{im_i,k}
&=
w'_{im_i,k},
\end{align*}
and the corresponding second derivatives are obtained by replacing
$w'_{ij,k}$ with $w''_{ij,k}$.
Differentiating \eqref{eq:ell_k} gives
\begin{align}
\ell_k'(\tilde\zeta_k)
&=
\frac{p_0-1}{\tilde\zeta_k}
-q_0
+
\frac12
\sum_{i=1}^{n}
\tilde\varphi_{i,k}
\sum_{j=2}^{m_i}
\frac{w'_{ij,k}}{w_{ij,k}}
-\frac12
\sum_{i=1}^{n}
\tilde\varphi_{i,k}
\Bigg[
\frac{\tilde a_k}{\tilde b_k}
\Bigg\{
\sum_{j=1}^{m_i}
d'_{ij,k}
\left(
y_{ij}-\mathbf x_{ij}^T\tilde{\boldsymbol\nu}_k
\right)^2
\nonumber\\
&\quad
+
2\sum_{j=2}^{m_i}
o'_{ij,k}
\left(
y_{ij}-\mathbf x_{ij}^T\tilde{\boldsymbol\nu}_k
\right)
\left(
y_{i,j-1}
-\mathbf x_{i,j-1}^T\tilde{\boldsymbol\nu}_k
\right)
\Bigg\}
\nonumber\\
&\quad
+
\sum_{j=1}^{m_i}
d'_{ij,k}
\mathbf x_{ij}^T
\tilde{\boldsymbol\Omega}_k
\mathbf x_{ij}
+
2\sum_{j=2}^{m_i}
o'_{ij,k}
\mathbf x_{ij}^T
\tilde{\boldsymbol\Omega}_k
\mathbf x_{i,j-1}
\Bigg],
\end{align}
and
\begin{align}
\ell_k''(\tilde\zeta_k)
&=
-\frac{p_0-1}{\tilde\zeta_k^2}
+
\frac12
\sum_{i=1}^{n}
\tilde\varphi_{i,k}
\sum_{j=2}^{m_i}
\frac{
w''_{ij,k}w_{ij,k}
-(w'_{ij,k})^2
}{
w_{ij,k}^2
}
-\frac12
\sum_{i=1}^{n}
\tilde\varphi_{i,k}
\Bigg[
\frac{\tilde a_k}{\tilde b_k}
\Bigg\{
\sum_{j=1}^{m_i}
d''_{ij,k}
\left(
y_{ij}-\mathbf x_{ij}^T\tilde{\boldsymbol\nu}_k
\right)^2
\nonumber\\
&\quad
+
2\sum_{j=2}^{m_i}
o''_{ij,k}
\left(
y_{ij}-\mathbf x_{ij}^T\tilde{\boldsymbol\nu}_k
\right)
\left(
y_{i,j-1}
-\mathbf x_{i,j-1}^T\tilde{\boldsymbol\nu}_k
\right)
\Bigg\}
\nonumber\\
&\quad
+
\sum_{j=1}^{m_i}
d''_{ij,k}
\mathbf x_{ij}^T
\tilde{\boldsymbol\Omega}_k
\mathbf x_{ij}
+
2\sum_{j=2}^{m_i}
o''_{ij,k}
\mathbf x_{ij}^T
\tilde{\boldsymbol\Omega}_k
\mathbf x_{i,j-1}
\Bigg].
\label{eq:ell_k_dprime}
\end{align}
The mode $\hat\zeta_k$ is obtained numerically by solving
$\ell_k'(\hat\zeta_k)=0$. The parameters
$\tilde r_k$ and $\tilde s_k$ are then obtained from
$\ell_k''(\hat\zeta_k)$.

\paragraph{Required expectations.}

The Gamma approximation for $\tilde\zeta_k$ is also used to evaluate the
nonlinear expectations required in the other variational updates.
In particular, Appendix~B.1.2 requires moments involving the whitened
quadratic forms, and the update of $z_i$ requires the log-Jacobian
expectation. Define
\begin{align}
A_{0,ij,k}
&=
\mathbb E_{\mathcal Q_{\mathrm{OU}}}[w_{ij,k}],
\nonumber\\
A_{1,ij,k}
&=
\mathbb E_{\mathcal Q_{\mathrm{OU}}}
[\zeta_{ij,k}w_{ij,k}],
\nonumber\\
A_{2,ij,k}
&=
\mathbb E_{\mathcal Q_{\mathrm{OU}}}
[\zeta_{ij,k}^2w_{ij,k}]
\nonumber\\
L_{ij,k}
&=
\mathbb E_{\mathcal Q_{\mathrm{OU}}}
[\log(1-\zeta_{ij,k}^2)].
\label{eq:A_defs}
\end{align}
We first derive $A_{0,ij,k}$ and $A_{1,ij,k}$.
The quantity $A_{2,ij,k}$ then follows immediately from
$A_{2,ij,k}=A_{0,ij,k}-1$. Let
\[
\mathcal H_k(t)
=
\mathbb E_{\mathcal Q_{\mathrm{OU}}}
[\exp(-t\tilde\zeta_k)]
=
\left(
\frac{\tilde s_k}{\tilde s_k+t}
\right)^{\tilde r_k}
\]
denote the Laplace transform under
$\tilde\zeta_k\sim\operatorname{Gamma}(\tilde r_k,\tilde s_k)$.
Using the geometric expansions
\[
\frac{1}
{1-\exp(-2\tilde\zeta_k\Delta_{ij})}
=
\sum_{\ell=0}^{\infty}
\exp(-2\ell\tilde\zeta_k\Delta_{ij}),
\]
and
\[
\frac{
\exp(-\tilde\zeta_k\Delta_{ij})
}{
1-\exp(-2\tilde\zeta_k\Delta_{ij})
}
=
\sum_{\ell=0}^{\infty}
\exp\{-(2\ell+1)\tilde\zeta_k\Delta_{ij}\},
\]
we obtain
\begin{align*}
A_{0,ij,k}
&=
\sum_{\ell=0}^{\infty}
\mathcal H_k(2\ell\Delta_{ij})
=
\sum_{\ell=0}^{\infty}
\left(
\frac{\tilde s_k}
{\tilde s_k+2\ell\Delta_{ij}}
\right)^{\tilde r_k},
\\
A_{1,ij,k}
&=
\sum_{\ell=0}^{\infty}
\mathcal H_k((2\ell+1)\Delta_{ij})
=
\sum_{\ell=0}^{\infty}
\left(
\frac{\tilde s_k}
{\tilde s_k+(2\ell+1)\Delta_{ij}}
\right)^{\tilde r_k},
\\
A_{2,ij,k}
&=
\sum_{\ell=1}^{\infty}
\mathcal H_k(2\ell\Delta_{ij})
=
A_{0,ij,k}-1.
\end{align*}
Define
$q_{ij,k}
=
{\tilde s_k}/{2\Delta_{ij}}$.
Then $A_{0,ij,k}$ and $A_{1,ij,k}$ can be expressed using the Hurwitz
zeta function
$\zeta(s,q)=\sum_{\ell=0}^{\infty}(\ell+q)^{-s}$:
\begin{align*}
A_{0,ij,k}
&=
q_{ij,k}^{\tilde r_k}
\zeta(\tilde r_k,q_{ij,k}),
\\
A_{1,ij,k}
&=
q_{ij,k}^{\tilde r_k}
\zeta\left(
\tilde r_k,q_{ij,k}+\frac12
\right).
\end{align*}
However, when $\Delta_{ij}$ is very small, $q_{ij,k}$ can become extremely large.
The factor $q_{ij,k}^{\tilde r_k}$ may then cause numerical overflow.
In this regime, we instead use the Euler--Maclaurin asymptotic expansion
\citep{euler_asym}
\[
\zeta(r,q)
=
\frac{q^{1-r}}{r-1}
+
\frac{1}{2q^r}
+
\frac{r}{12q^{r+1}}
+
O(q^{-r-3}),
\qquad
q\to\infty.
\]
This gives
\begin{align*}
A_{0,ij,k}
&\approx
\frac{q_{ij,k}}{\tilde r_k-1}
+
\frac12
+
\frac{\tilde r_k}{12q_{ij,k}},
\\
A_{1,ij,k}
&\approx
\frac{
q_{ij,k}^{\tilde r_k}
(q_{ij,k}+1/2)^{1-\tilde r_k}
}{
\tilde r_k-1
}
+
\frac12
\left(
\frac{q_{ij,k}}{q_{ij,k}+1/2}
\right)^{\tilde r_k}
+
\frac{\tilde r_k}{12}
q_{ij,k}^{\tilde r_k}
(q_{ij,k}+1/2)^{-\tilde r_k-1}.
\end{align*}
If neither the Hurwitz zeta representation nor the Euler--Maclaurin
approximation is numerically reliable, the corresponding infinite series
is evaluated directly.

For the log-Jacobian term,
\[
-\log(1-e^{-x})
=
\sum_{\ell=1}^{\infty}
\frac{e^{-\ell x}}{\ell},
\]
which gives
\begin{align*}
L_{ij,k}
&=
\mathbb E_{\mathcal Q_{\mathrm{OU}}}
\left[
\log
\left\{
1-\exp(-2\tilde\zeta_k\Delta_{ij})
\right\}
\right]
=
-
\sum_{\ell=1}^{\infty}
\frac{1}{\ell}
\mathcal H_k(2\ell\Delta_{ij})
=
-
\sum_{\ell=1}^{\infty}
\frac{1}{\ell}
\left(
\frac{\tilde s_k}
{\tilde s_k+2\ell\Delta_{ij}}
\right)^{\tilde r_k}.
\end{align*}
This infinite series is evaluated numerically. When $q_{ij,k}$ is large,
the terms decay slowly and direct summation becomes inefficient. We then use
the first-order approximation
\[
\log\left(
1+\frac{\ell}{q_{ij,k}}
\right)
\approx
\frac{\ell}{q_{ij,k}},
\]
so that
\[
\left(
\frac{q_{ij,k}}
{q_{ij,k}+\ell}
\right)^{\tilde r_k}
\approx
\exp\left(
-\frac{\tilde r_k\ell}{q_{ij,k}}
\right).
\]
Substitution into the infinite series gives
\[
L_{ij,k}
\approx
-
\sum_{\ell=1}^{\infty}
\frac{1}{\ell}
\left\{
\exp\left(
-\frac{\tilde r_k}{q_{ij,k}}
\right)
\right\}^{\ell}.
\]
Using the Mercator series
$-\log(1-x)=\sum_{\ell=1}^{\infty}x^\ell/\ell$, this reduces to
\[
L_{ij,k}
\approx
\log\left[
1-
\exp\left(
-\frac{\tilde r_k}{q_{ij,k}}
\right)
\right].
\]
Otherwise, the original series is evaluated by direct truncated summation.
Finally, the expectations required in the remaining updates follow directly
from \eqref{eq:A_defs}. With the conventions
$A_{0,i1,k}=A_{0,i,m_i+1,k}=1$, we have
\begin{align*}
\mathbb E_{\mathcal Q_{\mathrm{OU}}}[d_{ij,k}]
&=
A_{0,ij,k}
+
A_{0,i,j+1,k}
-1,
\qquad
j=1,\ldots,m_i,
\\
\mathbb E_{\mathcal Q_{\mathrm{OU}}}[o_{ij,k}]
&=
-A_{1,ij,k},
\qquad
j=2,\ldots,m_i,
\\
\mathbb E_{\mathcal Q_{\mathrm{OU}}}[\log w_{ij,k}]
&=
-L_{ij,k},
\qquad
j=2,\ldots,m_i.
\end{align*}

\subsubsection*{B.1.6 Updating $z_i$}

After collecting the terms in the log-joint density that involve
$z_i$ and taking the expectation with respect to all other variational
factors, we obtain, for $k=1,\ldots,K$,
\begin{align*}
\tilde q_i^{6}(z_i=k)
&\propto
\exp\Bigg\{
\mathbb E_{\mathcal Q_{\mathrm{OU}},-z_i}
\Bigg[
-\frac{\phi_k}{2}
\left\|
\mathbf y_{i,k}^*
-\mathbf X_{i,k}^*\boldsymbol\beta_k
\right\|_2^2
+
\frac{m_i}{2}\log\phi_k
+
\frac12
\sum_{j=2}^{m_i}\log w_{ij,k}
+
\log\pi_k(\mathbf v)
\Bigg]
\Bigg\}.
\end{align*}
The term $\mathbb E_{\mathcal Q_{\mathrm{OU}},-z_i}
\big[
-\frac{\phi_k}{2}
\bigl\| \mathbf y_{i,k}^* - \mathbf X_{i,k}^*\boldsymbol\beta_k \bigr\|_2^2
\big]$
can be evaluated using \eqref{eq:ou_quadratic_expectation}, followed by taking
the expectation with respect to $\tilde\zeta_k$. This gives
\begin{align}
\mathbb E_{\mathcal Q_{\mathrm{OU}},-z_i}
\left[
-\frac{\phi_k}{2}
\left\|
\mathbf y_{i,k}^*
-\mathbf X_{i,k}^*\boldsymbol\beta_k
\right\|_2^2
\right]
\nonumber
&=
-\frac12
\frac{\tilde a_k}{\tilde b_k}
\Bigg[
\sum_{j=1}^{m_i}
\mathbb E_{\mathcal Q_{\mathrm{OU}}}[d_{ij,k}]
\left(
y_{ij}
-\mathbf x_{ij}^T\tilde{\boldsymbol\nu}_k
\right)^2
\\
&\quad
-
2\sum_{j=2}^{m_i}
A_{1,ij,k}
\left(
y_{ij}
-\mathbf x_{ij}^T\tilde{\boldsymbol\nu}_k
\right)
\left(
y_{i,j-1}
-\mathbf x_{i,j-1}^T\tilde{\boldsymbol\nu}_k
\right)
\Bigg]
\nonumber\\
&\quad
-\frac12
\Bigg[
\sum_{j=1}^{m_i}
\mathbb E_{\mathcal Q_{\mathrm{OU}}}[d_{ij,k}]
\mathbf x_{ij}^T
\tilde{\boldsymbol\Omega}_k
\mathbf x_{ij}\\
&\quad
-
2\sum_{j=2}^{m_i}
A_{1,ij,k}
\mathbf x_{ij}^T
\tilde{\boldsymbol\Omega}_k
\mathbf x_{i,j-1}
\Bigg].
\label{eq:z_residual_marginalized}
\end{align}
Substituting this result into the coordinate update gives
\begin{align*}
\tilde q_i^{6}(z_i=k)
&\propto
\exp\Bigg\{
-\frac12
\frac{\tilde a_k}{\tilde b_k}
\Bigg[
\sum_{j=1}^{m_i}
\mathbb E_{\mathcal Q_{\mathrm{OU}}}[d_{ij,k}]
\left(
y_{ij}
-\mathbf x_{ij}^T\tilde{\boldsymbol\nu}_k
\right)^2
\\
&\quad
-
2\sum_{j=2}^{m_i}
A_{1,ij,k}
\left(
y_{ij}
-\mathbf x_{ij}^T\tilde{\boldsymbol\nu}_k
\right)
\left(
y_{i,j-1}
-\mathbf x_{i,j-1}^T\tilde{\boldsymbol\nu}_k
\right)
\Bigg]
\\
&\quad
-\frac12
\Bigg[
\sum_{j=1}^{m_i}
\mathbb E_{\mathcal Q_{\mathrm{OU}}}[d_{ij,k}]
\mathbf x_{ij}^T
\tilde{\boldsymbol\Omega}_k
\mathbf x_{ij}
-
2\sum_{j=2}^{m_i}
A_{1,ij,k}
\mathbf x_{ij}^T
\tilde{\boldsymbol\Omega}_k
\mathbf x_{i,j-1}
\Bigg]
\\
&\quad
+
\frac{m_i}{2}
\left\{
\Psi(\tilde a_k)-\log\tilde b_k
\right\}
-
\frac12
\sum_{j=2}^{m_i}
L_{ij,k}
+
\big(\Psi(\tilde \gamma_{k,1})-\Psi(\tilde \gamma_{k,1}+\tilde \gamma_{k,2})\big)
\\
&\quad
+
\textstyle\sum_{h<k}\big(\Psi(\tilde \gamma_{h,2})-\Psi(\tilde \gamma_{h,1}+\tilde \gamma_{h,2})\big)
\Bigg\}.
\end{align*}
Therefore,
\[
\tilde{\mathcal Q}_i^{6}(z_i)
=
\operatorname{Discrete}
(\tilde\varphi_{i,1},\ldots,\tilde\varphi_{i,K}),
\]
where
\begin{align*}
\tilde\varphi_{i,k}
&\propto
\exp\Bigg\{
-\frac12
\frac{\tilde a_k}{\tilde b_k}
\Bigg[
\sum_{j=1}^{m_i}
\mathbb E_{\mathcal Q_{\mathrm{OU}}}[d_{ij,k}]
\left(
y_{ij}
-\mathbf x_{ij}^T\tilde{\boldsymbol\nu}_k
\right)^2
-
2\sum_{j=2}^{m_i}
A_{1,ij,k}
\left(
y_{ij}
-\mathbf x_{ij}^T\tilde{\boldsymbol\nu}_k
\right)
\left(
y_{i,j-1}
-\mathbf x_{i,j-1}^T\tilde{\boldsymbol\nu}_k
\right)
\Bigg] \\
&\quad
-\frac12
\Bigg[
\sum_{j=1}^{m_i}
\mathbb E_{\mathcal Q_{\mathrm{OU}}}[d_{ij,k}]
\mathbf x_{ij}^T
\tilde{\boldsymbol\Omega}_k
\mathbf x_{ij}
-
2\sum_{j=2}^{m_i}
A_{1,ij,k}
\mathbf x_{ij}^T
\tilde{\boldsymbol\Omega}_k
\mathbf x_{i,j-1}
\Bigg]
\\
&\quad
+
\frac{m_i}{2}
\left\{
\Psi(\tilde a_k)-\log\tilde b_k
\right\}
-
\frac12
\sum_{j=2}^{m_i}
L_{ij,k}
+
\big(\Psi(\tilde \gamma_{k,1})-\Psi(\tilde \gamma_{k,1}+\tilde \gamma_{k,2})\big)
\\
&\quad
+
\textstyle\sum_{h<k}\big(\Psi(\tilde \gamma_{h,2})-\Psi(\tilde \gamma_{h,1}+\tilde \gamma_{h,2})\big)
\Bigg\},
\qquad
k=1,\ldots,K,
\end{align*}
normalized so that $\sum_{k=1}^{K}\tilde\varphi_{i,k}=1$.

\subsection*{B.2 Evidence lower bound}
\label{app:ou:elbo}

The ELBO
$\mathcal L(\mathcal Q)
=
\mathbb E_{\mathcal Q}[\log p(\boldsymbol\theta,\mathbf y)]
-
\mathbb E_{\mathcal Q}[\log q(\boldsymbol\theta)]$
decomposes under the mean-field factorization as
\begin{align*}
\mathcal L_{\mathrm{OU}}
&=
\sum_{k=1}^{K-1}
\left(
\mathbb E_{\mathcal Q_{\mathrm{OU}}}[\log p(v_k)]
-
\mathbb E_{\mathcal Q_{\mathrm{OU}}}[\log \tilde q_k^1(v_k)]
\right)
\\
&\quad+
\sum_{k=1}^{K}
\left(
\mathbb E_{\mathcal Q_{\mathrm{OU}}}
[\log p(\boldsymbol\beta_k,\phi_k\mid\boldsymbol\tau_k)]
-
\mathbb E_{\mathcal Q_{\mathrm{OU}}}
[\log \tilde q_k^2(\boldsymbol\beta_k,\phi_k)]
\right)
\\
&\quad+
\sum_{k=1}^{K}\sum_{j=1}^{D-4}
\left(
\mathbb E_{\mathcal Q_{\mathrm{OU}}}
[\log p(\tau_{k,j}\mid\lambda_k)]
-
\mathbb E_{\mathcal Q_{\mathrm{OU}}}
[\log \tilde q_{k,j}^3(\tau_{k,j})]
\right)
\\
&\quad+
\sum_{k=1}^{K}
\left(
\mathbb E_{\mathcal Q_{\mathrm{OU}}}[\log p(\lambda_k)]
-
\mathbb E_{\mathcal Q_{\mathrm{OU}}}[\log \tilde q_k^4(\lambda_k)]
\right)
\\
&\quad+
\sum_{k=1}^{K}
\left(
\mathbb E_{\mathcal Q_{\mathrm{OU}}}[\log p(\tilde\zeta_k)]
-
\mathbb E_{\mathcal Q_{\mathrm{OU}}}[\log \tilde q_k^5(\tilde\zeta_k)]
\right)
\\
&\quad+
\sum_{i=1}^{n}
\left(
\mathbb E_{\mathcal Q_{\mathrm{OU}}}
[\log p(z_i\mid\mathbf v)]
-
\mathbb E_{\mathcal Q_{\mathrm{OU}}}
[\log \tilde q_i^6(z_i)]
\right)
\\
&\quad+
\sum_{i=1}^{n}
\mathbb E_{\mathcal Q_{\mathrm{OU}}}
[\log p(\mathbf y_i
\mid z_i,\boldsymbol\beta,\boldsymbol\phi,\tilde{\boldsymbol\zeta})].
\end{align*}
We derive each of these seven contributions below.

\subsubsection*{B.2.1 Bound for $v_k$}

The two expectations required for the ELBO contribution of $v_k$ are given by
\begin{align*}
\mathbb E_{\tilde{\mathcal Q}}[\log p(v_k)]
&=\log\Gamma(\alpha+1)-\log\Gamma(\alpha)
+(\alpha-1)\Big(
\Psi(\tilde\gamma_{k,2})
-\Psi(\tilde\gamma_{k,1}+\tilde\gamma_{k,2})
\Big),\\
\mathbb E_{\tilde{\mathcal Q}}[\log \mathcal Q_k^1(v_k)]
&=\log\Gamma(\tilde\gamma_{k,1}+\tilde\gamma_{k,2})
-\log\Gamma(\tilde\gamma_{k,1})
-\log\Gamma(\tilde\gamma_{k,2})\\
&\quad
+(\tilde\gamma_{k,1}-1)\Big(
\Psi(\tilde\gamma_{k,1})
-\Psi(\tilde\gamma_{k,1}+\tilde\gamma_{k,2})
\Big)\\
&\quad
+(\tilde\gamma_{k,2}-1)\Big(
\Psi(\tilde\gamma_{k,2})
-\Psi(\tilde\gamma_{k,1}+\tilde\gamma_{k,2})
\Big).
\end{align*}
Therefore, the ELBO contribution associated with $v_k$ is
\begin{align*}
\mathbb E_{\mathcal Q_{\mathrm{OU}}}[\log p(v_k)]
-
\mathbb E_{\mathcal Q_{\mathrm{OU}}}[\log \tilde q_k^1(v_k)]
&=
\log\Gamma(\alpha+1)-\log\Gamma(\alpha)
-\log\Gamma(\tilde\gamma_{k,1}+\tilde\gamma_{k,2})
\\
&\quad
+\log\Gamma(\tilde\gamma_{k,1})
+\log\Gamma(\tilde\gamma_{k,2})
\\
&\quad
+
(1-\tilde\gamma_{k,1})
\left(
\Psi(\tilde\gamma_{k,1})
-\Psi(\tilde\gamma_{k,1}+\tilde\gamma_{k,2})
\right)
\\
&\quad
+
(\alpha-\tilde\gamma_{k,2})
\left(
\Psi(\tilde\gamma_{k,2})
-\Psi(\tilde\gamma_{k,1}+\tilde\gamma_{k,2})
\right).
\end{align*}

\subsubsection*{B.2.2 Bound for $\boldsymbol\beta_k,\phi_k$}

The two expectations required for the ELBO contribution of $\boldsymbol\beta_k,\phi_k$ are given by
\begin{align*}
\mathbb E_{\tilde{\mathcal Q}}[\log p(\boldsymbol{\beta}_k,\phi_k)]
&=
-\frac{D}{2}\log(2\pi)
+\Big(a_0+\frac{D}{2}-1\Big)
\Big(\Psi(\tilde a_k)-\log\tilde b_k\Big)\\
&\quad
-\frac12
\mathbb E_{\mathcal Q_{\mathrm{OU}}}
[\log|\mathbf C_{\boldsymbol\tau_k}|]
-\frac{1}{2}\Big(
\mathrm{tr}\!\big(\mathbb E_{\tilde{\mathcal Q}}[\mathbf C_{\boldsymbol\tau_k}^{-1}]
\,\tilde{\mathbf\Omega}_k\big)
+\frac{\tilde a_k}{\tilde b_k}
\tilde{\boldsymbol\nu}_k^{\top}
\mathbb E_{\tilde{\mathcal Q}}[\mathbf C_{\boldsymbol\tau_k}^{-1}]
\tilde{\boldsymbol\nu}_k
\Big)\\
&\quad
+a_0\log b_0-\log\Gamma(a_0)-b_0\frac{\tilde a_k}{\tilde b_k},\\[2ex]
\mathbb E_{\tilde{\mathcal Q}}[\log \mathcal Q_k^2(\boldsymbol{\beta}_k,\phi_k)]
&=
-\frac{D}{2}\log(2\pi)
+\Big(\tilde a_k+\frac{D}{2}-1\Big)
\Big(\Psi(\tilde a_k)-\log\tilde b_k\Big)\\
&\quad
-\frac{1}{2}\log|\tilde{\mathbf\Omega}_k|
-\frac{D}{2}
+\tilde a_k\log\tilde b_k
-\log\Gamma(\tilde a_k)
-\tilde a_k,
\end{align*}
where $\mathbb E_{\tilde{\mathcal Q}}[\mathbf C_{\boldsymbol\tau_k}^{-1}]$ is as
given in Appendix~A.1, \eqref{eq:Ctau_expectation}, with tildes on the
variational parameters. Therefore, the ELBO contribution associated with
$\boldsymbol\beta_k,\phi_k$ is
\begin{align*}
\mathbb E_{\tilde{\mathcal Q}}[\log p(\boldsymbol{\beta}_k,\phi_k)]
-\mathbb E_{\tilde{\mathcal Q}}[\log \mathcal Q_k^2(\boldsymbol{\beta}_k,\phi_k)]
&=
-\frac12
\mathbb E_{\mathcal Q_{\mathrm{OU}}}
[\log|\mathbf C_{\boldsymbol\tau_k}|]
+\frac{1}{2}\log|\tilde{\mathbf\Omega}_k|\\
&\quad
-\frac{1}{2}\Big(
\mathrm{tr}\!\big(\mathbb E_{\tilde{\mathcal Q}}[\mathbf C_{\boldsymbol\tau_k}^{-1}]
\,\tilde{\mathbf\Omega}_k\big)
+\frac{\tilde a_k}{\tilde b_k}
\tilde{\boldsymbol\nu}_k^{\top}
\mathbb E_{\tilde{\mathcal Q}}[\mathbf C_{\boldsymbol\tau_k}^{-1}]
\tilde{\boldsymbol\nu}_k
\Big)
+\frac{D}{2}\\
&\quad
+a_0\log b_0-\tilde a_k\log\tilde b_k
-\big(\log\Gamma(a_0)-\log\Gamma(\tilde a_k)\big)\\
&\quad
+(a_0-\tilde a_k)\Psi(\tilde a_k)
-b_0\frac{\tilde a_k}{\tilde b_k}
+\tilde a_k.
\end{align*}

\bigskip
\subsubsection*{B.2.3 Bound for $\tau_{k,j}$}

The two expectations required for the ELBO contribution of $\tau_{k,j}$ are
given by
\begin{align*}
\mathbb E_{\mathcal Q_{\mathrm{OU}}}
[\log p(\tau_{k,j}\mid\lambda_k)]
&=
\Psi(g_0+D-4)-\log\tilde h_k
-
\frac{g_0+D-4}{\tilde h_k}
\mathbb E_{\mathcal Q_{\mathrm{OU}}}[\tau_{k,j}],
\\
\mathbb E_{\mathcal Q_{\mathrm{OU}}}
[\log\tilde q_{k,j}^3(\tau_{k,j})]
&=
\frac14
\log\left(
\frac{\tilde c_{\tau_k}}{\tilde f_{\tau_{k,j}}}
\right)
-
\log\left[
2K_{1/2}
\left(
\sqrt{\tilde c_{\tau_k}\tilde f_{\tau_{k,j}}}
\right)
\right]
\\
&\quad
-\frac12
\mathbb E_{\mathcal Q_{\mathrm{OU}}}[\log\tau_{k,j}]
-\frac12
\left[
\tilde c_{\tau_k}
\mathbb E_{\mathcal Q_{\mathrm{OU}}}[\tau_{k,j}]
+
\tilde f_{\tau_{k,j}}
\mathbb E_{\mathcal Q_{\mathrm{OU}}}[\tau_{k,j}^{-1}]
\right].
\end{align*}
The required moments of $\tau_{k,j}$ are given in
Appendix~B.1.3. Therefore, the ELBO contribution associated with $\tau_{k,j}$ is
\begin{align*}
\mathbb E_{\mathcal Q_{\mathrm{OU}}}
[\log p(\tau_{k,j}\mid\lambda_k)]
-
\mathbb E_{\mathcal Q_{\mathrm{OU}}}
[\log\tilde q_{k,j}^3(\tau_{k,j})]
&=
\Psi(g_0+D-4)-\log\tilde h_k
-
\frac{g_0+D-4}{\tilde h_k}
\mathbb E_{\mathcal Q_{\mathrm{OU}}}[\tau_{k,j}]
\\
&\quad
-\frac14
\log\left(
\frac{\tilde c_{\tau_k}}{\tilde f_{\tau_{k,j}}}
\right)
+
\log\left[
2K_{1/2}
\left(
\sqrt{\tilde c_{\tau_k}\tilde f_{\tau_{k,j}}}
\right)
\right]
\\
&\quad
+\frac12
\mathbb E_{\mathcal Q_{\mathrm{OU}}}[\log\tau_{k,j}]
\\
&\quad
+\frac12
\left[
\tilde c_{\tau_k}
\mathbb E_{\mathcal Q_{\mathrm{OU}}}[\tau_{k,j}]
+
\tilde f_{\tau_{k,j}}
\mathbb E_{\mathcal Q_{\mathrm{OU}}}[\tau_{k,j}^{-1}]
\right].
\end{align*}

\bigskip
\subsubsection*{B.2.4 Bound for $\lambda_k$}
The two expectations required for the ELBO contribution of $\lambda_k$ are
given by
\begin{align*}
\mathbb E_{\mathcal Q_{\mathrm{OU}}}[\log p(\lambda_k)]
&=
g_0\log h_0-\log\Gamma(g_0)
\\
&\quad
+(g_0-1)
\left[
\Psi(g_0+D-4)-\log\tilde h_k
\right]
-h_0\frac{g_0+D-4}{\tilde h_k},
\\
\mathbb E_{\mathcal Q_{\mathrm{OU}}}[\log\tilde q_k^4(\lambda_k)]
&=
(g_0+D-4)\log\tilde h_k
-\log\Gamma(g_0+D-4)
\\
&\quad
+(g_0+D-5)
\left[
\Psi(g_0+D-4)-\log\tilde h_k
\right]
-(g_0+D-4).
\end{align*}
Therefore, the ELBO contribution associated with
$\lambda_k$ is
\begin{align*}
\mathbb E_{\mathcal Q_{\mathrm{OU}}}[\log p(\lambda_k)]
-
\mathbb E_{\mathcal Q_{\mathrm{OU}}}[\log\tilde q_k^4(\lambda_k)]
&=
g_0(\log h_0 - \log \tilde h_k)
-
(D-4)\Psi(g_0+D-4)
\\
&\quad
+\log\Gamma(g_0+D-4)
- \log\Gamma(g_0) + g_0+D-4 
\\
&\quad
- h_0\frac{g_0+D-4}{\tilde h_k}.
\end{align*}

\bigskip
\subsubsection*{B.2.5 Bound for $\tilde\zeta_k$}

The two expectations required for the ELBO contribution of
$\tilde\zeta_k$ are given by
\begin{align*}
\mathbb E_{\mathcal Q_{\mathrm{OU}}}
[\log p(\tilde\zeta_k)]
&=
p_0\log q_0
-\log\Gamma(p_0)
+(p_0-1)
\left[
\Psi(\tilde r_k)-\log\tilde s_k
\right]
-q_0\frac{\tilde r_k}{\tilde s_k},
\\
\mathbb E_{\mathcal Q_{\mathrm{OU}}}
[\log\tilde q_k^5(\tilde\zeta_k)]
&=
\tilde r_k\log\tilde s_k
-\log\Gamma(\tilde r_k)
+(\tilde r_k-1)
\left[
\Psi(\tilde r_k)-\log\tilde s_k
\right]
-\tilde r_k.
\end{align*}
Therefore, the ELBO contribution associated with $\tilde\zeta_k$ is
\begin{align*}
\mathbb E_{\mathcal Q_{\mathrm{OU}}}
[\log p(\tilde\zeta_k)]
-
\mathbb E_{\mathcal Q_{\mathrm{OU}}}
[\log\tilde q_k^5(\tilde\zeta_k)]
&=
p_0\log q_0
-\tilde r_k\log\tilde s_k
-\log\Gamma(p_0)
+\log\Gamma(\tilde r_k)
\\
&\quad
+(p_0-\tilde r_k)
\left[
\Psi(\tilde r_k)-\log\tilde s_k
\right]
+\tilde r_k
-q_0\frac{\tilde r_k}{\tilde s_k}.
\end{align*}

\bigskip
\subsubsection*{B.2.6 Bound for $z_i$}

The two expectations required for the ELBO contribution of $z_i$ are given by
\begin{align*}
\mathbb E_{\mathcal Q_{\mathrm{OU}}}
[\log p(z_i\mid\mathbf v)]
&=
\sum_{k=1}^{K-1}
\tilde\varphi_{i,k}
\left[
\Psi(\tilde\gamma_{k,1})
-\Psi(\tilde\gamma_{k,1}+\tilde\gamma_{k,2})
\right]
\\
&\quad
+
\sum_{k=1}^{K-1}
\left(
\sum_{h=k+1}^{K}\tilde\varphi_{i,h}
\right)
\left[
\Psi(\tilde\gamma_{k,2})
-\Psi(\tilde\gamma_{k,1}+\tilde\gamma_{k,2})
\right],
\\
\mathbb E_{\mathcal Q_{\mathrm{OU}}}
[\log\tilde q_i^6(z_i)]
&=
\sum_{k=1}^{K}
\tilde\varphi_{i,k}\log\tilde\varphi_{i,k}.
\end{align*}
Therefore, the ELBO contribution associated with
$z_i$ is
\begin{align*}
\mathbb E_{\mathcal Q_{\mathrm{OU}}}
[\log p(z_i\mid\mathbf v)]
-
\mathbb E_{\mathcal Q_{\mathrm{OU}}}
[\log\tilde q_i^6(z_i)]
&=
\sum_{k=1}^{K-1}
\tilde\varphi_{i,k}
\left[
\Psi(\tilde\gamma_{k,1})
-\Psi(\tilde\gamma_{k,1}+\tilde\gamma_{k,2})
\right]
\\
&\quad
+
\sum_{k=1}^{K-1}
\left(
\sum_{h=k+1}^{K}\tilde\varphi_{i,h}
\right)
\left[
\Psi(\tilde\gamma_{k,2})
-\Psi(\tilde\gamma_{k,1}+\tilde\gamma_{k,2})
\right]\\
&\quad
-
\sum_{k=1}^{K}
\tilde\varphi_{i,k}\log\tilde\varphi_{i,k}.
\end{align*}

\bigskip
\subsubsection*{B.2.7 Bound for $\mathbf y_i$}

Since $\mathbf y_i$ is observed, this term has no matching variational
factor. Its ELBO contribution is
\begin{align*}
\mathbb E_{\mathcal Q_{\mathrm{OU}}}
[\log p(\mathbf y_i
\mid z_i,\boldsymbol\beta,\boldsymbol\phi,\tilde{\boldsymbol\zeta})]
&=
\sum_{k=1}^{K}
\tilde\varphi_{i,k}
\Bigg[
-\frac{m_i}{2}\log(2\pi)
+
\frac{m_i}{2}
\left(
\Psi(\tilde a_k)-\log\tilde b_k
\right)
\\
&\quad
+
\mathbb E_{\mathcal Q_{\mathrm{OU}},-z_i}
\left[
-\frac{\phi_k}{2}
\left\|
\mathbf y_{i,k}^*
-\mathbf X_{i,k}^*\boldsymbol\beta_k
\right\|_2^2
\right]
-
\frac12
\sum_{j=2}^{m_i}
L_{ij,k}
\Bigg],
\end{align*}
where the residual expectation is given in
\eqref{eq:z_residual_marginalized}.

Substituting the seven contributions above into the decomposition
at the beginning of this section gives the complete ELBO:
\begin{align*}
\mathcal L_{\mathrm{OU}}
&=
\sum_{k=1}^{K-1}
\Bigg[
\log\Gamma(\alpha+1)
-\log\Gamma(\alpha)
-\log\Gamma(\tilde\gamma_{k,1}+\tilde\gamma_{k,2})
+\log\Gamma(\tilde\gamma_{k,1})
+\log\Gamma(\tilde\gamma_{k,2})
\\
&\quad
+
(1-\tilde\gamma_{k,1})
\left(
\Psi(\tilde\gamma_{k,1})
-\Psi(\tilde\gamma_{k,1}+\tilde\gamma_{k,2})
\right)
+
(\alpha-\tilde\gamma_{k,2})
\left(
\Psi(\tilde\gamma_{k,2})
-\Psi(\tilde\gamma_{k,1}+\tilde\gamma_{k,2})
\right)
\Bigg]
\\[1.5ex]
&\quad+
\sum_{k=1}^{K}
\Bigg[
a_0\log b_0
-\tilde a_k\log\tilde b_k
-\log\Gamma(a_0)
+\log\Gamma(\tilde a_k)
\\
&\quad
-\frac12
\left(
4\log\rho
+
\sum_{j=1}^{D-4}
\mathbb E_{\mathcal Q_{\mathrm{OU}}}
[\log\tau_{k,j}]
\right)
+
\frac12\log|\tilde{\boldsymbol\Omega}_k|
+
(a_0-\tilde a_k)
\Psi(\tilde a_k)
\\
&\quad
+
\frac D2
+
\tilde a_k
-
b_0\frac{\tilde a_k}{\tilde b_k}
-\frac12
\Bigg\{
\frac{\tilde a_k}{\tilde b_k}
\tilde{\boldsymbol\nu}_k^T
\mathbb E_{\mathcal Q_{\mathrm{OU}}}
\left[
\mathbf C_{\boldsymbol\tau_k}^{-1}
\right]
\tilde{\boldsymbol\nu}_k
+
\operatorname{tr}
\left(
\mathbb E_{\mathcal Q_{\mathrm{OU}}}
\left[
\mathbf C_{\boldsymbol\tau_k}^{-1}
\right]
\tilde{\boldsymbol\Omega}_k
\right)
\Bigg\}
\Bigg]
\\[1.5ex]
&\quad+
\sum_{k=1}^{K}
\sum_{j=1}^{D-4}
\Bigg[
\Psi(g_0+D-4)
-\log\tilde h_k
-
\frac{g_0+D-4}{\tilde h_k}
\mathbb E_{\mathcal Q_{\mathrm{OU}}}[\tau_{k,j}]
\\
&\quad
-\frac14
\log
\left(
\frac{\tilde c_{\tau_k}}
{\tilde f_{\tau_{k,j}}}
\right)
+
\log
\left[
2K_{1/2}
\left(
\sqrt{
\tilde c_{\tau_k}
\tilde f_{\tau_{k,j}}
}
\right)
\right]
+
\frac12
\mathbb E_{\mathcal Q_{\mathrm{OU}}}
[\log\tau_{k,j}]
\\
&\quad
+
\frac12
\left\{
\tilde c_{\tau_k}
\mathbb E_{\mathcal Q_{\mathrm{OU}}}[\tau_{k,j}]
+
\tilde f_{\tau_{k,j}}
\mathbb E_{\mathcal Q_{\mathrm{OU}}}[\tau_{k,j}^{-1}]
\right\}
\Bigg]
\\[1.5ex]
&\quad+
\sum_{k=1}^{K}
\Bigg[
g_0(\log h_0 - \log \tilde h_k)
-
(D-4)\Psi(g_0+D-4)
+\log\Gamma(g_0+D-4)
- \log\Gamma(g_0)
\\
&\quad + g_0+D-4 
- h_0\frac{g_0+D-4}{\tilde h_k}
\Bigg]
\\[1.5ex]
&\quad+
\sum_{k=1}^{K}
\Bigg[
p_0\log q_0
-\tilde r_k\log\tilde s_k
-\log\Gamma(p_0)
+\log\Gamma(\tilde r_k)
+
(p_0-\tilde r_k)
\left[
\Psi(\tilde r_k)-\log\tilde s_k
\right]
+
\tilde r_k
-
q_0\frac{\tilde r_k}{\tilde s_k}
\Bigg]
\\[1.5ex]
&\quad+
\sum_{i=1}^{n}
\Bigg[
\sum_{k=1}^{K-1}
\tilde\varphi_{i,k}
\left\{
\Psi(\tilde\gamma_{k,1})
-\Psi(\tilde\gamma_{k,1}+\tilde\gamma_{k,2})
\right\}
\\
&\quad
+
\sum_{k=1}^{K-1}
\left(
\sum_{h=k+1}^{K}
\tilde\varphi_{i,h}
\right)
\left\{
\Psi(\tilde\gamma_{k,2})
-\Psi(\tilde\gamma_{k,1}+\tilde\gamma_{k,2})
\right\}
-
\sum_{k=1}^{K}
\tilde\varphi_{i,k}
\log\tilde\varphi_{i,k}
\Bigg]
\\[1.5ex]
&\quad+
\sum_{i=1}^{n}
\sum_{k=1}^{K}
\tilde\varphi_{i,k}
\Bigg[
-\frac{m_i}{2}\log(2\pi)
+
\frac{m_i}{2}
\left(
\Psi(\tilde a_k)-\log\tilde b_k
\right)
-
\frac12
\sum_{j=2}^{m_i}
L_{ij,k}
\\
&\quad
-
\frac12
\frac{\tilde a_k}{\tilde b_k}
\Bigg\{
\sum_{j=1}^{m_i}
\mathbb E_{\mathcal Q_{\mathrm{OU}}}[d_{ij,k}]
\left(
y_{ij}
-\mathbf x_{ij}^T\tilde{\boldsymbol\nu}_k
\right)^2
-
2
\sum_{j=2}^{m_i}
A_{1,ij,k}
\left(
y_{ij}
-\mathbf x_{ij}^T\tilde{\boldsymbol\nu}_k
\right)
\left(
y_{i,j-1}
-\mathbf x_{i,j-1}^T\tilde{\boldsymbol\nu}_k
\right)
\Bigg\}
\\
&\quad
-
\frac12
\Bigg\{
\sum_{j=1}^{m_i}
\mathbb E_{\mathcal Q_{\mathrm{OU}}}[d_{ij,k}]
\mathbf x_{ij}^T
\tilde{\boldsymbol\Omega}_k
\mathbf x_{ij}
-
2
\sum_{j=2}^{m_i}
A_{1,ij,k}
\mathbf x_{ij}^T
\tilde{\boldsymbol\Omega}_k
\mathbf x_{i,j-1}
\Bigg\}
\Bigg].
\end{align*}

\bibliographystyle{apalike}
\setcitestyle{authoryear}
\bibliography{arxiv_ref}

\end{document}